\documentclass[rmp,aps,onecolumn]{revtex4-1}

\usepackage[caption=false]{subfig}
\usepackage{graphicx}
\usepackage{dcolumn}
\usepackage{bm}
\usepackage{xcolor}
\usepackage{braket}
\usepackage{mathtools}

\usepackage{tikz} 
\usetikzlibrary{positioning}
\usepackage{ dsfont } 

\usepackage{subfiles}

\newcommand{\lb}{\ensuremath{\overline{\lambda}}}
\newcommand{\zb}{\ensuremath{\overline{z}}}
\newcommand{\wf}{\ensuremath{D^\psi}}
\newcommand{\avwf}{\ensuremath{\overline{D}^\psi}}

\begin{document}

\title{The Fibonacci quasicrystal: case study of hidden dimensions and multifractality }

\author{Anuradha Jagannathan}
\affiliation{
Laboratoire de Physique des Solides, B\^at. 510, Universit\'{e} Paris-Saclay, 91405 Orsay, France
}%

\date{\today}

\begin{abstract}
 The distinctive electronic properties of quasicrystals stem from their long range structural order, with invariance under rotations and under discrete scale change, but without translational invariance. d-dimensional quasicrystals can be described in terms of lattices of higher dimension $D>d$, and many of their properties can be simply derived from analyses that take into account the extra ``hidden" dimensions. In particular, as recent theoretical and experimental studies have shown, quasicrystals can have topological properties inherited from the parent crystals. These properties are discussed here for the simplest of quasicrystals, the 1D Fibonacci chain. The Fibonacci noninteracting tight-binding Hamiltonians are characterized by multifractality of spectrum and states, which is manifested in many of its physical properties, notably in transport. Perturbations due to disorder and re-entrance phenomena are described, along with the crossover to strong Anderson localization. Perturbations due to boundary conditions also give information on the spatial and topological electronic properties, as is shown for the superconducting proximity effect. Related models including phonon and mixed Fibonacci models are discussed, as well as generalizations to other quasiperiodic chains and higher dimensional extensions. Interacting quasiperiodic systems and the case for many body localization are briefly discussed. Some experimental realizations of the 1D quasicrystal and their potential applications are described.

\end{abstract}

\maketitle
\tableofcontents

	\section{Introduction to the Fibonacci chain}
The Fibonacci chain is a one-dimensional quasiperiodic structure that is closely related to the three dimensional icosahedral quasicrystals discovered by Shechtman et al \cite{schechtman84}.  The study of electronic properties of quasicrystals thus logically begins with the study of electrons in a 1D Fibonacci chain. This ``fruitfly" of quasiperiodic studies is not only a theoretical construct but also can be experimentally realized in artificial atom chains or in heterostructures made from quasi-2D semiconducting layers, to give just two examples. Tight-binding Hamiltonians for the Fibonacci chain have been extensively investigated. Properties of the energy spectra and critical states of the chain have been studied using a variety of methods, exact solutions, perturbation theory and numerical analysis. Aperiodic Schr\"odinger operators in general and quasiperiodic systems in particular are an active subject of mathematical physics  \cite{damanik2014}, with a review devoted exclusively to the Fibonacci model \cite{damanikreview}. 

Quasiperiodic Hamiltonians are of growing interest for their nontrivial topological properties \cite{PhysRevLett.110.076403verbin, PhysRevB.100.085119huang,PhysRevLett.121.126401huang}. Like the well-known Aubry-André-Harper (AAH) model, with which it is often compared, the one-dimensional Fibonacci quasicrystal  possesses a higher dimensional ``parent" system, from which it inherits topological characteristics. Thus, in the Fibonacci chain, there appear topologically-protected boundary states equivalent to the edge states of the two-dimensional Integer Quantum Hall Effect \cite{PhysRevB.91.064201zilberberg}. Unlike the quasiperiodic AAH model, which is critical at a single point in its parameter space, the Fibonacci models which we discuss here are critical for $all$ values of the strength of the quasiperiodic modulation.

Multifractal states are omnipresent in the phase diagram of quasicrystals. Some recent works have given exact solutions for multifractal states in quasicrystals and, in particular for the $E=0$ central state of the off-diagonal model  \cite{mace2017critical,kaluginkatz}. To our knowledge there have not been attempts to solve for arbitrary states on the Fibonacci chain using exact methods along the lines of the Bethe Ansatz-based analysis presented in \cite{wiegmann} for the quasiperiodic AAH model. Fortunately, much valuable information can be obtained about the critical wavefunctions of the Fibonacci chain using a perturbative renormalization group treatment \cite{mace2016fractal}. Some of the physical manifestations of critical states have been discussed with relation to thermodynamic quantities such as in interesting proximity effects \cite{rai2019}, or in dynamical phenomena including growth of entanglement \cite{scipost}.

Studies of hyperuniformity in complex systems including quasicrystals and glasses have have received much recent interest, as a way to characterize spatial fluctuations in complex structures \cite{torquato}. The Fibonacci chain, as indeed all quasicrystals, possesses the hyperuniformity property \cite{hyperbaake}. This should have consequences for the electronic wavefunctions, since weaker geometric fluctuations can be expected to favor delocalization. The hyperuniformity should thus lead to distinctive spectral characteristics in the Fibonacci chain as compared, for example, to generic aperiodic chains not having this property. 

Some important conceptual, and experimentally pertinent, questions concern the role of perturbations. One can ask what the effects of disorder are in a quasicrystal, and how critical states are affected by randomness. Recent work on disorder and the approach to strong localization in the Fibonacci quasicrystal, showed re-entrant phenomena and the existence of a new crossover exponent \cite{tarziaEPJ}. These works concern single particle properties. Interacting quasiperiodic systems have been considered in a number of studies. In particular, many body localization (MBL) due to quasiperiodic potentials \cite{PhysRevB.87.134202iyer} has been an active topic of recent research. One of the questions addressed concerns differences in the critical behavior, if any, from MBL due to random potentials \cite{PhysRevLett.119.075702khemani}. It is becoming possible to study a number of models experimentally with cold atoms in optical potentials, possibly extending to realizations of generalized Fibonacci problems  \cite{PhysRevA.92.063426paramesh}. These could allow a new generation of experimental studies of interacting quasicrystals.

This brief outline of some of the interesting and not yet fully understood aspects of the Fibonacci model, will hopefully convince the readers that this ``toy model" is worth further study both by theory and experiments. This review voluntarily restricts the discussion to this simplest one dimensional case, allowing a reasonably detailed description of methods and presenting a state-of-the-art which should be useful to those wishing to work on these or related systems. The outline of the review is as follows: It begins with certain important structural properties of Fibonacci chains included here for completeness, since they are essential for the ensuing discussions of the tight-binding Hamiltonians.  Sec.\ref{sec.geometry} therefore focuses on geometrical aspects, introduces useful notations and properties of the Fibonacci chain and its approximants.  Sec.\ref{sec.overview} then presents the basic tight-binding models along with the principal spectral properties of the diagonal and off-diagonal Fibonacci models. Some important exact results are introduced in this section, in particular the well-known gap labeling theorem and an exact renormalization group (the trace map) method. Finally, an exact solution for the multifractal wave-function in the diagonal model for $E=0$ is given. Sec.\ref{sec.approximate} takes up approximate methods which have proven extremely useful, yielding many valuable insights into spectral and wave-function properties. In particular, this section introduces the perturbative renormalization group technique and its qualitative predictions. Sec.\ref{sec.rghopping} takes up the off-diagonal model in more detail, in order to illustrate the use of the perturbative RG technique in obtaining quantitative description of spectrum and critical wave-functions. Sec.\ref{sec.dynamics} is a first step to discussing physical observables. It introduces wave packet dynamics, time dependence of the correlation function and log-periodic behavior.  Sec.\ref{sec.transport} presents some results for transmission coefficient and chain conductances using Landauer formalism and Kubo-Greenwood approach. Sec.\ref{sec.perturbations} discusses effects due to disorder in the FC, and describes re-entrant phenomena and crossover to strong Anderson localization. The role of boundary conditions and the proximity effect when the FC is coupled to a superconductor are described. Sec.\ref{sec.mixed} presents some important  generalizations of the simple models hitherto considered, including phonon modes on the Fibonacci chain. The so-called ``mixed" models that combine diagonal and off-diagonal quasiperiodic modulations are discussed in this section. Sec.\ref{sec.other} briefly outlines a few other frequently encountered 1D quasiperiodic related to the Fibonacci chain. This section also describes some extensions to higher dimensions. Sec.\ref{subsec.interactions} lists results for interacting quasiperiodic systems. Finally, Sec.\ref{sec.experimental} gives some examples of experimental realizations of Fibonacci models in electronic, cold atom, phononic and photonic systems. Sec.\ref{sec.conclusions} concludes with a summary and outlook.

\section{Geometric properties of the Fibonacci chain}\label{sec.geometry}
The Fibonacci chain is a 1D quasicrystal according to the revised definition of the IUCr \cite{iucr}. The definition states that a quasicrystal, like a periodic crystal, is a material having a sharp diffraction pattern composed of Bragg peaks. The indexing of peaks proceeds as for crystals however it requires a set of $D$ reciprocal lattice vectors where $D$ is larger than the spatial dimension $d$. This distinguishes the quasicrystal from a periodic lattice, where the number of reciprocal lattice vectors is equal to $d$. This section describes how to generate the Fibonacci chain and reviews some of its geometrical and structural properties. We will introduce several methods, each of which is helpful in its way for a better understanding of electronic properties in this system.

\subsection{Substitution method} \label{subsec.inflation}
The substitution method explicitly introduces the notion of scale invariance of the quasicrystal, later used in the renormalization group transformation. The Fibonacci substitution rule, $\sigma$, acts on the two letters $A$ and $B$ and transforms these as follows:
\begin{equation}
\label{eq:FiboSub}
	\sigma: 
	\begin{cases}
		A \to AB\\
		B \to A.
	\end{cases}
\end{equation}
Letting the substitution act repeatedly on the letter $B$ generates a sequence of words $C_n = \sigma^n(B)$ of increasing length, as shown for the first few members in Table 1. These chains are finite approximants of the Fibonacci chain, which is obtained in the limit $n\rightarrow \infty$.  It is easy to see that the lengths of the words are equal to the Fibonacci numbers $F_n$, defined by the recursion relation $F_n=F_{n-1}+F_{n-2}$ with $F_0=F_1=1$. The ratio of two consecutive Fibonacci numbers tends to the golden mean as $n\rightarrow \infty$, 
\begin{eqnarray}
 \frac{F_{n-1}}{F_{n-2}} &=& \tau_n \nonumber \\ \lim_{n\rightarrow\infty} \tau_n&=& \tau   
\end{eqnarray}
where $\tau_n$ are the rational approximants of the golden mean $\tau=(1+\sqrt{5})/2$. The lengths of the chains are given by $F_n\sim \tau^n$ in the large $n$ limit.

\begin{center}
    \begin{tabular}{|clc|}
\hline 
   $n$  & $C_n$& $F_n$\\
   \hline 
   0 & B & 1 \\
   1 & A & 1 \\
   2 & AB & 2 \\
   3 & ABA & 3\\
   4 & ABAAB& 5 \\
   5 & ABAABABA& 8 \\
   6 & ABAABABAABAAB& 13 \\
   \hline
\end{tabular}
\vskip 0.5cm
{\small{Table 1. The first six approximants built using the substitution $\sigma$ (defined in Eq.\ref{eq:FiboSub})}}
\end{center}

\begin{figure}
		\includegraphics[width=0.4\textwidth]{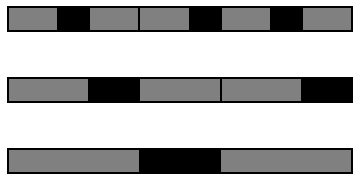}
\caption{Illustration of inflation transformations, going from the $C_5$ (top) to $C_4$ (middle) to $C_3$ (bottom) chain. A (resp. B) tiles are shown by grey (resp. black) rectangles}
\label{fig.inflation}
\end{figure}

\bigskip
\noindent
{\bf{Inflation/deflation of tiles}} The substitution method shows the hierarchical relations between the chains and suggests that problems on the chain could be tackled using renormalization group methods.  Consider a 1D tiling of A and B tiles such that the ratio of their lengths $l_A/l_B=\tau$. The approximant chains $C_n$ correspond to a series of finite tilings, which can be transformed into one another by so-called inflation and deflation operations.  Using the substitution \ref{eq:FiboSub} in reverse, one goes from a chain of $F_n$ tiles to a chain of $F_{n-1}$ tiles. Note that this corresponds to a ``site decimation" process which eliminates a certain subset of sites. Rescaling all the tiles by a factor of $\tau_n$ restores the length of the chain to its original value, as illustrated in Fig.\ref{fig.inflation}.  The infinite chain is invariant under inflation/deflation -- i.e. the FC has a discrete scale invariance. 
 
\bigskip
\noindent
{\bf{Concatenation}} From Fig.\ref{fig.inflation} (or Table 1) one sees that the $n$th chain $C_n$ can be obtained by the concatenation of two shorter chains $C_{n-1}$ and $C_{n-2}$. This property will be useful later for the transfer matrix method \ref{subsec.transfer}. Repeating this operation, one obtains a relation between the $n$th chain and $n-2$th and $n-3$th chains as follows :
\begin{align}
C_{n}  = C_{n-2}\oplus C_{n-3}\oplus C_{n-2}  
\end{align}
where the symbol $\oplus$ here denotes concatenation (joining the chains in the specified order from left to right). This recursive construction holds also for the energy spectrum in perturbative RG, as we will later see in Sec.\ref{sec.rghopping}.

{\it{Inflation matrix}}
Let $N^{(n)}_A$ and $N^{(n)}_B$ be the number of occurrences of A and B in an approximant chain $C_n$. From the 
substitution rule given in Eq.\ref{eq:FiboSub}, it is easy to see that $N^{(n+1)}_A = N^{(n)}_A + N^{(n)}_B $ and
$N^{(n+1)}_B = N^{(n)}_A$,  
with the initial condition $N^{(0)}_A=0$ and $N^{(0)}_B=1$. This relation can be put in a matrix form, 
\begin{equation}
\begin{bmatrix}
N^{(n+1)}_A \\
N^{(n+1)}_B
\end{bmatrix} = 
	\begin{bmatrix}
		1 & 1  \\
		1 & 0  
	\end{bmatrix} 
	\begin{bmatrix}
N^{(n)}_A \\
N^{(n)}_B
\end{bmatrix}
	\label{eq.substmatrix}
	\end{equation}
where the $2\times 2$ matrix is called the substitution matrix $M$.
The eigenvalues of $M$ are $\lambda_1 = \tau$, and $\lambda_2 =-\tau^{-1}$. The corresponding eigenvectors are $\{\tau, 1\}$ and $\{1, -\tau\}$. The first eigenvector gives the relative frequencies of the A- and B-tiles, by virtue of the Perron-Frobenius theorem. The ratio $N_A/N_B$ tends to $\tau$ when $n\rightarrow\infty$.
For more details on symbolic substitutions, see \cite{baake_grimm_2013}. 

At this point, it is useful to make a brief digression relating to other types of self-similar binary chains generated by the substitution method. 
\begin{enumerate}
\item The so-called silver mean chain can be obtained from repeated application of the rule 
\begin{equation}
\label{eq:silverSub}
	\sigma_{Ag}: 
	\begin{cases}
		A \to AAB\\
		B \to A.
	\end{cases}
\end{equation}
It can be easily checked, by writing the inflation matrix for this case, that the eigenvalues satisfy the equation $\lambda^2-2\lambda-1=0$. The Perron-Frobenius eigenvalue in this case is $\lambda_1=(\sqrt{2}+1)$. This type of rule can be generalized to yield a series of so-called ``metallic mean" chains, having the substitution rule $B\rightarrow A, A \rightarrow A^nB$ ($n\geq 1$). It is important to note that number theoretic properties enter crucially in aperiodic systems and lead to very different spatial and physical characteristics. The two chains described above are both based on irrational numbers which are Pisot numbers (a Pisot number is the root $\alpha$ of an $n$th degree monic polynomial equation with integer coefficients, such that $\alpha$ is greater than 1, while all other roots are of modulus less than 1). 

\item Consider next the substitution rule $B\rightarrow A, A \rightarrow ABBB$, giving rise to a self-similar aperiodic structure which is however $not$ a quasicrystal. This structure is non-Pisot  --  it can be checked that the two eigenvalues of the inflation $\lambda=(1\pm\sqrt{13})/2$, which are both greater than 1 in absolute value.  The spatial properties of this chain are very different from those of the golden and silver mean quasicrystals. In particular, the diffraction pattern of such a chain does not have Bragg peaks \cite{baake2019}, as we now explain, by considering the nature of geometric fluctuations.
\end{enumerate}

\bigskip
\noindent
{\bf{Fluctuations of geometry}} Consider a subsystem of $N$ letters from the infinite Fibonacci chain. For values of $N$ which are not Fibonacci numbers the number of A's and B's in the sample $\delta_N=N_A-N_B$ fluctuate around the mean value $\overline{\delta}$. The behavior of $\delta_N$ for large $N$ is described in terms of $\eta$, the so-called ``wandering exponent" defined  \cite{luckwandering} for one-dimensional chain as
\begin{eqnarray}
\delta_N-\overline{\delta} \sim N^{\eta}
\end{eqnarray}
In periodic systems, fluctuations are sub-extensive (are due to boundary effects) and $\eta=0$. 
Fluctuations are similarly negligible in the thermodynamic limit for Pisot aperiodic chains, including the Fibonacci chain. For these structures $\eta=0$, and as a consequence their diffraction spectrum is pure point, i.e. consists solely of Bragg peaks \cite{PhysRevB.45.176luckpisot}.\footnote{This result holds as well in higher dimensions of deterministic tilings such as the well-known Penrose 2D and 3D tilings, the Ammann-Beenker tiling etc.}  

It is instructive to compare this with geometry fluctuations in random systems. Consider a random sequence of the letters A and B for some fixed probability $p_A$ and $p_B=1-p_A$ of the letters. The law of large numbers leads to the exponent $\eta=1/2$ in this case. The huge fluctuations, divergent in the thermodynamic limit, reflect the fact that there exist rare regions in which the number of A's (for example) vastly exceeds the number of B's. For non-Pisot systems such as the ``3B" aperiodic structure defined earlier, fluctuations diverge with the system size, and the wandering exponent is given by
$\eta = \ln\vert\lambda_2\vert/\ln\lambda_1$ \cite{PhysRevB.45.176luckpisot,luckwandering}. 

This property of bounded geometrical fluctuations of the quasicrystal is the principal reason for electronic states in quasicrystals being relatively more extended, as compared with, for example, the critical states at the metal-insulator transition in disordered structures. This is also related to the hyperuniformity property, discussed in the subsection on the structure factor below. 

		\begin{figure*}
\includegraphics[width=0.5\textwidth]{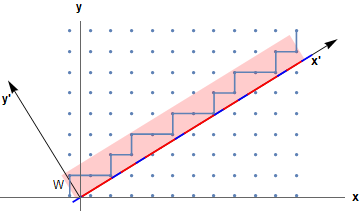}
\caption{Schema of the cut-and-project method. Selected points (joined by a broken line) of a 2D square lattice are projected onto the $x'$ axis, giving the binary quasiperiodic sequence of short red (light grey) and long blue (dark grey) tiles. The infinite selection strip $\mathcal{S}$ is colored red (grey).  }
\label{fig.cutproj}
\end{figure*}

\subsection{Higher dimensional representation of the FC}\label{subsec.cutproject}
Quasicrystals can be generated by projection from a higher dimensional periodic lattice by the cut and project method as we illustrate here for the in the case of the Fibonacci chain. 

{\subsubsection{Cut-and-Project method}} The parent system is a 2D square lattice, and the quasicrystal is obtained by projection onto the physical axis called $x'$ in Fig.\ref{fig.cutproj}. To be selected, a point must lie within the red strip $\mathcal{S}$ of slope given by
\begin{eqnarray}
\tan\theta=\omega
\end{eqnarray}
where the notation $\omega=1/\tau=(\sqrt{5}-1)/2$ is introduced for convenience. The width of the strip is chosen to span one unit cell of the 2d lattice. Upon projection onto the $x'$ axis (the physical axis), horizontal bonds and vertical bonds project onto the red and blue intervals respectively. The $y'$ axis is called the perpendicular (or internal) space.

Expressing all lengths in units of the square lattice parameter $a$, vertices of the square lattice are located at $\vec{R}_{mn}=m\vec{x} + n\vec{y}$ where $m$ and $n$ are integers. The $x'$-coordinate and the coordinate along the perpendicular direction $y'$ are given by
\begin{align}
x'= m \cos\theta + n \sin\theta \nonumber \\
y'= -m \sin\theta + n \cos\theta 
\label{eq.projections}
\end{align}
upto an overall shift. To be selected, the point must satisfy the condition $0\leq y'< W$, where $W=\sin\theta+\cos\theta$ is the cross-section of the strip $\mathcal{S}$. 

After projection on the $x'$ axis, the spacing between nearest neighbors can have two values:  $\cos\theta = 1/\sqrt{1+\omega^2}$  or  $\sin\theta =\omega/\sqrt{1+\omega^2}$, corresponding to the lengths of A and B tiles of the previous section.

\bigskip
\noindent
{\bf{Translational symmetries}} 
The origin of the $x'y'$ axes is arbitrary: i.e. shifting the strip perpendicularly to itself does not result in a new quasiperiodic structure. This statement requires a clarification of what is meant by equivalence between quasiperiodic tilings. Tilings are said to be equivalent (or locally isomorphic) if the same sequences of tiles -- which may be of arbitrarily large size -- can be found in both structures, and with the same frequency of occurrence. To say it differently, one can make the two tilings overlap out to arbitrarily large distances, by suitably translating one with respect to the other.

\begin{figure*}
\includegraphics[width=0.5\textwidth]{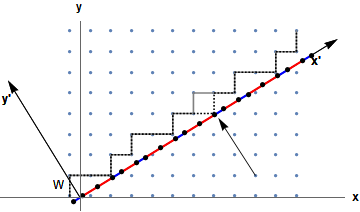}
\caption{Illustration of a phason flip or local rearrangement of the tiling when the lattice is slightly displaced with respect to the selection strip. The old (new) tilings are obtained by projecting the continuous(dashed) lines. The arrow shows the location of the phason flip where a long-short tile ordering has become an short-long ordering.}
\label{fig.phasonflip}
\end{figure*}

The higher dimensional representation of the FC shows that the structure has a translational symmetry with respect to displacements parallel to the physical space, but also in the perpendicular direction. This leads theoretically to the possibility of having two kinds of Goldstone modes : phonons and phasons. Phonons are 1D vibrational modes in the usual sense. To visualize phason modes, one can look at Fig.\ref{fig.phasonflip} which shows how the projected structure would change if the lattice were to be slightly displaced in the direction perpendicular to the strip. One of the vertices $(m,n)$ in the figure moves out of the strip while, simultaneously, the vertex $(m-1,n+1)$ enters the strip. The net result is a small discontinuous jump of one site, while the other points in its neighborhood remain unaffected. This so-called phason jump corresponds to exchanging the A and B tiles around a given vertex, AB $\rightarrow$ BA. The shift produces a new chain structure which is equivalent to the old one. In accepted terminology, a phason mode is a coherent excitation of the perfect Fibonacci chains, corresponding to long wavelength fluctuations. In contrast, when phason flips are introduced in a random uncorrelated fashion all along a chain, this would give rise to a geometrically disordered chain. In practice, however, spontaneous phason flips presumably have a significant energy cost, and are unlikely to be excited at low temperatures. 

\bigskip
\noindent
{\bf{Approximants} }
Periodic approximants of the FC are generated by taking a rational slope for the strip $S$. If the slope is chosen to be the ratio of two successive Fibonacci numbers
\begin{align*}
    \tan\theta_n = F_{n-2} /F_{n-1}
\end{align*}
then the projected chain has a repeating structure consisting of $F_{n-1}$ and $F_{n-2}$ tiles of type A and B respectively, and the total number of tiles is $F_n$. It is easily checked that one obtains the same approximant sequences already seen in Table 1. For example, for $n=3$ the strip has a slope equal to $\frac{1}{2}$, and one obtains a periodic repetition of the motif $ABA$.

\subsubsection{Structure factor}
As we have said at the beginning of this section, the defining characteristic of a quasicrystal is that it has a pure point diffraction pattern, i.e. Dirac delta-function peaks of the structure factor.  The peaks of the structure factor of the FC occur for q-vectors given by linear combinations of $two$ reciprocal lattice vectors. 

It is rather easy to see that the structure factor of the perfect FC has only Bragg peaks, in the higher dimensional representation. The position of the peaks of the structure factor of the FC can be easily deduced from those of the square lattice. The reciprocal space of the square lattice is given by $\vec{G}_{hk}=2\pi(h\vec{x}+k\vec{y})$. 
There are Bragg peaks of $S(q)$ ($q$ is the reciprocal space coordinate) corresponding to each of the $\vec{G}_{hk}$. Projecting each of the points onto the $q$-axis one sees that all Bragg peak positions $q$ are indexed by two integers $h$ and $k$  
\begin{eqnarray}
q = q(h,k) &=& 2\pi (h\cos\theta + k\sin\theta) \nonumber \\
&=& g (h +\omega k)
\label{eq.strucfac}
\end{eqnarray}
where $g=\frac{2\pi}{\sqrt{1+\omega^2}}$. That is, there are Bragg peaks at positions given by all the integer linear combinations of two incommensurate wave vectors $g$ and $\omega g$. 

\begin{figure*}
\includegraphics[width=0.4\textwidth]{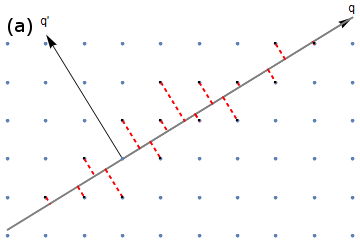}
\includegraphics[width=0.3\textwidth]{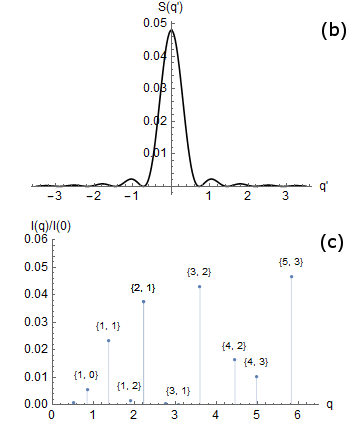}
\caption{a) Plot of the reciprocal lattice of the square lattice, showing the $(q_x,q_y)$ and $(q,q')$ axes and the projections of some representative points. b) Plot of the form factor corresponding to the selection strip c) Plot of relative intensities of the first few peaks for $q>0$ corresponding to the points outlined in bold in figure a). All $q$ values are shown in units of $(2\pi/a)$.}
\label{fig.reciprocal}
\end{figure*}

Fig.\ref{fig.reciprocal}a) shows some of the positions obtained by projecting the vertices of the reciprocal space lattice (shown in black) onto the $q$-axis. In Eq.\ref{eq.strucfac} note that $h$ and $k$ can take all possible integer values, resulting in a dense distribution of peaks of the structure factor along the $q$ axis. In practice the observable peaks of the structure factor are however far fewer, for
most of the peaks have intensities which are too small to observe. This occurs because the intensities of the peaks depend on a form factor, namely, the Fourier transform of the selection strip. The strip is described by the function $\mathcal{S}(y')=1$ for $0\leq y'\leq W$ and 0 elsewhere. As a result, peak intensities are modulated by the function 
\begin{eqnarray}
\vert \mathcal{S}(q')\vert^2 \propto \frac{\sin^2(\frac{Wq'}{2})}{q'^2 }  
\label{eq.airy}
\end{eqnarray}
where the perpendicular reciprocal space coordinate is defined by $q'=\frac{2\pi}{a}( -h\sin\theta + k\cos\theta)$. The $\mathcal{S}$ function is akin to the Airy function for diffraction through a slit in optics, and it is shown in Fig.\ref{fig.reciprocal}b.  The resulting variation of the peak intensities of the FC are shown in Fig.\ref{fig.reciprocal}c. 

To conclude this description of the structure factor, we point out one remarkable characteristic of the FC related to the small $q$ (large wave-length) behavior of $S(q)$. Notice that the Bragg peak intensities near the origin must correspond to large values of the perpendicular component $q'$, and therefore have vanishingly small intensities due to Eq.\ref{eq.strucfac}. Excluding the peak at $q=0$ for forward scattering, one concludes that the structure factor must tend to zero for small $q$, $\lim_{q\rightarrow 0}S(q) \rightarrow 0$. For a rigorous discussion of this property see \cite{hyperbaake}. Structures with this property are said to be hyperuniform \cite{torquato}. This is another way of expressing the fact that, for the FC, as indeed for all quasicrystals as mentioned before, the spatial fluctuations are bounded. As counterexamples of chains NOT possessing this property one can mention binary 1D structures obtained for non-Pisot substitution rules \cite{luckgodreche}. For the interested reader a discussion of diffraction patterns of aperiodic systems from a mathematical perspective can be found in \cite{baake_grimm_2013}.

\bigskip
\noindent
{\bf{On patterns and their probabilities}} The cut-and-project method provides a convenient way to compute the probability of a given pattern to occur in the FC. These probabilities are proportional to the length of a corresponding ``acceptance zone" in perpendicular space. To illustrate the method, consider the probability of occurrence of a 1-letter pattern: the B tile. To be selected, the corresponding vertical bond, of projected length $\cos\theta$ must lie within the window. The acceptance zone is $W-\cos\theta=\sin\theta$, and the probability of $B$ is $\sin\theta/W$. The probability of finding an A tile is proportional, by a similar argument, to  $\cos\theta/W$. The ratio of the probabilities is $p(A)/p(B)= \cot\theta=\tau$, as expected. 

Consider a 2-letter pattern, such as AA. This pattern corresponds, in the 2d square lattice, to 2 consecutive horizontal bonds sandwiched between two vertical bonds. The probability of the pattern AA is given by $(\cos\theta - \sin\theta)/W \approx 0.238$. The configurations $AB$ and $BA$ have equal probabilities of $\approx 0.382$.

\subsubsection{Conumbering scheme} \label{subsub.conum} 
\begin{figure}
\includegraphics[width=0.5\textwidth]{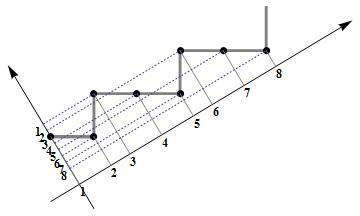}
\caption{The $n=5$ ($N=8$) approximant showing the real space labels of sites (along the chain) as well as their alternative labeling using their conumbers (along the axis perpendicular to the chain). }
\label{fig.conumbering}
\end{figure}

Sire and Mosseri  showed that in approximant chains it can be advantageous to work with an alternative ordering of sites, rather than the usual real space ordering $i=1,...,N$ \cite{siremoss90}. The so-called conumber of the site $i$ depends on its perpendicular space coordinate -- i.e. the distance along the $y'$ axis. This is illustrated in Fig.\ref{fig.conumbering} for a periodic approximant of 13 sites. The conumbers are given by $c(i) = \mathrm{Mod}[i\pm F_{n-2},F_n]$, upto a global cyclic permutation. 

This numbering orders the sites {\sl according to their local environments} as follows: 
\begin{itemize}
    \item sites with conumbers $1<c<F_{n-2}$ have an B-tile to the left and an A-tile to their right.  \item sites with $F_{n-2}+1<c<F_{n-1}$ have A-tiles both on the left and on the right. 
    \item  the remaining $F_{n-2}$ sites have an A-tile to the left and an B-tile to their right. 
    \end{itemize}
As we saw in Sec.\ref{subsec.inflation} the FC is a self-similar structure. Its inflation symmetry is coded by a hierarchical structure of the conumbering indices. Each of the three groups of sites listed above is, in turn, composed of three sub-groups having the same properties but on a bigger length scale. In particular, consider the central site, according to the conumber scheme, for an approximant for which $n$ is a multiple of 3. This site always has the same local environment (A-tiles on both sides) under inflation of $n\rightarrow n-3$, all the way down to $n=1$. The conumbering scheme will be useful later in Sec.\ref{sec.rghopping} for a compact representation of the spectrum and states of the hopping model. 

In the above, we have discussed the cut-and-project method to obtain the FC by projection from a two dimensional lattice. More generally, projections from higher dimensional lattices can yield Fibonacci chains as part of higher dimensional quasiperiodic structures. For example, 2D planes composed of parallel FC can be found in certain three dimensional structures generated by cut-and-projection from a 4D lattice \cite{benabra}. In those examples the projections involve the golden mean $\tau$, which is of course intimately linked to the Fibonacci sequence. The well-known 3D icosahedral tilings also depend on the golden mean, which is why Fibonacci modulation of atomic density can be seen in experimental studies of surfaces of 3D icosahedral quasicrystals, as one sees in Fig.\ref{fig.copper} taken from \cite{sharma,ledieu2014surfaces}.

\subsection{The characteristic function method}	
The Fibonacci quasicrystal corresponds to a special case of the family of Sturmian potentials, defined by $V(n)=\chi_{[1-\omega,1)}(n\omega +\varphi \mathrm{mod} 1)$, where $\varphi$ is an arbitrary phase. The characteristic function $\chi$ is non-zero and corresponds, say, to the letter A, when its argument lies in the interval $[1-\alpha,1)$. It takes the value 0 or the letter B otherwise.  \footnote{More generally, Sturmian potentials can be written in terms of two parameters, $\alpha$ and an interval $\Delta$.  For generic values of these, it has been shown that the resulting model is not integrable, and has large unbounded fluctuations, in contrast to the Fibonacci chain \cite{luckmoussa}.}
The characteristic function can be written out explicitly as $\chi_j=2([(j+1)\omega] - [j\omega])-1$, where $[X]$ stands for the integer part of $X$. In this formulation $\chi_j=-1$ stands for A and $\chi_j=1$ for B. The preceding form was modified to yet an alternative form using a cosine function by Kraus and Zilberberg, as a way to connect the Fibonacci and AAH models \cite{characteristicfn}.
The $j$th letter of the Fibonacci chain is obtained via the characteristic function $\chi_j$ defined by
\begin{align}
		\chi_j = \mathrm{sign}\left[\cos(2\pi j\omega + \varphi) - \cos(\pi\omega)\right] \label{eq.chij}
\end{align}	
where $\varphi$ is an arbitrary constant which one can term a phason angle. This function also serves to generate approximants by using a rational approximant of the golden mean in the expression Eq.\ref{eq:chi}. The approximants $C_{n}$ defined earlier are found by making suitable choices of $0\leq \varphi <2\pi$. The important feature to note is that, when the angle $\varphi$ is varied, phason flips occur -- a single flip at a time. In this way, one can generate a family of $F_{n+1}$ chains of length $F_n$, corresponding to different values of $\varphi$ (see Fig. 6). The phason angle will become a tuning parameter, later, to control the  edge modes in open finite approximant chains. 

\begin{figure}
\label{fig.chainsbyphi}
\includegraphics[width=0.3\textwidth]{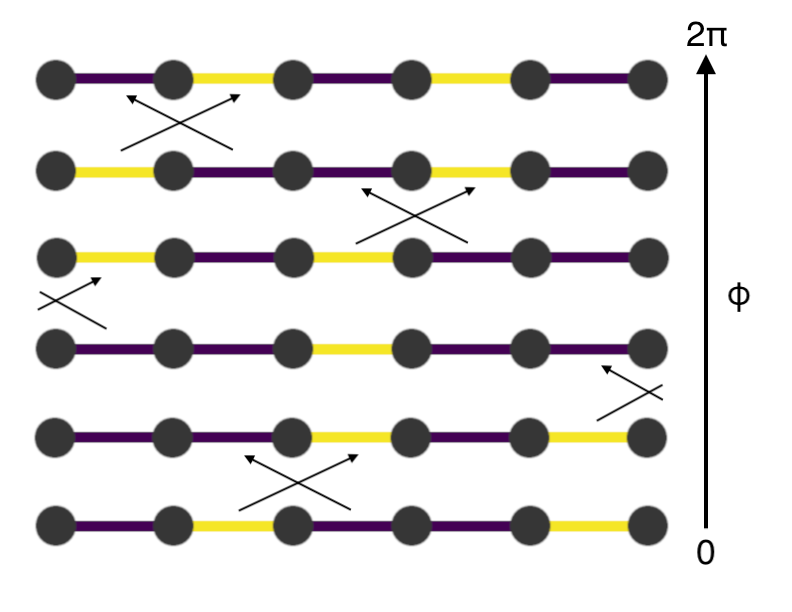}
\caption{Successive phason flips in the $n=4$ approximant ($N=5$, the first and last sites are equivalent under a translation) as $\phi$ is varied in Eq.\ref{eq.chij}. }
\end{figure}

\section{Tight-binding Models. Exact results}\label{sec.overview}
Much of the extensive literature on electronic properties of the FC is devoted to the study of tight-binding models of the form
\begin{align}
    H = \sum_n \epsilon_n c^\dag_{n}c_n -(t_{n} c^\dag_{n+1}c_n 
    + h.c.) 
    \label{eq.hamilt}
\end{align}
where the $\epsilon_n$ are the site energy at the nth site, while $t_{n}$ is the hopping 
amplitude between sites $n$
and $n +1$. We will assume that these parameters are defined by local rules, i.e., the values depend on the environment of each site. The first papers on this family of Hamiltonians appeared very shortly before the discovery of quasicrystals \cite{kohmoto1983,PhysRevLett.50.1873ostlund}, and were followed by many other groundbreaking papers in the next few years  \cite{kohmotooono,PhysRevB.29.1394pandit,kohmoto1986,kohmoto1987critical,kaluginlevitov,kohmotosutherland,luckPhysRevB.39.5834,evangelou}. When written for approximant chains with periodic boundary conditions, the Hamiltonian Eq.\ref{eq.hamilt} is invariant with respect to translations of the selection strip in the 2D space (Sec.\ref{subsec.cutproject}).

\subsection{Diagonal and off-diagonal Fibonacci models} The Hamiltonian of Eq.\ref{eq.hamilt} is termed mixed, as there can be spatially modulations in both diagonal and off-diagonal terms. However, the following simple cases contain all the essential new physics, namely:

\begin{enumerate}
\item {\sl {Off-diagonal case}} Site energies are assumed to be constant i.e. $\epsilon_n=\epsilon$, while the hopping amplitudes $t_{n}$ can can take the value $t_A$ or $t_B$, according to a Fibonacci sequence. This is also referred to as the pure-hopping Fibonacci Hamiltonian since the constant energy term can be dropped by a suitable redefinition of the energy.
\begin{align}
    H = -\sum_n t_{n} c^\dag_{n+1}c_n + h.c.  
    \label{eq.hophamilt}
\end{align}
Absorbing $t_B$ in the definition of the units of energy, leaves as sole parameter the ratio $\rho=t_A/t_B$, which controls all the properties of the chain. We will without loss of generality, henceforth assume both amplitudes to be positive, since these signs can be changed by a gauge transformation. 

    \item {\sl {Diagonal case}}
Here the quasiperiodicity is assumed to be present in the diagonal term, while the hopping amplitudes are assumed to be uniform, $t_{n,n+1}=t$, for all values of $n$ :
\begin{align}
    H = \sum_n \epsilon_n c^\dag_{n}c_n -t \sum_n  c^\dag_{n+1}c_n + h.c. 
    \label{eq.diaghamilt}
\end{align}
where the onsite potentials $\epsilon_n$ take on two discrete values $\epsilon_A$ and $\epsilon_B$, according to a Fibonacci sequence. As in model 1 above, there is only one nontrivial parameter in this model, and it depends on the energy difference, $\varepsilon=(\epsilon_A-\epsilon_B)/t$. 
 
\end{enumerate}

The above models are often compared with those of a particular quasiperiodic Aubry-Andre-Harper model \cite{aubry1980analyticity,harper,gordon1997} hereafter called simply the AAH model. The AAH model is equivalent to a tight-binding problem of an electron hopping in a 2D square lattice and subjected to a uniform magnetic field, with a flux per plaquette $\Phi=\omega\Phi_0$, where $\Phi_0=h/2e$ is the flux quantum. The resulting quasiperiodic AAH Hamiltonian is of the form
\begin{align}
    H = \sum_n t (c^\dag_{n+1}c_n + h.c.)  + 2V \cos\left(2\pi n\omega +\phi\right) c^\dag_{n}c_n
    \label{eq.aubry}
\end{align}
where the strength of the onsite potential energy $2V$ depends on the hopping amplitude along the direction transverse to the chain and $\phi$ is a phase.  A well-known duality transformation takes this Hamiltonian  Eq.\ref{eq.aubry} into a Hamiltonian of the same form but with the exchange $t \leftrightarrow V$. When $V=t$ the model is self-dual. This is the critical AAH model, having many properties in common with the Fibonacci model, as described below.

\subsection{Multifractal energy spectra}
Spectra can be classified into three types : continuous spectra associated with extended states (as in periodic solids), pure point spectra associated with localized states (as in disordered solids) and singular continuous spectra associated with multifractal states. If one defines the scaling of the integrated density of states, $N(E)$ (defined as the fraction of states of energy equal to or less than $E$) in the vicinity of the energy $E$  by
\begin{equation}
N(E+\Delta E)-N(E) \sim   \Delta E^\alpha  
\end{equation}
then the three cases correspond to $\alpha=1$, $\alpha=0$ and $0<\alpha<1$.  The wavefunctions typically associated with this last type of fractal spectrum are ``critical" states -- neither extended nor localized. This intermediate type of state appears to be rather generically found in quasiperiodic structures not only in 1D, but higher dimensions as well, as can be seen in the review of quasiperiodic tight-binding Hamiltonians by Grimm and Schreiber \cite{grimmreview}. 

For the AAH model, the nature of the spectrum depends on the parameter $V/t$: the spectrum is continuous for $V/t < 1$, pure point for $V/t > 1$ and singular continuous for $V/t = 1$. 

For the Fibonacci models in Eqs.\ref{eq.hamilt} and \ref{eq.hophamilt} the energy spectrum is singular continuous as soon as there is aperiodicity, however small \cite{delyonPhysRevLett.55.618,  sutospectra,bellissard1989spectral}. The situation is analogous to that of the Anderson model for 1D disordered metals, where the critical value for localization in 1D for disorder strength is zero. 

It is noteworthy that the spectra in all three cases are pure spectra -- meaning purely singular continuous, or pure point, or absolutely continuous. In general, however, models may have spectra with several different components, and there may be one or more mobility edges separating different regions. This is the case for the Anderson model in 3D for disorder strengths which are smaller than the critical value, for example. In one dimension also there can be mobility edges. This occurs in generalized Harper models, where the potential energy depends, for example, on two incommensurate wave vectors, as discussed by Hiramoto and Kohmoto \cite{kohmotohiramotoPRB,kohmotohira}, and reconsidered recently in, for example \cite{pixley2015,PhysRevB.37.1097dassarma,PhysRevB.91.014108ghosh}.

\begin{figure*}
\includegraphics[width=0.25\textwidth]{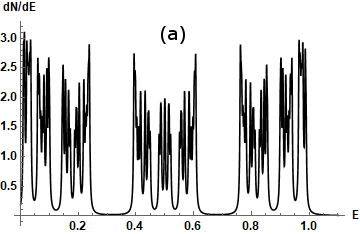} \hskip 0.5cm
\includegraphics[width=0.25\textwidth]{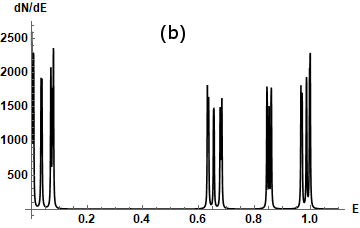}\hskip 0.5cm
\includegraphics[width=0.25\textwidth]{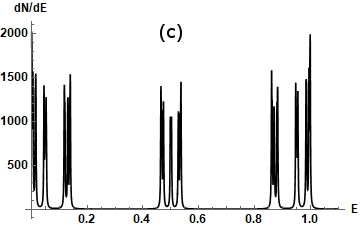}
\caption{Densities of states $dN/dE$ for three models a) the off-diagonal Fibonacci model Eq.\ref{eq.hophamilt} with $t_A/t_B=0.6$ b) Diagonal Fibonacci model  Eq.\ref{eq.diaghamilt} with $\varepsilon=4$ and c) AAH model   Eq.\ref{eq.aubry} at criticality. }
\label{fig.dos}
\end{figure*}

Fig.\ref{fig.dos} shows three typical forms of the densities of states for the three models. The left hand figure shows the DOS for the off-diagonal model for the case $t_A=0.6$, $t_B=1$ and periodic boundary conditions. This model has a chiral symmetry -- for each solution $\vert\psi\rangle$ of energy $E$, one has a solution $\vert\psi'\rangle$ of energy $-E$ such that $\langle n\vert \psi'\rangle = (-)^n \langle n\vert \psi\rangle$. The spectrum is therefore symmetric around 0, as can be seen in Fig.\ref{fig.dos}.  The horizontal axis represents the dimensionless energy $E/t_B$. The band structure is seen to be composed of three main clusters, each of which comprises three sub-clusters, and so on. Each level broadens into a band when periodic boundary conditions are assumed. The process of subdivision into smaller bands continues as one considers larger and larger approximant chains. 

The spectrum for the diagonal model for $\epsilon_A \neq \epsilon_B$ is very similar. Part b) of Fig.\ref{fig.dos} shows the DOS computed numerically for an approximant chain of $N=144$ sites. The parameters were taken to be $\epsilon_A=-\epsilon_B= 2t$, with periodic boundary conditions. The spectrum, which is asymmetric, has two main clusters. These clusters are in turn composed of three sub-clusters, which trifurcate into three clusters and so on.   

Finally, in part c) of the figure, we show the DOS for the AAH model computed at criticality $V=1$, with $\tau=\tau_n$ in the cosine term, for a system of $N=144$ sites with periodic boundary conditions. For the AAH model at criticality as for the off-diagonal model, the bands divide into three sub-bands when going from one approximant to the next \cite{kohmotohiramotoPRB}.

\begin{figure}
		\includegraphics[width=0.5\textwidth]{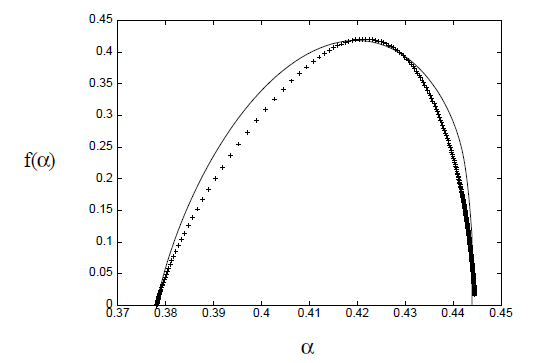}
\caption{$f(\alpha)$ computed numerically (crosses) for the off-diagonal Fibonacci model for $\rho=0.2$. The solid line corresponds to an analytic expression obtained using the trace map (see text) (figure reprinted from R\"udinger and Pi\'echon \cite{rudinger98} with permission) }
\label{fig.falpha}
\end{figure}

The figures in Fig.\ref{fig.dos} correspond to a given system size. When the size is increased, one observes a characteristic feature of multifractal structures, namely local power law singularities of the DOS, $N(E+\Delta E)-N(E) \sim \Delta E^{-\alpha(E)}$. As the system size gets bigger, band widths shrink, each one scaling with different exponent $\alpha$. To fully describe this multifractal spectrum, one needs the full set of exponents $\alpha$ and their densities $f(\alpha)$. 
This can be done numerically by standard methods of multifractal analysis \cite{Halsey1986}. 
The method consists of defining the ``partition function"
 \begin{equation}
\label{eq:gamma_spec}
	\Gamma_n(q,\tau) = \sum_{E} \frac{\left( 1/F_n \right)^q}{(\Delta_n(E))^\tau}
\end{equation}
where $\Delta_n(E)$ is taken to be the width of the energy band associated to the energy level labelled $E$. One determines $\tau$ as a function of $q$ by requiring that $\Gamma\sim 1$ as $n\rightarrow \infty$. The function $f(\alpha)$ is the Legendre transform of $\tau(q)$, given by

\begin{equation}
	\alpha_q = \frac{d (q-1)d_q}{d q}
\end{equation}
and
\begin{equation}
	f(\alpha_q) = q \alpha_q - \tau_q
\end{equation}
The function $f(\alpha)$ gives the fraction of sites around which the DOS scales with the power $\alpha$.
$f(\alpha)$ is typically a convex curve extending between the extremal values $\alpha_{min}$ and $\alpha_{max}$. For the periodic crystal when the spectrum is continuous, this curve reduces to just two points: $\alpha=1$ describing the interior of the band, and $\alpha=1/2$, due to van Hove singularities at the band edges, as at the bottom of the band where $dN(E) \sim (E-E_{min})^{1/2}$. In the quasiperiodic case, singularity strengths vary, depending on the energy. Fig.\ref{fig.falpha}, taken from \cite{rudinger98} shows $f(\alpha)$ values (indicated by crosses) computed numerically for $t_A/t_B=0.2$.
Exact expressions can be obtained for scaling exponents $\alpha(E)$ for two special energies, as we will explain later using the trace map method.

\begin{figure*}
		\includegraphics[width=0.5\textwidth]{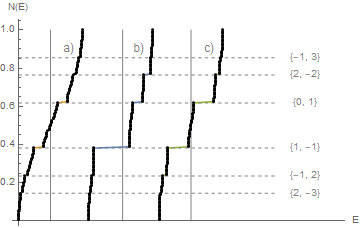}
\caption{Integrated DOS $N(E)$ plotted versus the energy $E$ (in dimensionless units) for three topologically equivalent models a) off-diagonal model Eq.\ref{eq.hophamilt} ($\rho=0.7$) b) diagonal model Eq.\ref{eq.diaghamilt} ($\varepsilon=4$) and c) critical AAH model Eq.\ref{eq.aubry}. In all the plots, energies were shifted and normalized such that the total band widths are equal to 1 (system size $N=144$, periodic boundary conditions).}
\label{fig.threemodels}
\end{figure*}

\subsection{Gap labeling and topological indices}\label{subsec.gaplabel}
In one dimensional problems, the IDOS $N(E)$, defined as the fraction of states of energy less than $E$, is also equal to the number of changes of sign (nodes) of the wavefunction per unit length. One can thus introduce a ``wave number" $k(E)=N(E)/2$ corresponding to a state of energy $E$. For the $n$th periodic approximant, the spectrum consists of $F_n$ bands, each corresponding to wave vectors $k_j$, such that the IDOS between two bands $j$ and $j+1$ has the value $j/F_n$.
Note that the IDOS is a more tractable quantity than the DOS, which fluctuates violently as a function of the given energy, when system size is increased. The IDOS, in contrast, has plateaux whose positions are well-defined as system sizes are increased. This has its importance for calculations in which the chemical potential enters as a parameter. The IDOS curve approaches the devil's staircase form in the limit of infinite size. 

The gap labeling theorem \cite{bellissardgaps,bellissardgaps1} states that, of the class of models given by Eq.\ref{eq.hamilt}, the IDOS $N(E)$ within the gaps must take values given by 
 \begin{equation}
     N(E) = \mathrm{p} + \frac{\mathrm{q}}{\tau}
     \label{eq:gaplabels}
 \end{equation}
where $\mathrm{p}$ and $\mathrm{q}$ (not to be confused with the reciprocal space vector or the multifractal parameter $q$ defined in the last section) are integers. The pair of indices $(\mathrm{p},\mathrm{q})$ label all of the possible gaps, but the gap labeling theorem does not specify whether or not a given gap is actually opened. It suffices to specify only one integer, $\mathrm{q}$, as $\mathrm{p}$ is then fixed by the condition $0\leq N(E)\leq 1$. 

The mapping between the FC quasicrystal and a 2D quantum Hall system is the subject of numerous papers by Zilberberg and collaborators and is reviewed in \cite{zilberreview}. Arguments can be made thus that $\mathrm{q}$ is a topological quantity, a Chern number inherited from the 2D parent model. As expected for a topological invariant, it is robust under local perturbations due to disorder or other scattering that preserves symmetry. The gap label $\mathrm{q}$ represents a winding number that describes, for example, the variations of the edge modes in a system which has interfaces. This bulk-edge correspondence will be discussed in more detail in Sec.\ref{subsec.topological}). 

Fig.\ref{fig.threemodels} shows the IDOS for each of the three models plotted as a function of the dimensionless energy $E/t$. The horizontal lines indicate values of IDOS given by Eq.\ref{eq:gaplabels} for gaps corresponding to values of $-3 \leq \mathrm{q} \leq 3 $. It shows for the three cases a, b and c, that they have identical positions of the gaps, as given by the gap labeling theorem. This topological equivalence can shown by an explicit mapping between models \cite{PhysRevLett.109.106402kraus,PhysRevLett.110.076403verbin}. In an elegant experiment using a photonic waveguide array, the topological equivalence of the Fibonacci and Harper models was explicitly shown by \cite{PhysRevB.91.064201zilberberg}.

\subsection{Trace map method}\label{subsec.transfer}
The trace map analysis is a powerful technique which has led to many results for quasiperiodic Hamiltonians. The starting point is the definition of transfer matrices relating the wavefunction amplitudes on the ${n+1}$th site to the amplitudes on sites $n$ and $n-1$. From Eq.\ref{eq.hamilt} one has the tight-binding equations
\begin{align}
    (E-\epsilon_n) \psi_n + t_{n}\psi_{n+1}++ t_{n-1}\psi_{n-1} = 0
\end{align}
This relation can be re-expressed in terms of a $2\times 2$ matrix equation as follows
\begin{eqnarray}
	\begin{bmatrix}
		\psi_{n+1}  \\
	\psi_{n}  
	\end{bmatrix} &=&
	\begin{bmatrix}
	\frac{(E-\epsilon_n)}{t_{n}} & - \frac{t_{n-1}}{t_{n}} \\
	1 & 0
	\end{bmatrix} 
	\begin{bmatrix} \psi_{n}  \\
	\psi_{n-1}  
	\end{bmatrix} \nonumber \\
	& =& \left( \prod_{k=2}^{n} T_{k,k+1}\right)
	\begin{bmatrix} \psi_{1}  \\
	\psi_{0}  
	\end{bmatrix} \nonumber \\
	&=& M_n \begin{bmatrix} \psi_{1}  \\
	\psi_{0}  
	\end{bmatrix}
	\label{eq.transfermatrix}
\end{eqnarray}
where we have introduced the local transfer matrix $T_{k,k+1}$, and the global transfer matrix $M_n$ which is a product of $n-1$ such transfer matrices, with the order of multiplication of the matrices given by $M_n=.....T_{2,3}T_{1,2}$. 

\begin{enumerate}
    \item {\it{Diagonal model.}}
The local transfer matrix depends on the onsite energy $\epsilon_n$ and the amplitudes for hopping onto the sites to the left and to the right of site $n$. For the diagonal model, there are only two different possible transfer matrices, namely 
\begin{eqnarray}
    T_A=\begin{bmatrix} \frac{(E-\epsilon_A)}{t} & -1  \\
	1 & 0 
	\end{bmatrix}, \nonumber \\
	T_B=\begin{bmatrix} \frac{(E-\epsilon_B)}{t} & -1  \\
	1 & 0 
	\end{bmatrix}
\end{eqnarray}
Let us now consider the  approximant chain of length $N=F_n$, and let $x_n=\frac{1}{2} \mathrm{Tr} M_n$ be the half-trace of the transfer matrix. For an energy $E$ to correspond to an allowed (normalizable) wavefunction, the half-trace of $M_N$ must satisfy the condition $\vert x_n\vert \leq 1$. Thanks to the concatenation property of chains already mentioned in Sec.\ref{subsec.inflation}, namely $C_{n+1}=C_n\oplus C_{n-1}$, the global transfer matrices for successive approximants satisfy \cite{PhysRevLett.50.1873ostlund,kohmoto1983}
\begin{eqnarray}
    M_{n+1} = M_{n-1}M_n 
    \label{eq.matrixrecursion}
\end{eqnarray}
Given Eq.\ref{eq.matrixrecursion}, it can be shown that the half-traces satisfy a three term recursion relation \cite{kohmoto1983}:
\begin{eqnarray}
    x_{n+1} = 2x_n x_{n-1} - x_{n-2}
    \label{eq.tracemap}
\end{eqnarray}
with the initial conditions 
\begin{eqnarray}
x_{-1}=1,  \nonumber \\ x_0=\frac{(E-\epsilon_B)}{2},  x_1=\frac{(E-\epsilon_A)}{2} 
\end{eqnarray}
To determine which energies belong in the spectrum one computes the iterates $x_n$ using Eq.\ref{eq.tracemap} and checks to see whether they remain bounded and within the interval $(-1,1)$. 

\item {\it{Off-diagonal model.}} A similar set of relations obtains in the case of the off-diagonal model. One starts by introducing three different transfer matrices, $T_{AA}$, $T_{AB}$ and $T_{BA}$, for the three bond configurations AA,AB and BA which are possible in the FC. These matrices are defined by 
\begin{eqnarray}
    T_{AA}&=&\begin{bmatrix} E/t & -1  \\
	1 & 0 
	\end{bmatrix}, \nonumber \\
	T_{AB}&=&\begin{bmatrix} E/t & -t_B/t_A  \\
	1 & 0 
	\end{bmatrix}, \nonumber \\
	T_{BA}&=&\begin{bmatrix} E/t & -t_A/t_B  \\
	1 & 0 
	\end{bmatrix}
\end{eqnarray}
The problem of writing the global transfer matrix can be simplified \cite{kohmoto1987critical}, as the hopping on Fibonacci chains can be described with just two matrices $T_{AA}$ and $T_{AB}T_{BA}$. If we rename these transfer matrices as $T_B$ and $T_A$ respectively, the global transfer matrices for approximant chains can be written exactly as in the diagonal case. 

The recursion relations Eq.\ref{eq.matrixrecursion} and Eq.\ref{eq.tracemap} therefore hold also for the off-diagonal model. 
The initial conditions for the half traces in the off-diagonal case are
\begin{eqnarray}
    x_{-1}&=&\frac{1}{2}\left( \frac{t_B}{t_A} +  \frac{t_A}{t_B}\right), \nonumber \\ 
    x_0&=& \frac{E}{2t_B}, x_1 = \frac{E}{2t_A} 
\end{eqnarray}

\end{enumerate}
The recursion relation for the traces Eq.\ref{eq.tracemap} constitute a dynamical system in a three-dimensional space. Defining the variables $x=x_{n-1}$, $y=x_n$ and $z=x_{n+1}$,  Eq.\ref{eq.tracemap} maps a given point as follows:
\begin{eqnarray}
x \rightarrow x'=y  \nonumber \\
y \rightarrow y'=z \nonumber \\
z \rightarrow z'=2yz -x
\label{eq.mapping}
\end{eqnarray}
One of the invariants of the dynamical system Eq.\ref{eq.mapping} is the quantity \cite{kohmoto1983, kohmotooono,kohmoto1987critical}:
\begin{eqnarray}
I &=& x^2+y^2+z^2 -2xyz-1 \nonumber \\
 &=& \frac{1}{4}(\epsilon_A-\epsilon_B)^2 \qquad \text{(diagonal model)}\nonumber \\
 &=& \frac{1}{4}\left( \frac{t_A}{t_B} - \frac{t_B}{t_A}\right) \qquad \text{(off-diagonal model)}
\label{eq.invariantI}
\end{eqnarray}
where the last two equations were written using the initial conditions for the diagonal and off-diagonal model respectively. For a proof of the invariance of $I$, a so-called Fricke character, see \cite{baake93}. Under the dynamical map, points move on the surface $I=const$, in the three dimensional space. Details of the form of these surfaces for different values of the parameters and of different kinds of orbits are given in \cite{kaluginlevitov}. Orbits which escape to infinity, such that $\lim_{n\rightarrow \infty} x_n$ is infinite, correspond to energies which are not in the spectrum. Numerically, this is found to be the case of almost all energies, consistent with the fact that the spectrum has a Lebesgue measure of zero. Periodic orbits with $\lim_{n\rightarrow \infty} x_n \leq 1$ correspond to allowed energies. \footnote{There are in principle two other more ``exotic" possibilities iii) aperiodic and bounded orbits and iv) recurrent orbits where the point returns to the allowed region and for which $\lim_{n\rightarrow \infty} x_n \leq 1$.} It can be shown, by tracing orbits for successive periodic approximants, that the band structure has a self-similar structure -- is a Cantor set.

Kohmoto, Kadanoff and Tang obtained exact results for two cases where the trace map leads to a periodic orbit \cite{kohmotooono,PhysRevLett.50.1873ostlund, kohmoto1987critical,kohmotobanavar}. By considering the linearized map around these special points of the spectrum, they found the scaling exponents for the corresponding bands in terms of the ``escape rates" of the dynamical map. The band widths are given by $\Delta \sim \omega^\epsilon$, where $\epsilon$ is the minimal eigenvalue of the linearized map as given below for two special cases.  

\begin{enumerate}
    \item Solution for the band center. The trace map has a periodic orbit consisting of the six-cycle $(0,0,a) \rightarrow (-a,0,0)\rightarrow (0,-a,0) \rightarrow (0,0,-a) \rightarrow (a,0,0) \rightarrow (0,a,0) \rightarrow (0,0,a)$, where $a=\sqrt{I+1}$.
The scaling exponent for this band can be expressed in terms of the eigenvalue $\epsilon_6$ of the linearized equation around this six-cycle. The result thus obtained for $\alpha_{ctr}$ \cite{kohmotooono} is
\begin{eqnarray}
\alpha_{ctr} &=& \ln \tau^6/\ln\epsilon_6 \nonumber \\
\epsilon_6 &=& [\sqrt{1+4(1+I)^2}+2(1+I)]^2
\label{eq.alphactr}
\end{eqnarray}

\item Solution for band edges. These correspond to two-cycles of the trace map, $(a,b,b) \rightarrow (b,a,a)\rightarrow (a,b,b)$ where $a=J+\sqrt{J^2-J} $ and $b=J-\sqrt{J^2-J}$ where $J=\frac{1}{8}[3+\sqrt{25+16I}]$. The scaling exponent for these bands is expressed in terms of $\epsilon_2$, the eigenvalue of the linearized map, as follows
\begin{eqnarray}
\alpha_{edge} &=& \ln \tau^2/\ln\epsilon_2 \nonumber \\
\epsilon_2 &=& [8J-1  +\sqrt{(8J-1)^2-4}]/2
\label{eq.alphaedge}
\end{eqnarray}
\end{enumerate}

It was conjectured in \cite{kohmoto1987critical} that the $\alpha_{ctr}$ and $\alpha_{edge}$ values correspond to the extremal values, namely, $\alpha_{min}$ and $\alpha_{max}$. In fact, however, R\"udinger and Pi\'echon \cite{rudinger98} showed that this is not always the case, by analyzing the trace map in the vicinity of a 4-cycle that governs scaling for IDOS value $N(E)=\frac{1}{3}$ and $\frac{2}{3}$. Their analysis showed that the maximal value of $\alpha$ occurs at these points when $\rho$ is small. Another conjecture concerned the possibility that the spectrum becomes monofractal when $\rho=\rho_c$ \cite{zhongbellmoss}. This conjecture was based on the fact that $\alpha_{ctr}$ is equal to $\alpha_{edge}$ for the hopping ratio $\rho_c \approx 0.0944$ (as deduced from Eqs.\ref{eq.alphactr} and \ref{eq.alphaedge}). However, R\"udinger and Pi\'echon's calculation showed that for $\rho=\rho_c$, $\alpha_{1/3}$ is different from (more precisely is larger than) the other two exponents. Thus the spectrum is not a monofractal. An approximate analytical expression for $f(\alpha)$ derived in \cite{rudinger98} is shown in Fig.\ref{fig.falpha}, along with the numerical data for $\rho=0.2$. 

The fractal exponent for the self-similar $E=0$ wavefunction in the hopping model is found by the trace map calculation to be $\vert\ln\rho\vert/\ln\tau^3$, in good agreement with numerical calculations of the wavefunction \cite{kohmotobanavar}.

Discussions of trace maps and their dynamical properties can be found in the reviews by \cite{baake93, damanik2014}. Generalizations of the trace map for mixed Hamiltonians with two and more parameters will be briefly discussed in Sec.\ref{sec.mixed}. As one might expect, these generalized models have a larger parameter space, with more possibilities for the spectra. They can admit, for example, extended states for special energies.

\subsection{Log-periodic oscillations} \label{subsec.logperiod}
\begin{figure}
		\includegraphics[width=0.4\textwidth]{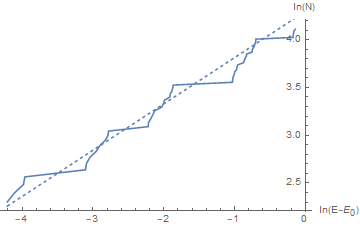}
		\includegraphics[width=0.4\textwidth]{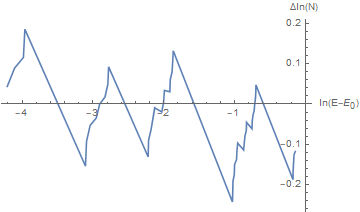}
\caption{ (top) Log-log plot showing the numerically computed IDOS versus energy for the hopping model in the vicinity of $E=E_{min}$ (for $\rho=0.4$). Dashed line indicates an averaged behavior. (bottom) Expanded plot of the fluctuations around the dashed line, $\Delta\ln(N)$, showing the main period of the oscillation and some smaller periods.}
\label{fig.idoszoom}
\end{figure}

Systems with discrete scale invariances can display log-periodic oscillations in thermodynamic properties, noted in treatments of critical phenomena \cite{nauenberg,derrida}. To cite a more recent study, Gluzmann and Sornette  considered an observable $f(x)$ in a system close to criticality, where $f(x) = \mu^{-1} f(\gamma x)$ under a renormalization transformation \cite{PhysRevE.65.036142sornette} . They showed that $f(x)$ has a power law scaling ``decorated" by a log-periodic function $f(x)=x^m P(x)$. The power is given by $m=\ln \mu/\ln\gamma$ and the period of the oscillations is $\log\gamma$. This is indeed what one observes for the IDOS $N(E)$ in the Fibonacci model. Fig.\ref{fig.idoszoom}a) shows in a log-log plot the IDOS versus energy in the bottom of the band. The points are obtained by numerical diagonalization, and the straight dashed line indicates the average IDOS. Fig.\ref{fig.idoszoom}b) shows the fluctuations of ln(N) around the average value. The period of the oscillations, close to 1, corresponds to the inflation factor which in this case is $\gamma=\tau^2$ for the side bands (for a description of the renormalization transformation of the Hamiltonian see Sec.\ref{sec.approximate}). The main period and some smaller ones, indicating a fractal structure can be seen.

\subsection{The wavefunction for $E=0$}\label{subsec.wf0}

This subsection discusses an exact solution for one of the wavefunctions of the hopping model. It provides a rare example of a non trivial case where multifractal properties can be computed analytically as a function of $\rho$.

\begin{figure}
		\includegraphics[width=0.45\textwidth]{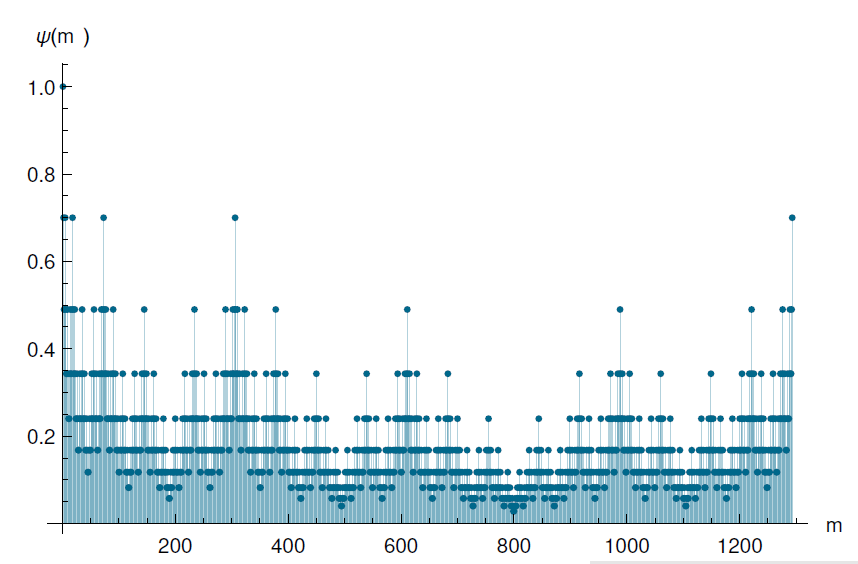}
\caption{ Center state onsite probabilities $\vert\psi_i(E=0)\vert^2$ plotted versus site index $i$ (numerically computed for $n=12$ chain with $\rho=0.6$ and periodic boundary conditions) }
\label{fig.wavefnE0}
\end{figure}

The wavefunction at the band center $E=0$ (shown in Fig.\ref{fig.wavefnE0} for a finite approximant) can be determined exactly, by a recursive construction. It turns out to be a particularly simple form of the critical states proposed by Kalugin and Katz (KK) for the ground state of tilings \cite{kaluginkatz}.
These authors argued that for a family of quasiperiodic Hamiltonians that includes all the standard ones, the ground state $\ket{\psi}$ can be written as a product of two factors. The amplitude on site $i$ is given by
\begin{equation}
\label{eq:KK}
	\braket{i | \psi} = C(i) e^{\kappa h(i)}
\end{equation}
where $\kappa$ is a (real) constant. The prefactors $C(i)$ depend on the local configuration of the atoms around site $i$. Long range correlations between sites are given by the exponent $h$, called the \emph{height field}, which is the integral of a quasiperiodic function. \footnote{ It is interesting to note that the expression \ref{eq:KK} can be considered as a generalization of the usual Bloch form for the wavefunctions in a periodic lattice, which can be written as a product of a periodic function $u_n(x)$ (in 1D) and an exponential $e^{ih(x)}$ where $h(x)=kx$ is the integral of the Bloch wave vector $k$.}
The $E=0$ wave function of the Fibonacci chain has the form of Eq.\ref{eq:KK} with the particular choice $C(i)=\pm 1$ depending on the sublattice to which the site $i$ belongs. This solution for the 1D chain is a relatively tractable case study to illustrate some of the properties of the KK eigenstates which are of course more complex in higher dimensional quasicrystals.

\begin{figure}
		\includegraphics[width=0.35\textwidth]{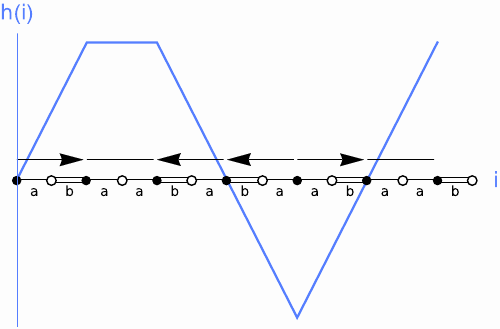}
\caption{Definition of the local arrow function and its integral, the height function (figure reprinted from \cite{mace2017critical}) }
\label{fig.heightsfig}
\end{figure}

From the Hamiltonian Eq.\ref{eq.hophamilt}, the following relation obtains for the $E=0$ wave function amplitudes on
$i$ and $i+2$
\begin{equation}
	t_{i+1} \psi_{i+2} + t_{i} \psi_{i} = 0
\end{equation}
There are two independent $E=0$ solutions, one for each sublattice. The two being equivalent in the limit of the infinite FC let us henceforth consider the even sub-lattice solution, for sites $i=2m$. There are three possibilities for the bond configurations between sites $i$ and $i+1$, namely, $AA,AB$ or $BA$. The three cases are given by the relation
\begin{equation}
	\psi_{m+1} = -\rho^{A(m)} \psi_{m}
	\label{eq.recurpsi}
\end{equation}
where $\rho=t_A/t_B$ and the A (for \emph{arrow}) function is defined locally according to the configuration of the bonds between the two sites
\begin{equation}
\label{eq:arrows}
	\begin{cases}
		A(m) & = +1 \qquad (AB) \\
		A(m) & = -1 \qquad (BA) \\
		A(m) & = 0 \qquad (AA)
	\end{cases}
\end{equation}
Fig.\ref{fig.heightsfig} shows the arrow function for the even sites of a small chain segment. The figure shows the arrows corresponding to the three bond configurations which are possible using the conventions: $\rightarrow$ (rightarrow) for the bond sequence $AB$ , $\leftarrow$ (left arrow) for the bond sequence $BA$ and no arrow for the bond sequence $AA$.  

\begin{figure}
		\includegraphics[width=0.4\textwidth]{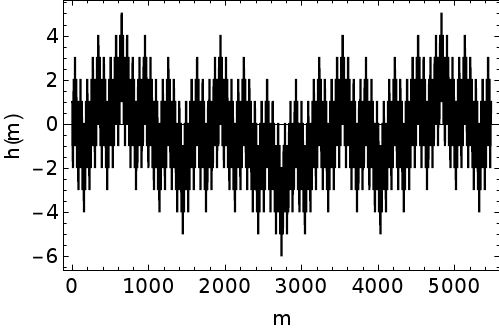}
\caption{ The height function plotted for a long chain, showing its fluctuations at large distances (figure reprinted from \cite{mace2017critical}) }
\label{fig.heights}
\end{figure}

Repeating the recursion relation Eq.\ref{eq.recurpsi}, one obtains 
\begin{eqnarray}
	\psi_{m} &=& (-1)^m \rho^{h(m)} \psi_0 \nonumber\\
	&=& (-1)^m e^{\kappa h(m)} 
 \label{eq.wf0}
\end{eqnarray}
where  $h(m)=\sum_0^m A(j)$, $\kappa=\ln\rho$ and  $\psi_0$ was set equal to $1$. This expression is of the KK form, Eq.\ref{eq:KK}, with a constant prefactor on all sites. The function $h$ in the exponent is an integral of a quasiperiodic function, $A(m)$.
Fig.\ref{fig.heightsfig} shows the height function for the first few even sites. As the length of the chain gets larger, the height function fluctuates more and more. 
Fig.\ref{fig.heights} shows the height function calculated for a long segment of the Fibonacci chain. The properties of the wave function can be determined when the distribution of heights is known. One can show by explicit calculation, that the wave function is multifractal, and express all of its generalized dimensions $D^\phi_q$ in terms of $\rho$. This can be done by introducing inflation matrices to relate the heights in the $C_n$ and $C_{n-2}$ chains. In the large $n$ limit, the height distribution $P(h)$ satisfies a diffusion equation as a function of $t$ (the number of inflations), as follows
\begin{equation}
	P^{(t)}(h) \sim \frac{1}{\sqrt{4 \pi D t}} \exp\left(-\frac{h^2}{4 D t}\right)
	\label{eq:heightP}
\end{equation}
where the ``diffusion coefficient" is given by $D=\frac{1}{2\sqrt{5}}$. We refer the reader to \cite{mace2017critical} for details.

Since $P(h)$ is a symmetric distribution around $h=0$, the resulting form of the wave function has a symmetry between peaks and valleys, as can be seen in Fig.\ref{fig.wavefnE0}. The  typical value of $h$ in a chain of $N=\tau^n$ sites is given by the standard deviation of the gaussian after a ``time" $t=n$, $h_{typ} \sim \sqrt{2Dn}$. This leads to a typical value of $\psi$ which falls off with the chain length, $N=\tau^n$, as $\psi(2N) \sim e^{- cst \sqrt{2D \ln N}}$. The spatial decay of the wavefunction is thus faster than power law but slower than exponential. In contrast, note that, for the randomly disordered off-diagonal model, a similar argument gives the wavefunction at $E=0$ to be $\psi(2N) \sim e^{- cst \sqrt{2DN}}$ \cite{theodorou_extended_1976,economou_static_1981,PhysRevB.49.3190inui}. This is a stretched exponential function, decaying much faster than the $E=0$ wavefunction of the Fibonacci chain. 

The fractal dimensions of $\psi$ can be exactly computed. These quantities are deduced from the scaling of the moments of the wavefunctions, which are defined as follows.
Let the $q$-weight of the wavefunction $\psi$ be defined by:
\begin{equation}
\label{eq:chi}
	\chi_q(\psi, \mathcal{R}) = \frac{\sum_{i \in \mathcal{R}} |\psi_i|^{2q}}{\left(\sum_{i \in \mathcal{R}} |\psi_i|^2\right)^q}
\end{equation}
where the sums run over all sites in a given region $\mathcal{R}$.
The $q$-weight is a measure of the fraction of the presence probability contained inside region $\mathcal{R}$. 

Consider a sequence of regions $\mathcal{R}_n$ whose radius grows to infinity as $n\rightarrow\infty$.
The $q^\text{th}$ fractal dimension, $\wf_q(\psi)$, is the scaling of the $q$-weight with the volume of the region:
\begin{equation}
\label{eq:dq}
	\wf_q(\psi) = \lim_{n \to \infty} \frac{-1}{q-1} \frac{\log \chi_q(\psi, \mathcal{R}_t)}{\log \Omega(\mathcal{R}_n)}
\end{equation}
where $\Omega$ is the number of sites inside region $\mathcal{R}$. As we have already seen for the DOS in Sec.\ref{subsec.spectral}, one can then compute the Legendre transform of the fractal dimensions, $f_{E=0}(\alpha)$. This function gives the fraction of sites for which the wavefunction scales with the power $\alpha$. 

\begin{figure}
		\includegraphics[width=0.4\textwidth]{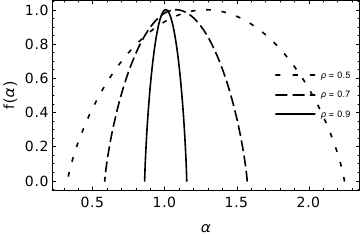}
\caption{ $f_{E=0}(\alpha)$ spectrum as given by Eq.\ref{eq:dq} for the $E=0$ wave function for different values of the hopping ratio $\rho$. Note the symmetry of $f_{E=0}$ around its maximum value. (Reproduced with permission from \cite{mace2017critical}).}
\label{fig.falphae0}
\end{figure}

$f_{E=0}(\alpha)$ curves obtained by the exact calculation for different values of $\rho$ are plotted in Fig.\ref{fig.falphae0}. As the figure shows, in each case the $\alpha$ values lie within a finite interval, indicating that $\psi$ is multifractal. As $\rho$ approaches 1, the support of the function shrinks to a single point, $\alpha=1$, corresponding to the extended state. Another point to note: thanks to the symmetry of the heights distribution,
the function $f_{E=0}(\alpha)$ is $symmetric$ around its maximum, as was already observed in numerical studies of this wave function \cite{evangelou,PhysRevB.40.7413fujiwara}. In other words, as the system size increases, the minima/maxima of $\psi$ scale to zero/infinity in the same manner. 

The transmission coefficient for this $E=0$ wave function can also be calculated exactly. It is interesting to note that this quantity (which will be discussed in section Sec.\ref{sec.transport}) has the exact value of 1 (i.e. transmission is perfect), for certain well defined distances along the chain, out to arbitrarily large distances.   

{\subsection{ Chern numbers. Bulk-edge correspondence}} \label{subsec.topological} The close connection between topological phases and quasicrystals which are described in a higher dimensional space has been pointed out by many authors \cite{PhysRevLett.109.106402kraus,PhysRevLett.110.076403verbin, PhysRevB.100.085119huang,PhysRevLett.121.126401huang}. These show that, for 1D quasicrystals, there are 2D Chern numbers and that there are topologically protected boundary states similar to those in a 2D quantum Hall system. 

\begin{figure}[htp]
	\centering
	\includegraphics[width=.45\textwidth]{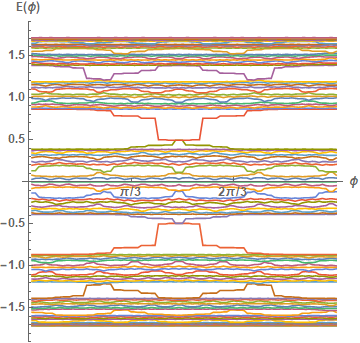}
	\caption{\small{Plot of the energy spectrum  versus $\phi$ for energy levels of the off-diagonal model in an open 89-site chain ($\rho=0.7$) . The number of gap-crossings of states is most easily counted for the largest gaps ($\vert \mathrm{q}\vert \leq 4$)}}
	\label{fig.crossings}
\end{figure}

It is expected that edge modes should be present in the quasicrystal, analogous to those in the AAH model. Just as changing the arbitrary phase $\phi$ in the AAH model Eq.\ref{eq.aubry} leads to tuning the edge mode energy, one can tune edge modes in FC approximants by varying the arbitrary phase $\varphi$ in Eq.\ref{eq.chij}.  This is seen in Fig.\ref{fig.crossings}, which shows for an 89-site chain the energy levels as a function of the parameter $0\leq \varphi \leq 2\pi$ (the figure was rendered symmetric with respect to $\pi$ by shifting the angle by $\varphi_0=-\omega\pi (N+1))$. It can be seen that the levels remain flat as $\phi$ is varied until a sudden phason flip occurs somewhere along the chain. There are, in all, $N$ such flips in the interval. The label of the gap $q$ gives the number of gap-crossings of the states, which are seen most clearly in the main gaps of the spectrum in Fig.\ref{fig.crossings}.

Experimental studies of the hopping Hamiltonian using a polaritonic cavity modes to detect eigenmodes and their energies can be performed, as shown in \cite{tanese2014fractal,baboux2017measuring}.
In the experiments, nearest neighbor cavities were spaced so as to be linked by a strong or a weak coupling, following the Fibonacci sequence. Chains of given length $N$, one for each $\phi$-value, were fabricated in a Fabry-Perot geometry, resulting in edge modes located at the central mirror symmetric position. For a given gap of label $\mathrm{q}$, the corresponding edge mode was observed to cross the gap $\mathrm{q}$ times, confirming that this is indeed a winding number. The sign of $\mathrm{q}$ determines the sense of the gap crossing (from the upper to lower edge, or vice versa).

\begin{figure}[htp]
	\centering
	\includegraphics[width=.2\textwidth]{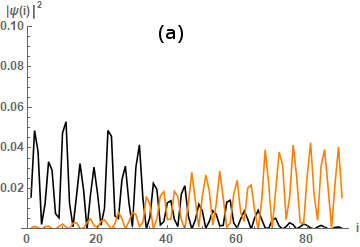}
	\includegraphics[width=.2\textwidth]{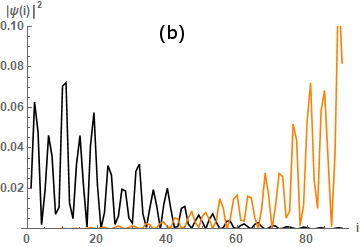}
	\includegraphics[width=.2\textwidth]{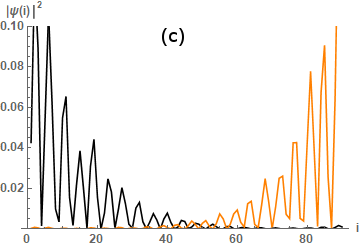}
	\includegraphics[width=.2\textwidth]{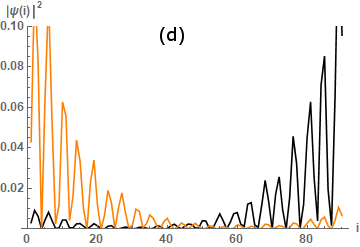}
	\includegraphics[width=.2\textwidth]{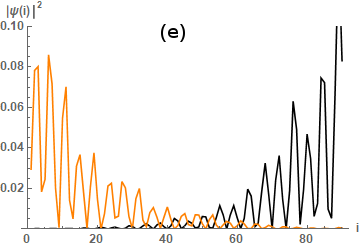}
	\includegraphics[width=.2\textwidth]{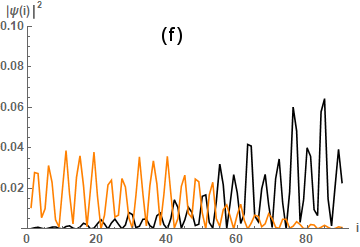}
	\caption{\small{Evolution of eigenstates on either side of the $\mathrm{q}=2$ gap showing the changes as $\phi$ is varied. In this series of figures the upper band edge state probability colored orange (light grey) and the lower edge state probability shown in black evolve progressively from extended in-band states, to localized edge states, which then exchange positions (hop between the edges in opposite directions), and finally become in-band states again. From a) through f) the values of $\phi$ are equal to $0.22\pi, 0.33\pi,0.44\pi,0.51\pi,0.62\pi$ and $0.71\pi$}}
	\label{fig.wfjumpings}
\end{figure}
One can use the winding property of states to ``pump" charge adiabatically across the chain, as was done in experiments in photonic quasicrystals by Kraus et al   \cite{PhysRevLett.109.106402kraus,PhysRevLett.110.076403verbin}. The ``jumping" of states from one edge to the other as the $\phi$ parameter is changed, is shown in Fig.\ref{fig.wfjumpings} for the $\mathrm{q}=2$ gap states (where there are two such jumps in the 2$\pi$ interval). This topological pumping of charge can have consequences for physical properties.  We will see later one consequence for the proximity effect in a chain coupled to a superconductor, discussed in Sec.\ref{sec.perturbations}.

R\"ontgen et al studied real space bond configurations for the appearance of edge modes in \cite{PhysRevB.99.214201rontgen}. They showed that the edge modes are linked to local ``resonators" -- this term used by them to denote clusters of bonds which are symmetric under reflection. The localization length and energy of the edge state depend on the resonator ``size" (i.e. the distance out to which they possess reflection symmetry). 

Chern numbers have, interestingly, also been observed by a light diffraction experiment \cite{PhysRevLett.119.215304dareau}. In the experiment, a digital micromirror device was used to realize a set of approximant chains of fixed length and different values of $\phi$ in Eq.\ref{eq.chij}. The behavior of the diffraction peak at different wave vectors $k$ was shown to depend on the associated topological number $\mathrm{q}$. 


\section{Approximate methods }\label{sec.approximate}
\subsection{Perturbation theories}
Many different kinds of perturbative calculations have been done to study electronic properties of the FC, both in real and reciprocal space.  

Luck \cite{luckPhysRevB.39.5834} carried out a perturbation expansion for the diagonal model in terms of the Fourier components of the potential. Taking the onsite energies to be $\epsilon_A=V$ and $\epsilon_B=-V$ by a suitable shift of the origin, one can compute the Fourier transform of the potential, $VG_N(k)$, and the structure factor, $S_N(k)=V^2\vert G_N(k)\vert^2$. In the thermodynamic limit the structure factor can be shown to have power law singularities at a dense set of reciprocal space vectors $k_0$ (as seen in Sec.\ref{subsec.cutproject}). In the vicinity of each of the peaks, one has
\begin{eqnarray}
S(k) dk \sim \vert k-k_0\vert^\alpha
\end{eqnarray}
For a general potential the singularity strength $\alpha$ can vary depending on the peak, whereas for the quasicrystal $\alpha=1$ for all peaks according to the arguments we presented in Sec.\ref{sec.geometry}. Luck showed for the general case that the width of the gap opened at the unperturbed energy $\epsilon(K=k_0/2)$ is related to $\alpha$, the singularity at $k=k_0$. Specifically, for the gap where the IDOS $N(E)=k_0/2\pi$,  the gap width is given by 
\begin{eqnarray}
\Delta \sim V^\beta
\end{eqnarray}
where $\beta=2/(2-\alpha)$. In the case of the quasicrystal, $\beta=1$ for all the gaps.
This analysis shows that, for weak quasiperiodic potentials, the plateaux of the IDOS are related in a natural way to the module of wave vectors. It should be noted that this perturbation theory does not converge, even for arbitrarily weak potentials, as pointed out by Kalugin et al \cite{kaluginlevitov}, because of the nature of the Fourier module of the quasicrystal -- consisting of a dense distribution of peaks. Nevertheless, the indexing of gaps using this method is robust. The gap labeling theorem \cite{bellissardgaps} provides the rigorous justification that the indexing continues to hold for arbitrarily strong potentials. 
Many other perturbative approaches have been proposed for the off-diagonal model, including a real space perturbation theory starting from the periodic limit \cite{sireperturb,rudingersireperturb}. The method of Sire and Mosseri when applied to the Fibonacci chain yields exact expressions for positions of the gaps and associated gap labels, and perturbative results for gap widths. The latter hold well for the main gaps even for moderately large perturbation, but not for small gaps.  For the diagonal model, Barache and Luck \cite{luckPhysRevB.49.15004} have introduced a perturbation theory which starts from a strong atomic limit $V_i = \pm V$, where the onsite potential strength $V$ is large compared to hopping amplitude $t$. The spectrum and density of states were computed in degenerate perturbation theory, and the gap structure deduced for this case was shown to be consistent with the gap labeling theorem.

\subsection{Approximate renormalization group}
This subsection describes the main ideas behind a perturbative real space RG method due to Niu and Nori \cite{niu1990spectral,PhysRevLett.57.2057niuprl} and Kalugin, Kitaev and Levitov \cite{kaluginlevitov,levitov89}. This approach has been extremely fruitful for describing a great number of static and dynamic properties of the FC. In this section we will describe the basic notions of this RG for the off-diagonal and the diagonal models. 

\bigskip
\noindent
{\bf {RG for off-diagonal model (1)} }
We begin with details of the RG for the off-diagonal model  Eq.\ref{eq.hophamilt}, where $t_B > t_A$. The hopping ratio $\rho=t_A/t_B$ lies in the range $0 \leq \rho \leq 1$. Recall that one can assume that both $t_A$ and $t_B$ are positive for, if not, the solutions can be found from our model by a suitable mapping - or ``local gauge transformation" of the wavefunctions. The goal of this RG is to obtain a description of the spectrum and states perturbatively in $\rho$. 

\begin{figure}[htp]
	\includegraphics[width=.5\textwidth]{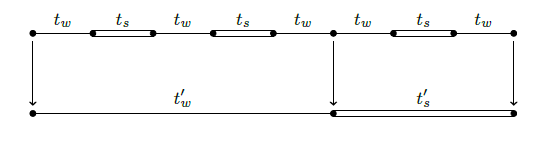}
	\caption{\small{Illustration of the molecular deflation rule: the fifth approximant is transformed to the third (figure reproduced from \cite{mace2016fractal}).}}
\label{fig:mol_defl}
\end{figure}

For $\rho=0$, the chain breaks up into disconnected groups of sites which can be classified as follows:\\
atom sites -- the sites sandwiched between A bonds \\
molecule sites -- pairs of sites linked by a $B$ bond \\
The spectrum, in this limit, consists of only three discrete degenerate levels: the $E=0$ level of the atoms, and $E=\pm t_B$ for the molecular bonding/antibonding levels. For a chain of $N=F_{n}$ sites, the degeneracy of the $E=0$ level is  given by the number of atoms, $F_{n-3}$. The degeneracy of the $E=\pm t_B$ levels is given by the number of molecules, $F_{n-2}$, in the chain.  
For small non-zero $\rho \ll 1$, these three levels split into three clusters of levels, the molecular bonding (+m) and antibonding (-m) bands and the atomic (0) cluster. The separations between these three clusters for small $\rho$ is roughly $t_B$. In perturbation theory, the three clusters do not mix, and can be treated as three independent systems for the calculation of the effective Hamiltonians. 
\\
{\it{Molecular RG} } Consider the Hamiltonian $H$ for a Fibonacci chain of length $F_n$ and consider the lowest molecular bonding level  located at the energy $-t_B$ and having a wavefunction which is non-zero on the molecule sites. It is easy to check that the chain formed by molecules is precisely the $n-2$th approximant chain. It can be shown using degenerate perturbation theory  \cite{PhysRevLett.57.2057niuprl,niu1990spectral,kaluginlevitov} that, upto an overall constant shift, the new effective Hamiltonian $H'$ is again a Fibonacci hopping Hamiltonian, with the renormalized hopping amplitudes   $t_A'$ and $t_B'$. The old chain and the new chain after decimation of atoms are indicated in Fig.\ref{fig:mol_defl}, along with the two new hopping amplitudes. The renormalized hopping amplitudes are given to lowest order in $\rho$ by
\begin{eqnarray}
t_A' = z t_A  \nonumber \\ t_B' = z t_B 
\end{eqnarray}
where $z=\rho/2$. Note that the hopping ratio is unchanged to lowest order under RG, as the new weak and strong hopping amplitudes satisfy $t'_A/t'_B=\rho'=\rho$. To summarize, the effective Hamiltonian for the bonding set of levels is, up to a global shift, that of the $n-2$ approximant chain with renormalized hoppings. The original level located at $E=-t_B$ is split into 3 levels, which can be labeled $-+$, $-0$ and $--$, separated by gaps of width $t_B'$.

A similar analysis shows that the effective Hamiltonian for the antibonding levels ``+" is an identical FC of $F_{n-2}$ sites with hopping amplitudes given by $t_A' = z t_A $ (weak) and $ t_B' = -z t_B$ (strong). The original level located at $E=-t_B$ is split into 3 levels, labeled $++$ $+0$ and $+-$.

\begin{figure}[htp]
	\includegraphics[width=.5\textwidth]{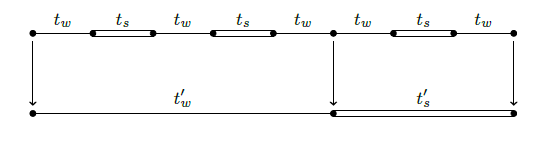}
	\caption{\small{Illustration of the atomic deflation rule: here the fifth approximant is transformed to the second (figure reproduced from \cite{mace2016fractal}).}}
\label{fig:at_defl}
\end{figure}

{\it{Atomic RG } } Consider the central band located around $E=0$. The levels around $E=0$ correspond to wavefunctions which are largest on the atom sites, of which there are $F_{n-3}$. It is easy to check that the new chain formed by these atom sites is nothing but the $n-3$th approximant chain.
Degenerate perturbation theory shows that the effective Hamiltonian $H'$ depends on two new renormalized hopping amplitudes. The new strong and weak bonds, $t_A''$ and $t_B''$, are given to lowest order in $\rho$ by
\begin{eqnarray}
t_A'' = \overline{z} t_A \nonumber \\ 
t_B'' = \overline{z} t_B \end{eqnarray}
where $\overline{z}=\rho^2$. As for the molecular RG, the hopping ratio is preserved, since the new weak and strong hopping amplitudes satisfy $\rho''=t_A''/t_B''=\rho$. To summarize, the effective Hamiltonian for the atom set of levels is the Hamiltonian of a chain of $F_{n-3}$ sites and with renormalized hoppings. The original level is therefore split into 3 levels, labeled $0+$, $00$ and $0-$. These are separated by gaps of width  $t_B''$.

\begin{figure}
		\includegraphics[width=0.4\textwidth]{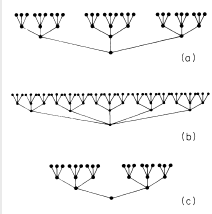}
\caption{ Cluster and subclustering structures for the three different models described in Sec.\ref{sec.approximate} (figure reprinted from \cite{niu1990spectral} with permission) }
\label{fig.clusters}
\end{figure}

This process can be repeated until one reaches the three first chains. The result for the clustering structure is a succession of trifurcations as illustrated in Fig.\ref{fig.clusters}a). One can reverse the process, alternatively, and track each band as it splits into three sub-bands when $n$ increases by 2 or by 3. Doing this, one sees that each of the $F_n$ levels of the spectrum follows a unique path under successive renormalizations  (termed renormalization path)  of the form $\{c_1c_2c_3....\}$ where $c_i$ can take the three values $0,\pm 1$. Note that this trifurcation scheme also holds for the critical AAH model \cite{kohmotohiramotoPRB}.

\bigskip
\noindent
{\bf{RG for off-diagonal model (2) }} Consider the FC in which the strong bond corresponds to A-bonds. Again, one can assume, as for the preceding model that the amplitudes are positive without loss of generality, with $t_A \ll t_B$ . The perturbation theory is now carried out in powers of $t_B/t_A$. Inspection of the FC shows that in the limit $t_B=0$ the chain breaks up into diatomic and triatomic molecules. These molecules give rise to five energy levels $E/t_A= \pm \sqrt{2}, \pm 1, 0$. It can be shown that the new effective Hamiltonians within each of the five bands is a Fibonacci hopping Hamiltonian  \cite{niu1990spectral} of the type discussed above (model A). Thus each of the five levels trifurcates, and continue thereafter to trifurcate under successive RG steps. This is illustrated in Fig.\ref{fig.clusters}b.

\bigskip
\noindent
{\bf{RG for diagonal model}} For the diagonal model Eq.\ref{eq.diaghamilt}, a perturbation theory in $t/(\epsilon_A-\epsilon_B)$ shows once again the recursive structure of the energy spectrum. For $t=0$, one has two isolated levels, $E=\epsilon_A$ (degeneracy $F_{n-1}$) and $E=\epsilon_B$ (degeneracy $F_{n-2}$). For small non-zero $t$, one can compute the new effective Hamiltonians in perturbation theory. It is found that these are given, once again, by two hopping parameters, one strong and one weak. Thus after one RG step, we are led back to the Fibonacci hopping model, leading to a splitting into three levels, and thereafter, with each successive RG step, trifurcations.  This is illustrated in Fig.\ref{fig.clusters}c.

\section{Multifractal spectrum and states of the off-diagonal model}\label{sec.rghopping}
In this section we will review the use of the perturbative RG method described above to obtain a variety of multifractal properties of the diagonal (pure hopping) model. Sec.\ref{subsec.spectral} shows how RG methods introduced in the last section can be applied to this case to compute spectral properties as done in \cite{PhysRevA.35.1467zheng,piechon1995analytical}. Gap structures are taken up in detail in Sec.V.B. Wavefunctions are considered in Sec.V.C.  

\subsection{Multifractality of the energy spectrum}\label{subsec.spectral}

\begin{figure}
		\includegraphics[angle=270,width=0.4\textwidth]{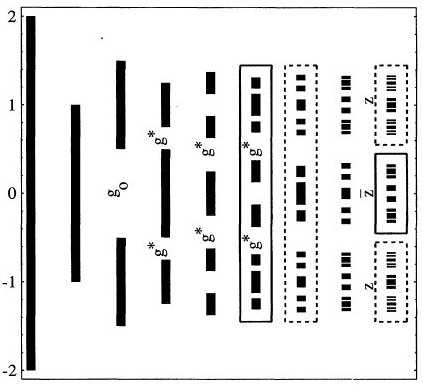}
\caption{ Schematic view of the recursive construction of spectra using RG. The first three spectra for $n=1,2$ (single band) and $n=3$ (two bands) are shown at the top of the figure. The $n$th spectrum is obtained by the union of the $n-2$th and $n-3$th spectra multiplied by the RG factors $z$ and $\overline{z}$, and shifted as described in Eq.\ref{eq:recur_ham}. The labels $g_0$ and $g^*$ refer to transient and stable gaps, see text (figure reprinted from \cite{piechon1995analytical}) }
\label{fig.recursionbands}
\end{figure}

The RG method described in the preceding section showed that the spectrum of a chain of number of sites equal to $F_n$ can be mapped, after one RG step, to the spectra of two shorter chains $F_{n-2}$ and $F_{n-3}$. We now apply this to obtain quantitative information on the spectral properties, following Zheng \cite{PhysRevA.35.1467zheng} and Pi\`echon et al  \cite{piechon1995analytical}. This can be formally expressed by the relation
\begin{equation}
\label{eq:recur_ham}
	H_n = \underbrace{\left( z H_{n-2} - t_s \right)}_{\text{bonding levels}} \oplus \underbrace{\left( \zb H_{n-3} \right)}_{\text{atomic levels}} \oplus \underbrace{\left( z H_{n-2} + t_s \right)}_{\text{antibonding levels}} + \mathcal{O}(\rho^4)
\end{equation}
to lowest order in $\rho$. Thus, given the first three spectra $W^{(n)}$ corresponding to $n=0,1,2$, one can construct all the $n>2$th generation spectra. The spectra of the first three chains are simple to obtain. For $n=0$ and $n=1$ for example, the approximant chains have only one hopping amplitude, $t_B$ or $t_A$. Applying periodic boundary conditions the spectrum for $n=0$ is the band $-2t_B< E < t_B$, with a band width of $\Delta^{(0)}=2t_B$. The spectrum of the $n=1$ chain is a narrower band $-2t_A < E < t_A$. The $n=2$ chain is an alternating sequence of $t_A$ and $t_B$, thus the spectrum has two bands separated by a gap as shown in Fig.\ref{fig.recursionbands}. One can now proceed to construct the spectrum for $n=3$. $W^{(3)}$ is composed of two (bonding and antibonding) lateral bands and one central (atom) band. The side bands are simply $W^{(1)}$ multiplied by the factor $z$ and translated in energy by $\pm t_B$. The central band is
$W^{(0)}$ multiplied by the factor $\overline{z}$. This procedure can be used to construct all the spectra shown in the figure.

\begin{figure}[htp]
\centering
\includegraphics[width=.4\textwidth]{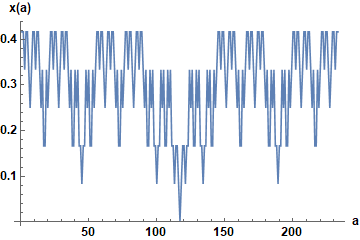}
\caption{\small{ $x(a)$ plotted versus the level index $a$ for each of the energy levels $E_a$ of the $n=12$ chain. Lines are drawn to guide the eye.}}
\label{fig:xvals}
\end{figure}

In each RG step, the band widths are reduced by the factors $z$ or $\overline{z}$. The resulting band width $\Delta$ of a given level in the $n$th approximant depends on the sequence of RG steps that were taken. A simple example is the two bands at the top and bottom edges of the spectrum, which remain always molecular, throughout successive RG transformations. They have the RG paths $111...$ and $\overline{1}\overline{1}\overline{1}...$ having $\sim n/2$ steps. Thus the band width of the first level and its scaling exponent $\alpha_{edge}$, defined through $\Delta_{edge}\sim F_n^{-1/\alpha} = \omega^{n/\alpha}$ are given by 
\begin{eqnarray}
\Delta_{edge} = z^{\frac{n}{2}}t_A \nonumber \\
\alpha_{edge} = \log(\omega^2)/\log(z)
\end{eqnarray}
to lowest order in $\rho$. The second simple case concerns the atomic level at $E=0$, which has a RG path of $000...$ having $n/3$ steps (taking $n$ a multiple of 3). Thus its band width and its scaling exponent $\alpha_{ctr}$ are given by 
\begin{eqnarray}
\Delta_{ctr} =\overline{z}^{\frac{n}{3}}t_A \nonumber \\
\alpha_{ctr} = \log(\omega^3)/\log(\overline{z})
\end{eqnarray}
Comparison with the exact results of Kohmoto et al for these two exponents obtained using the trace map, one sees that these values of $\alpha$ represent the first terms of an expansion in $\rho$.
Other levels have a mixture of atom and molecular RG, so that the scaling is given in general by $z^{n_m}\overline{z}^{n_a}$ where $n_m(a)$ is the number of molecular(atomic) RG steps in its RG path. These numbers are not independent as they must satisfy the condition $n \sim 2n_m+3n_a$. It is useful now to introduce the variable $x=n_m/n$ as a measure of the degree to which a given RG path has molecular character. For very long chains, $x$ is a continuous variable in the interval $(0,\frac{1}{2})$. The smallest value, $x=0$, of a purely atom state, corresponds to the energy in the middle of the spectrum. The maximum value $x=1/2$ corresponds to the levels at band edges which are molecular states at every stage of the RG. Fig.\ref{fig:xvals} shows the values of $x$ versus the level index, for the $n=12$ approximant. Note that, in general, many different levels can share a given value of $x\neq 0$, whereas the value $x=0$ corresponds to a single state which occurs only in every third chain ($n$ a multiple of 3).

The exponent $\alpha$ can be computed for a given value of given $x$, and is given by
\begin{eqnarray}
\alpha &=& -\frac{\log F_n}{\log \Delta} \nonumber \\
&=& \frac{n\log \omega}{(n_m \log z + n_a \log \overline{z})}\nonumber \\
&=& \frac{\log \omega}{x \log z/\overline{z}^{2/3} + \log \overline{z}^{1/3}}
\label{eq.alphax}
\end{eqnarray}

Let the number of levels scaling with a power $\alpha$ be $N \sim F_n^{f(\alpha)}$, thus defining the function $f(\alpha)$. This is just the number of levels corresponding to a given value of $n_a$ and $n_m$, which is $2^{n_m} n_m!/(n_m+n_a)!$. With the use of Stirling's approximation  this leads to $N(x) \sim F_n^{g(x)}$ with

\begin{eqnarray}
g(x) = \frac{1}{\log \omega}(\frac{x}{2} \log 3x + \frac{1+x}{3} \log (1+x) \nonumber \\
+\frac{1-2x}{3} \log (1-2x))
\label{eq.gx}
\end{eqnarray}
Eqs.\ref{eq.alphax} and \ref{eq.gx} determine the function $f(\alpha)$, and describe the multifractal scaling of the spectrum. The exponent for any band can be computed if its RG path is specified. The expressions for $\alpha_{ctr}$ and $\alpha_{edge}$ are the leading terms in an expansion in small $\rho$ of the exact formulae Eqs.\ref{eq.alphactr} and \ref{eq.alphaedge} obtained by Kohmoto et al. We note finally that according to the results in Eqs.\ref{eq.alphax} and \ref{eq.gx} there is a special point when $\rho=1/8$. At this point $z=\overline{z}^{2/3}$, and the spectrum becomes mono-fractal, because all bands scale in the same way according to our perturbation theory. However, in fact, as we have mentioned in the discussion at the end of Sec.\ref{subsec.transfer}, exact results show that the spectrum is not a monofractal for any $\rho$. Higher order terms must be considered, in order to resolve the apparent discrepancy. A similar observation holds for the wavefunctions  (subsection C) where the lowest order calculation yields an ordinary fractal, with multifractality appearing only at the next order in the perturbative RG.

The scaling of the total band width, $B^{(n)}=\sum_j^{F_n} \Delta^{(n)}_j$ with the system size can now be determined. From Eq.\ref{eq:recur_ham}, one can deduce a recursion equation relating the total band width of the $n$th chain to those of the $n-2$ and $n-3$ chains, as follows:
\begin{eqnarray}
B^{(n)} =2z B^{(n-2)} + \overline{z}B^{(n-3)} 
\label{eq:recurwidth}
\end{eqnarray}
Defining the exponent $b$ by $B^{(n)}\sim F_n^{-b}$ for large $n$, from Eq.\ref{eq:recurwidth} one sees that $b$ must satisfy the equation
\begin{eqnarray}
1 =2z \omega^{-2b} + \overline{z}\omega^{-3b} 
\label{eq:measure}
\end{eqnarray}
Recursive relations can be written likewise for all of the moments of the density of states (i.e. the inverse band width $\Delta_j^{-1}$). Recall that the generalized dimensions for the spectrum are the exponents corresponding to the $q$th moment of the DOS. Following the thermodynamical formalism introduced before, one defines the partition function $\Gamma^{(n)}(q,\tau) = {F_n}^{-q}\sum_j^{F_n} \Delta_j^{-\tau} $, where $\tau_q =D_q(q-1)$.  This partition function obeys the recursion relation 
\begin{eqnarray}
\Gamma^{(n)}(q,\tau)  = 2\frac{\omega_n^{2q}}{z^\tau} \Gamma^{(n-2)}(q,\tau) + \frac{\omega_n^{3q}}{\overline{z}^\tau} \Gamma^{(n-3)}(q,\tau)
\label{eq:partfn}
\end{eqnarray}
For each $q$, the corresponding $\tau$ value is obtained by requiring that $\Gamma$ be stationary. This results in the condition
\begin{eqnarray}
1 =2 \omega^{2q}z^{(1-q)D_q} + \omega^{3q} \overline{z}^{(1-q)D_q}
\label{eq:tauofq}
\end{eqnarray}

{\it {Relations for generalized dimensions}}  
The Hausdorff dimension $D_F$ of the spectrum is given by $D_0$, which satisfies the equation 
\begin{eqnarray}
2z^{D_F} + \overline{z}^{D_F}=1
\end{eqnarray}
Although derived in the limit of small values $\rho\ll 1$, this relation nevertheless gives a rather good value even for relatively large values of $\rho$ -- one obtains $D_F=0.76$ for $\rho=0.5$. \footnote{Strictly speaking, this approach is valid only for strong quasiperiodic modulations. However, these results remain pertinent even for moderate to weak quasiperiodicity, and calculations on finite chains show that, when the gaps are opened, they persist for all $\rho$, closing only in the periodic case.} The information dimension, $D_1$, enters in an inequality for the diffusion exponent, discussed in Sec.\ref{sec.dynamics}. The exponent $D_2$ is also of special interest, in particular, for dynamics, as will be shown later in Sec.\ref{sec.dynamics}. One sees, from Eqs.\ref{eq:tauofq} and \ref{eq:measure} that the band width exponent $b$ is related to the $D_q$ via $D_\delta=1/(1+\delta)$.

\subsection{Gaps, stable gaps and topological numbers} 
In the construction of spectra with the RG recursion scheme, it becomes apparent that two kinds of gaps appear in the spectra of approximant chains: there are transient gaps and stable gaps. To understand these notions consider the spectra of the first few approximants shown in Fig.\ref{fig.recursionbands}: one sees that the spectrum for $n=2$ has a gap labeled $g_0$ which disappears for $n=3$ and $n=4$, reappearing as a smaller gap for $n=5$, and going to zero as $n$ tends to infinity. This is an example of a transient gap. Stable gaps are descendants of the gaps marked $g^*$ whose widths remain finite. Writing recursion relations for the stable gap distribution, $P(g)$, one can show that it has a power law form
\begin{eqnarray}
P(g) \sim g^{-(1+D_F)}
\end{eqnarray}
The limiting value as $n$ tends to infinity of the two main gaps, $g^*$, and the width of the spectrum $\Delta^*$ have also been computed in terms of $\rho$ (see \cite{piechon1995analytical} for details).

\begin{figure}[htp]
	\centering
	\includegraphics[width=.45\textwidth]{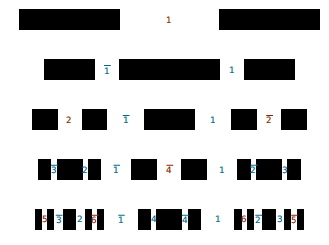}
	\caption{\small{Recursive construction of the spectra of approximant chains showing the gap structure. Labels give the values of $\mathrm{q}$ determined according to Eq.\ref{eq:gaplabels} in red for transient, and black for stable gaps (overbars indicate negative sign). Figure reprinted from \cite{mace2017gap}}}
	\label{fig:gaplabels}
\end{figure}

{\it {Gap labeling}} Given the recursive structure of the spectrum, it is easy to determine the labels for each of the gaps of the system of $N=F_n$ levels. For the $j$th plateau of the IDOS $N(E)=N_j/N$, one must solve the relation $N_j=\mathrm{Mod}[\mathrm{q} F_{n-1},F_n]$ to obtain $\mathrm{q}$.
Stable gaps and transient gaps are indicated by different colors for their $\mathrm{q}$ labels  in Fig.\ref{fig:gaplabels} (black for stable and red for transient). Stable gaps have the lowest values of $\mathrm{q}$ and these are stable gap labels, i.e. independent of system size. In contrast transient gaps have large values of $\mathrm{q}$ which depend on the system size, and these gaps vanish in the infinite size limit.

\begin{figure}
		\includegraphics[width=0.5\textwidth]{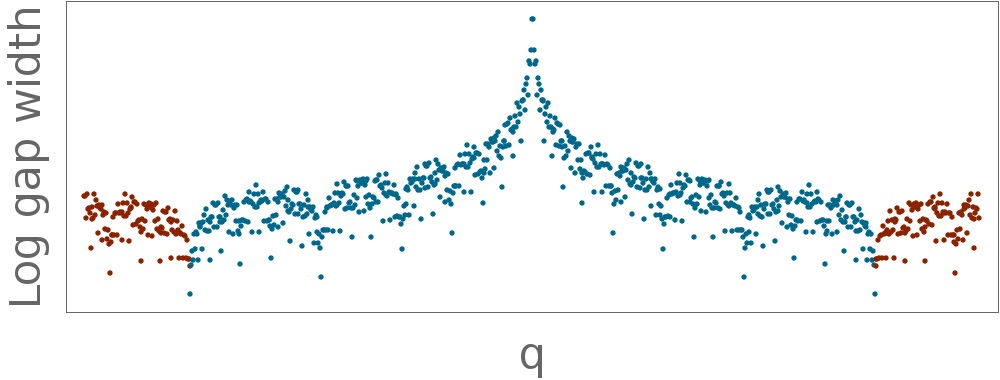}
\caption{Gap widths plotted on a log scale versus the topological index $\mathrm{q}$ for the $n=16$ approximant in the off-diagonal model. Red (darker grey) data points stand for transient gaps (see text) (figure reprinted from \cite{mace2017gap}) }
\label{fig.gaps}
\end{figure}

The gap widths tend to decrease with $\mathrm{q}$, although not monotonically. This is shown in Fig.\ref{fig.gaps} which shows the result of gap widths plotted against $\mathrm{q}$, for the $n=16$ approximant. These were computed for the pure-hopping model using an approximate renormalization group, as outlined in Sec.\ref{sec.approximate}. The black (red) data points correspond to stable (transient) gaps respectively (these are gaps which tend to finite (zero) width as $n$ tends to infinity). The gap widths have log-periodic oscillations in $\mathrm{q}$, and one sees a self-similar structure. It is also interesting to note that the smallest values of $\mathrm{q}$ up to a certain maximum correspond to stable gaps. For $\mathrm{q}$ above a certain value, there are only transient gaps.

\subsection{Multifractality of  wavefunctions}\label{subsec.wavefuncs}
The RG construction of the energy spectrum in Sec.\ref{subsec.spectral} has its parallel for the construction of wavefunctions, as was already noticed by \cite{PhysRevLett.57.2057niuprl}. Consider the wavefunction $\psi(E)$ for an allowed energy $E$. If the energy is located in a side-band, the support of  $\psi(E)$ is  concentrated on the molecular sites. If the energy is located in the middle-band, the support of  $\psi(E)$ is  concentrated on atom sites. Under RG, the initial chain is transformed to a shorter chain, and the site $i$ maps to site $i'$ of the new chain. One can introduce, as we did for the energy recursion relations, wavefunction renormalization factors $\lambda$ and $\overline{\lambda}$, and write 
\begin{eqnarray}
			|\psi_i^{(n)}(E)|^2 = \lb |\psi_{i'}^{(n-3)}(E')|^2 \text{~if $E$ is atomic} \nonumber \\
		|\psi_{l,r}^{(n)}(E)|^2 = \lambda |\psi_{i'}^{(n-2)}(E')|^2 \text{~if }E\text{~is molecular}
		\label{eq:wfrg}
\end{eqnarray}
where $E'$ denotes the energy after renormalization. In the case of molecular RG (second line), there are two sites $l$ (left) and $r$(right) forming the molecule  corresponding to the site $i'$. The wavefunction renormalization factors are given by $\lambda\approx 1/2$ and $\lb \approx 1$, to lowest order in $\rho$. The higher order corrections are important to keep for a correct description of multifractality, as shown in \cite{mace2016fractal}.  Given the RG path of a state, with the help of Eqs.\ref{eq:wfrg}, one can reconstruct the corresponding $\psi(E)$. 

 		\begin{figure}
		\includegraphics[width=0.3\textwidth]{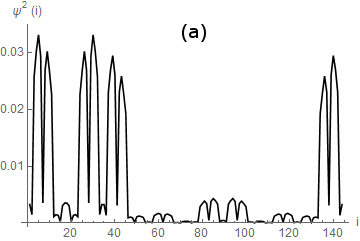}
		\includegraphics[width=0.3\textwidth]{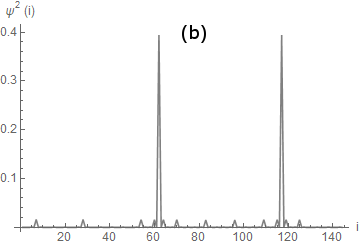}
		\includegraphics[width=0.3\textwidth]{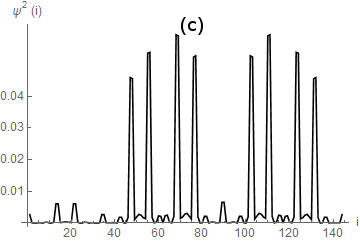}
\caption{(top) Ground state probabilities $\psi_i^2$ plotted versus site index $i$. (middle) $E=0$ state probabilities $\psi_i^2$ plotted versus site index $i$. (bottom) Probabilities $\psi_i^2$ plotted versus site index $i$ for state $a=34$. Data obtained by diagonalization for N=144 chain with PBC for $\rho=0.2$. }
\label{fig.wavefns}
\end{figure}

To illustrate the different structures that are obtained, let us consider some examples. The RG path of the lowest level of the spectrum is $\{1111...\}$, and the wavefunction constructed recursively has the largest amplitudes on pairs of sites which derive from molecules on a larger length scale, which derive from still bigger molecules and so on. The top part of Fig.\ref{fig.wavefns}, shows the wavefunction for a chain of 144 sites ($n=11$) as computed numerically for a value of $\rho=0.2$. One sees here the characteristic double-peak composed of double-peak structure on several scales. The figure shows deviations from the RG construction that we outlined: there are for example small peaks which we ignored in our lowest order approximation. There are asymmetries in the amplitudes of the double peaks. These are due to higher order terms in $\rho$, which in our example, is not particularly small.  The middle figure shows the wavefunction in the same chain for the energy $E=0$, whose RG path is $\{0,0,0,...\}$. The peaks are this time primarily localized on two atom sites. Finally the bottom figure shows an example of a randomly chosen mixed wavefunction that has both atomic and molecular components in its construction.

\bigskip
\noindent
{\bf{Hamiltonian in the conumber basis}} In preparation for the discussion of wavefunctions, we discuss the representation of the Hamiltonian in the conumbering basis introduced in Sec.\ref{subsub.conum}. In this basis, the hopping Hamiltonian takes the form of a T\"oplitz matrix, where the non-zero elements lie at the distance $F_{n-2}$ and $F_{n-1}$ from the principal diagonal \cite{siremoss90}. For example, the Hamiltonian for the $n=5$ chain ($N_n=8$) can be written as follows
\[ 
 H = 
 \begin{bmatrix*}[r]
    .&.&.& t_A &.& t_B&.&. \\
    .&.&.&.& t_A &.& t_B&.\\
   .&.&.&.&.&. t_A &.& t_B\\
  t_A &.&.&.&.&.& t_A &.\\
    .&t_A&.&.&.&.&.& t_A \\
   t_B.& .&t_A&.&.&.&.&.\\
   .&t_B.& .&t_A&.&.&.&.\\
   .&.&t_B.& .&t_A&.&.&.
  \end{bmatrix*}
\]
As can be read off directly from the matrix, sites numbered 1 through 3 and 6 through 9 have a strong bond $t_B$ and thus form molecules in the first RG step, while sites 4 and 5 have weak bonds to either side and are atom sites. Within the two groups of $F_{n-2}$ molecular sites, under a second RG step, one further has a subgrouping into $F_{n-4}$ molecules and $F_{n-5}$ atoms of ``second generation". This sub-grouping occurs in the middle block as well. {\it{The conumbering scheme thus automatically classifies sites according to the same rules as the band structure.}} This remark has its importance for the discussion of wavefunctions which follows next.

\begin{figure}[htp]
\centering
  \includegraphics[width=.45\textwidth]{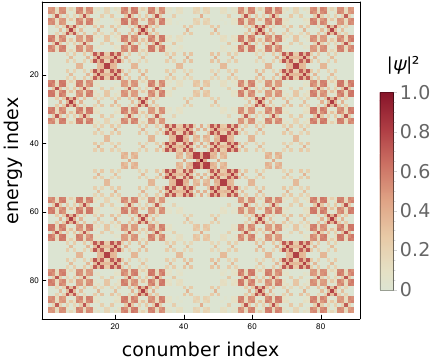}
    \vskip 1cm
  \includegraphics[width=.4\textwidth]{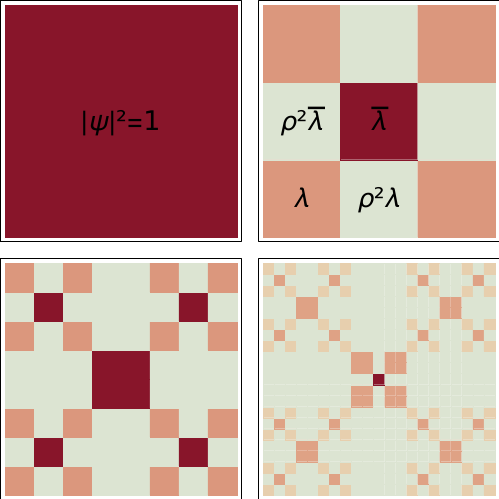}

\caption {\small {Upper figure: the numerically computed intensity plot of wavefunctions for the $N=89$ approximant. Intensities are shown for each of the sites in the conumber basis (x-axis) and for every level (y-axis). Lower figure: the first few steps of the geometrical construction of the wavefunctions according to perturbation theory Eq.\ref{eq:wfrg}. Note the symmetry under reflection with respect to the diagonal, which holds only for our simplified RG treatment. (Figures reproduced from \cite{mace2016fractal})}}
\label{fig:wf_iDoS}
\end{figure}

\bigskip
\noindent
{\bf {Energy-position symmetry}} As we have said, the construction of states and of the spectrum follow the same schema. Classifying sites according to their conumber corresponds, in fact, exactly to way energies are ordered. This leads to a remarkable approximate symmetry between states and energies. Fig.\ref{fig:wf_iDoS}a shows an intensity plot of the numerically calculated (for an 89 sites chain with periodic boundary conditions) values of $\psi(E_a)j^2$ plotted against the conumber $j$ for each allowed energy $E_a$. Fig.\ref{fig:wf_iDoS}b shows the result for the wavefunctions after four RG steps. Two observations can be made: i) the similarity of the RG-constructed and the numerical data is manifest and ii) the figures show a reflection symmetry with respect to the diagonal, i.e. if $i$ and $a$ represent the position and the energy, then
\begin{eqnarray}
\vert\psi_{i,a}\vert^2=\vert\psi_{a,i}\vert^2
\end{eqnarray}
to lowest order in $\rho$. This striking symmetry between spatial and spectral variables holds for sufficiently small $\rho$.

\bigskip
\noindent
{\bf{Multifractal exponents for wavefunctions}}

\begin{figure}[htp]
\centering
\includegraphics[width=.45\textwidth]{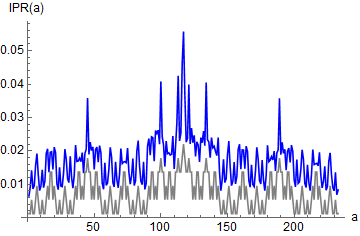}
\caption{\small{Shown in blue (dark grey) are values of the IPR ($\chi_2$) computed for all states of the hopping model for an $n=12$ chain ($\rho=0.25$). The light grey line is obtained by reflecting the plot of x-values shown in Fig.\ref{fig:xvals}. An arbitrary scale factor and shift were applied to the gray curve to facilitate comparison with the IPR.}}
\label{fig:ipr}
\end{figure}

The recursion relations for  wavefunctions are analogous to those presented for the spectrum. These relations involve the two different rescaling factors in the recursion formulae, $\lambda$ and $\lb$. Thus all the wavefunctions are multifractal, in the same way that the spectrum is multifractal due to the presence of two distinct shrinking factors $z$ and $\zb$. Defining the fractal dimensions $\wf_q$ for the wavefunction $\psi(E)$ by

\begin{equation}
\label{eq.chifn}
	\chi_q^n(E) = \sum_i |\psi_i^n(E)|^{2q} \sim \left( \frac{1}{F_n} \right)^{(q-1)\wf_q(E)}
\end{equation}
For $q=2$ the quantity $\chi_2(E)$ is nothing but the inverse participation ratio (IPR). For a given system size, the inverse of the IPR provides an indication of the spatial spread of the wavefunction. The scaling of the IPR as a function of the system size determines whether or not a state is localized. The exponent $\wf_2(E)$ is an often used indicator helping to locate the  metal-insulator transition in, for example, the 3D Anderson model. 
To recall, the value $D^\psi_2(E) = 1$ indicates that the state $E$ is extended, while $D^\psi_2(E) = 0$ characterizes a localized state.
Intermediate values $0 < D^\psi_2(E) < 1$ are a signature of a critical state. Fig.\ref{fig:ipr} shows the IPR computed numerically for $\rho=0.5$  for the $n=12$ approximant (blue curve). The gray curve is obtained by reflecting Fig.\ref{fig:xvals} with respect to a horizontal axis and translating up. The IPR-curve clearly tracks the curve corresponding to inversed $x$ values. This anti-correlation can be qualitatively explained for sufficiently small $\rho$ by the observation that firstly, molecular states are more delocalized (has lower IPR) than atom states, and secondly, the number of molecular RG steps is given by the variable $x$. Therefore the higher the $x$-value, the lower the IPR of the state. 

On the quantitative level, the full set of exponents $D^\psi_q(E)$ can be 
computed in the perturbative RG scheme \cite{mace2016fractal}. To lowest order, these are given by the value of $x$ (which measures the extent to which a given state is of molecular type), as follows
\begin{equation}
\label{eq:dqpsi0}
	\wf_{q,0}(E) =- x(E) \frac{\log 2}{\log \omega} + \mathcal{O}(\rho^2)
\end{equation}
This simple result for very small $\rho$ gives a monofractal, since the fractal dimensions do not depend on $q$ at leading order. For larger $\rho$, however, the higher order corrections show that the wavefunctions are indeed multifractal (see Mace et al \cite{mace2016fractal}). For small $\rho$, Eq.\ref{eq:dqpsi0} shows that the smaller $x$, the smaller the fractal dimension.The most extended states, according to Eq.\ref{eq:dqpsi0} are those at the spectrum edges, where $x=\frac{1}{2}$. The equation also predicts that $\wf_{q,0}(E)=0$ upto higher order corrections for the level in the center of the band where $x=0$. This indicates that the state is localized or close to being localized. However, as the exact calculations in the preceding section Sec.\ref{subsec.wf0} showed, it is in fact a critical state.  The discrepancy is corrected by including the higher order corrections which are lacking in Eq.\ref{eq:dqpsi0}, as was shown in \cite{mace2016fractal}.

\begin{figure}[htp]
	\centering
	\includegraphics[width=.45\textwidth]{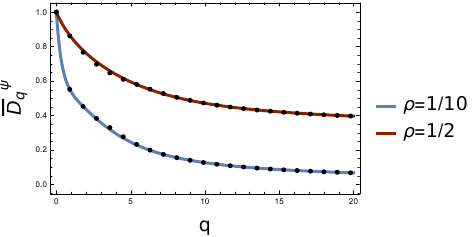}
	\caption{\small{The averaged fractal dimensions of the wavefunctions $\avwf_q$ as a function of $q$, for $\rho = 0.1, 0.5$. Dots: numerical results, solid line: theoretical predictions (Figure reproduced from \cite{mace2016fractal}).}}
	\label{fig:avwf}
\end{figure}

\bigskip
\noindent
{\bf{Spectrally averaged wavefunction dimensions}} In certain contexts, one may need to know, not the behavior of a single eigenstate, but the average behavior of wavefunctions close to a certain energy (such as the Fermi energy). One can for example wish to determine the average value of the fractal dimension $D^\psi_2(E)$, within some energy interval $\Delta E$. This averaged exponent occurs for example in some rigorous inequalities for dynamical quantities as described in the next section. The averaged wavefunction exponents could also be relevant for other physical properties, such as the Kondo screening of impurities. These exponents can be calculated by considering a generalization of the $\chi_q$-function in Eq.\ref{eq.chifn} to include a sum over all energies. The results for the averaged dimensions $\overline{D}^\psi_2(E)$, for two different values of $\rho$, are shown in Fig.\ref{fig:avwf} (taken from \cite{mace2016fractal}).

\section{Dynamical properties}\label{sec.dynamics}
The quantum diffusion of wavepackets is determined by the spectral properties of the underlying Hamiltonian. Extended wavefunctions and continuous spectra typically lead to ballistic motion, where the particle moves with a well-defined group velocity. Singular continuous spectra such as that of the FC result in more complex behaviors. Some of the characteristics of wave packet diffusion are discussed in this section. 

\subsection{The diffusion exponent }
The mean square displacement in a time $t$ of an electron starting from the site $i_0$ of the chain is given by
\begin{eqnarray}
d^2(i_0,t) = \sum_j (i-i_0)^2 p (i,i_0,t)
\end{eqnarray}
where $p (i,i_0,t)$ is the probability to be on site $i$ at time $t$ for a given starting position $i_0$, normalized such that $\sum_i p(i,i_0,t)=1$. The time dependence of $d^2(i_0,t)$ is in principle extremely complex, due to the multifractality both of the density of states and of the wavefunctions. The exponent $\beta$ which can depend on the initial position $i_0$, describes its leading long time behavior after smoothing out the fluctuations, i.e.
\begin{eqnarray}
d(i_0,t) \sim t^{\beta(i_0)}
\end{eqnarray}
for sufficiently long times $t$. In the simplest cases, values of the exponent are well-known: for electrons in a periodic crystal there is ballistic propagation with a constant group velocity. The system has translational invariance and $\beta(i_0)=\beta$ independently of the initial position. For localized electrons, $d(t)$ tends to a constant at large times, and $\beta=0$. For standard diffusion, as in the case of electrons in a moderately disordered crystal, $\beta=1/2$. 

For quasicrystals which are invariant under translations, the $\beta$ values depend on the choice of origin $i_0$ of the wavepacket.  One can define an effective averaged diffusion exponent $\beta$ by considering the averaged quantity  $d(t)=\langle d(i_0,t)\rangle$ where the brackets denote the average over initial positions $i_0$. Then $d(t) \sim t^{\beta}$ defines the globally averaged value of the diffusion exponent $\beta$. 

\begin{figure}[htp]
	\centering
	\includegraphics[width=.45\textwidth]{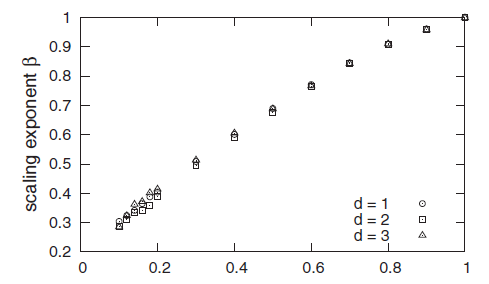}
	\caption{\small{Plots of the diffusion exponent $\beta$ as a function of the hopping ratio for the Fibonacci chain and its $d$-dimensional generalizations as described in Sec.\ref{sec.other} (reproduced with permission from \cite{PhysRevB.85.224205thiemschr})}}
	\label{fig.betavals}
\end{figure}

Numerical results for the values of $\beta$ in the FC for different values of $w=t_A/t_B$  are shown by circular symbols in Fig.\ref{fig.betavals} (taken from Thiem and Schreiber \cite{thiemschrJPC}). One sees that the diffusion exponent increases in value monotonically with $t_A/t_B$ and reaches the expected value of 1 in the limit of the periodic chain. Data for $d$ dimensional product lattices (see Sec.\ref{sec.other}) also included in the figure, show that the exponent $\beta$ does not depend on the dimensionality in this class of models.

It can be noted that this exponent enters in many phenomenological theories of transport in quasicrystals. From a dimensional analysis the diffusivity $D \sim d^2(\tau)/\tau$, and the conductivity $\sigma$ which is proportional to $D$ by  the Einstein relation, should scale as $\tau^{2\beta-1}$, where $\tau$ is some characteristic cut-off time for diffusion. For sub-diffusive motion, i.e. $\beta<\frac{1}{2}$, the electrical conductivity $\sigma$ would then $decrease$ with increasing $\tau$. This behavior is the opposite of what one would observe in metals, where conductivity increases linearly with $\tau$ according to the Drude formula. In real quasicrystals, experiments show that conductivity decreases when structural quality of the sample is improved for example by annealing \cite{PhysRevLett.70.3915berger} -- which could be interpreted to mean that $\beta$ is smaller than $\frac{1}{2}$.

\begin{figure}[htp]
	\centering
	\includegraphics[width=.45\textwidth]{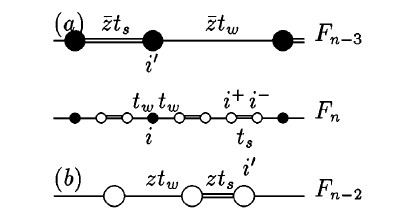}
	\caption{\small{(a) Relation between atom $i$ of $F_n$ and atom $i'$ of $F_{n-3}$. (b) Relation between a molecule $(i^+,i^-)$ of $F_n$ and its corresponding site $i'$ of $F_{n-2}$. Figure reproduced from \cite{piechon1996} with permission. }}
	\label{fig.piechon} 
\end{figure}

\bigskip
\noindent
{\bf{RG method for diffusion}} Studies of dynamics using the approximate RG scheme have been carried out by Abe and Hiramoto \cite{PhysRevA.36.5349abe}, Pi\'echon \cite{piechon1996} and Thiem and Schreiber \cite{PhysRevB.85.224205thiemschr}. These studies use a recursive approach to computing $d(i_0,t)$ as well as all of the generalized moments of the displacement,
\begin{eqnarray}
d_q(i_0,t) = 1/N \sum_j (i-i_0)^q p (i,i_0,t)
\end{eqnarray}
To lowest order, one can write two different recursive relations for the probability $p^{(n)}$, depending on whether the initial site is an atomic or a molecular site. Following the approach in \cite{piechon1996}, let the initial site be labeled $i_0$,  and $i$ refer to a site of the $n$th chain. In the renormalized chain, similarly, the initial site has the label $i'_0$ and $i'$ refers to a site of the new chain (see Fig.\ref{fig.piechon}). The length scale renormalization factors is either $\omega^3$ or $\omega^2$ depending on the type of RG transformation (atom or molecule), while the corresponding energy-time renormalization factors are $z$ and $\zb$. The relations between probabilities defined on the old and new chains can be stated as follows :
\begin{eqnarray}
p^{(n)}(i,i_0,t) \vert_{ato} \approx \omega_n^3p^{(n-3)}(i',i'_0,\zb t) \nonumber \\
p^{(n)}(i,i^{\pm}_0,t) \vert_{mol} \approx \omega_n^2p^{(n-2)}(i',i'_0,zt) 
\label{eq.recurdyn}
\end{eqnarray}
In the second relation, the initial site in the case of the molecule are further labeled with a $\pm$ standing for the left and right atoms of that molecule.  The first of these relations says that the probability to go from site $i_0$ to site $i$ in a time $t$ in the $n$th chain is reduced by a factor $\omega^3$ with respect to the probability to go from site $i'_0$ to site $i'$ (a distance shorter by $\omega^3$) in a time $\zb t$ in the $n-3$th chain. For molecules, a similar statement applies, with the additional assumption that left and right atoms play a symmetric role in the propagation.  
 
Note two simple extreme cases:
\begin{enumerate}
    \item 
If the origin of the wavepacket is chosen such that one obtains a purely atomic type diffusion at every RG step, then the result would be a power law $d(t) \sim t^{\beta_{ato}}$ with
$$\beta_{ato} = \ln\omega^3/\ln\zb$$
\item In the opposite situation of a pure molecular diffusion process, one obtains another exponent
$$\beta_{mol} = \ln\omega^2/\ln z $$
\end{enumerate}

Thiem and Schreiber defined an average value of $\beta$ using the relative fraction of atom and molecular sites, as follows:
\begin{eqnarray}
\overline{\beta} = \frac{\tau -1}{\tau +1}\beta_{ato}  + \frac{2}{\tau +1}\beta_{mol}  
\end{eqnarray}
They noted that this value fits the dynamics well for small $\rho$, up to about $\rho=0.02$ (when the approximations made  for the wavefunctions  become inadequate). 

Using the recursion relations in Eq.\ref{eq.recurdyn}, Pi\'echon derived recursion formulae for the moments $d_q(i_0,t)$, and for their averages $d_q(t) = \langle d_q(i_0,t)\rangle$, as follows 
\begin{eqnarray}
d^{n}_q(t) = \omega^{3(1-q)}_n d^{n-3}_q(\zb t) + 2\omega^{2(1-q)}_n d^{n-2}_q(zt) 
\label{eq.momentsdyn}
\end{eqnarray}

Let us suppose there is a fixed point solution of the probability, $p^*$, in the limit $n\rightarrow\infty$. Eq.\ref{eq.momentsdyn} implies a self-consistency condition :
\begin{eqnarray}
p^*(r,t) = \omega^6p^*(\omega^3r, \zb t) +2\omega^4p^*(\omega^2r,zt) 
\label{eq:pstar}
\end{eqnarray}

Note that the probability depends on two scaling factors leading to a multiscaling property: the function $p^*(r,t)$ cannot be written as a function of a single variable. Consider now the probability $p(r=0,t)$. Letting $p(0,t)^* \sim t^{-\gamma}$ and comparing Eq.\ref{eq:pstar} with Eq.\ref{eq:tauofq} one sees that $\gamma=D_2$, one of the spectral dimensions introduced earlier. 

More generally, if one assumes that higher moments of the diffusion distance scale as $d_q(t) \sim  t^{q\sigma_q}$, then one finds that $\sigma_q = D_{1-q}$. This result shows that the diffusion exponents $\sigma_q\geq D_1$, thus satisfying the Guarneri inequalities \cite{guarneri} for all values of $q$. This formalism, and similar conclusions are applicable to other quasiperiodic chains such as the AAH model studied by \cite{evangeloukatsanos,ketzmerick}. 

\begin{figure}[htp]
	\centering
	\includegraphics[width=.4\textwidth]{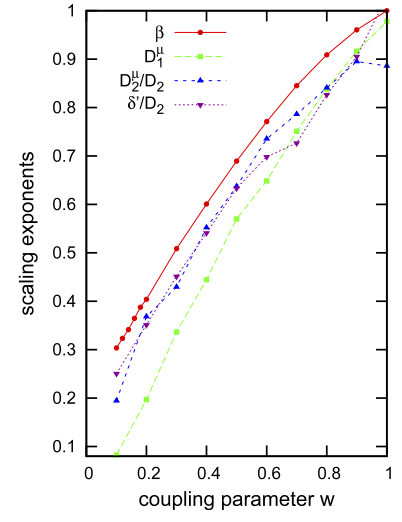}
	\caption{\small{Comparative plot of diffusion exponent, and theoretical lower bounds involving generalized spectral dimensions and wavefunction dimensions. Exponents were computed numerically for different values of the hopping ratio $w=t_A/t_B$ (reproduced with permission from \cite{thiemschrJPC})}}
	\label{fig.exponents}
\end{figure}

We have seen already in Sec.\ref{subsec.spectral} that in perturbation theory the spectrum was approximately monofractal for $z=\zb^{2/3}$. This simplification occurs for the wavepacket dynamics as well --  the exponent $\sigma_q=D_F$ for all $q$, where $D_F=\ln \omega^2/\ln z$. In this case, the dynamics can be expected to be simple diffusion with a single exponent However this is only true to the extent of the approximations made (see the caveat based on the trace map analysis Sec.\ref{sec.overview}). 

\bigskip
\noindent
{\bf{Exponent relations and inequalities}} Guarneri \cite{guarneri} derived an inequality stating that $\beta \geq D_1$, where $D_1$ is the earlier mentioned information dimension of the spectrum.  Another inequality derived by Ketzmerick et al \cite{ketzmerick} involves the average wavefunction exponent $D^\psi_2$ and it states that 
$\beta \geq D_2/D^\psi_2$. Both these inequalities are satisfied in the FC, as can be seen from Fig.\ref{fig.exponents}. For the diagonal Fibonacci model,  \cite{damanikphilmag,damaniktcherem} showed that $\beta$ must satisfy certain bounds that depend on the onsite energies $\epsilon_A$ and $\epsilon_B$.

\subsection{Autocorrelation function}
\begin{figure}[htp]
	\centering
	\includegraphics[width=.4\textwidth]{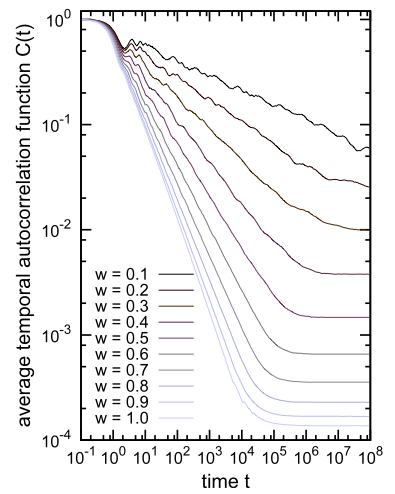}
	\caption{\small{Log-log plot of the smoothed autocorrelation function for different values of the hopping ratio ($w=t_A/t_B$), showing the fit to power law. (reproduced with permission from \cite{thiemschrJPC})}}
	\label{fig.exponentsb}
\end{figure}

As for the mean square distance, the time dependent correlation function in quantum systems with Cantor spectra is expected to have a power law decay, falling off as $t^{-\delta}$ at long times. Ketzmerick, Petschel and Geisel have argued \cite{PhysRevLett.69.695ketzmerick92} that the exponent  $\delta$ should be equal to $D_2$. 
Several authors \cite{PhysRevLett.69.695ketzmerick92, PhysRevB.85.224205thiemschr,zhongmoss,thiemgrimm} have studied the behavior of the smoothed autocorrelation function given by, 
\begin{eqnarray}
C(i_0,t) = \frac{1}{t}\int_0^t P(i_0,i_0,t') dt'
\label{eq.correln}
\end{eqnarray}
which gives the integrated probability up to time $t$ for the particle to be found at the initial position.  Averaging over all initial positions, one obtains $C(t) =(1/N) \sum_i C(i,t) \sim t^{-\delta'}$. This quantity is easier to fit than the autocorrelation function. Note that $\delta'$ may be different from $\delta$ due to logarithmic corrections coming from the integral in Eq.\ref{eq.correln}. Fig.\ref{fig.exponentsb} taken from \cite{thiemschrJPC} shows the smoothed autocorrelation function for different values of $t_A/t_B$ (the variable $w$ in the figure).  The exponent should tend to the expected value 1 in the periodic limit when the spectrum becomes continuous ($t_A=t_B$ or $\epsilon_A=\epsilon_B$). \footnote{In practice, however, as pointed out in Yuan et al \cite{PhysRevB.62.15569yuan}, it is hard to get convergence in numerical studies, thus some early numerical works obtained incorrect values -- which were smaller than the expected value of 1, in a periodic system.} 

\subsection{Log-periodic oscillations} 
\begin{figure}[htp]
	\centering
	\includegraphics[width=.3\textwidth]{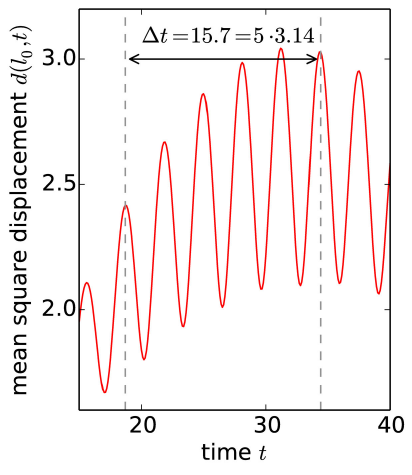}
\includegraphics[width=.3\textwidth]{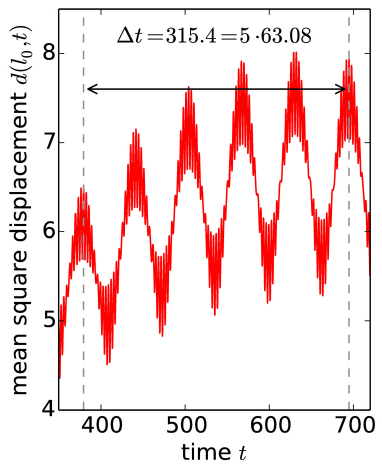}
\includegraphics[width=.3\textwidth]{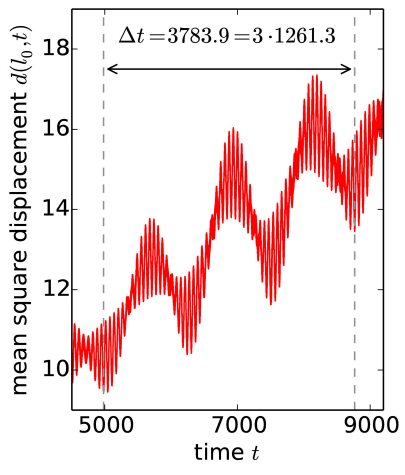}
	\caption{\small{Log-log plots of $d(t)$ versus time for short time scales (left figure) and intermediate time scales (center figure), showing the two shortest periods in the oscillations. A third period (right figure) governs yet larger time scales. Data correspond to diffusion from a fixed (molecular) site of the $n=14$ approximant chain, with $t_A=0.1$ and $t_B=1$. (Figure taken from \cite{Thiem2015} with permission)}}
	\label{fig.logperiodic}
\end{figure}

Hiramoto and Abe observed that there are oscillations superposed on top of the average power law behavior of the rms distance $d(i_0,t)$. These oscillations were studied in more detail by Lifshitz and Even-Dar Mandel \cite{lifshitzdarmandel}. A similar oscillatory behavior is seen for the return probability $p(0,t)$. 
Fig.\ref{fig.logperiodic} taken from Thiem \cite{Thiem2015},  shows $\ln d(t)$ plotted versus $\ln t/t_B$ for a wave packet diffusing from an initial site chosen to be of molecule type. The two shortest periods, corresponding to very short times and to intermediate times, can be clearly seen in these plots. The system size considered was sufficiently large ($n=14$ approximant chain) that a third period was observed, for even longer times (see \cite{Thiem2015}).  

The empirical form that was used to fit these data is $d(i_0,t) \sim t^{\beta(i_0)}(1 + \alpha e^{i f\ln t})$. The frequencies $f$ are different, depending on the time scale that is considered, and they follow a hierarchical rule. 
The shortest times correspond to the fastest oscillations, whose frequency depends on the nature of the initial site $i_0$ -- whether it is atom or molecule. At the shortest time scale the oscillations have a period (in dimensionless units) of $2\pi$ for an initial site of type atom, and $\pi$ when the initial site is molecule type. Going to longer times the oscillations have frequencies which are smaller by factors $\zb$ and $z$ respectively. Thiem has given a quantitative account of these oscillations in terms of the perturbative RG theory \cite{Thiem2015} and has argued that they stem, at each length scale of the RG process, from resonances due to the molecular energy level splitting.  
The author notes that such oscillations are not observed for the quasiperiodic critical AAH model \cite{Thiem2015}. This may be attributed to an essential difference between the two potentials -- since the potential energy is a continuous-valued function in the AAH model, there are no molecular clusters in its RG scheme, and thus no characteristic resonance frequencies.

\section{Transport Properties}\label{sec.transport}

Some frequently asked questions concern the resistivity of a quasicrystal: are these materials intrinsically metallic or insulating ? How does transport depend on the sample size, disorder, temperature, on external magnetic fields, etc ? The following studies of transport properties of 1D Fibonacci chains attempt to shed light on these questions.

\subsection{An exact result for $E=0$ transmission}
The $E=0$ state transmission coefficient in the hopping model can be calculated from the exact solution given in Sec.\ref{subsec.wf0}. This transmission coefficient is proportional to the zero temperature conductivity at half-filling, when the Fermi energy of the system is $E_F=0$. Let us consider a FC of length $L=2n$ attached to periodic chains (input and output chains) at either end. The transmission coefficient is given by \cite{RevModPhys.69.731beenakker,economou_static_1981} 
\begin{eqnarray}
\mathcal{T}_n = \frac{4}{(x_n+x_n^{-1})^2}
\end{eqnarray}
 where $x_n = \vert\psi(2n)/\psi(0)\vert$.  Using the expression eq.\ref{eq.wf0} one obtains \cite{mace2017critical}
 \begin{eqnarray}
\mathcal{T}_n = \frac{1}{\cosh^2(\kappa (h(n)-h(0))}
\end{eqnarray}
where $h$ the height function, it will be recalled, depends solely on the geometry. This expression shows that the transmission $\mathcal{T}_n=1$ -- there is perfect transmission --  when the heights  $h(n)$ and $h(0)$ are equal. This type of ``intermittent" transparency occurs for sites separated by distances that can tend to infinity. 

\begin{figure}[htp]
	\centering
	\includegraphics[width=.45\textwidth]{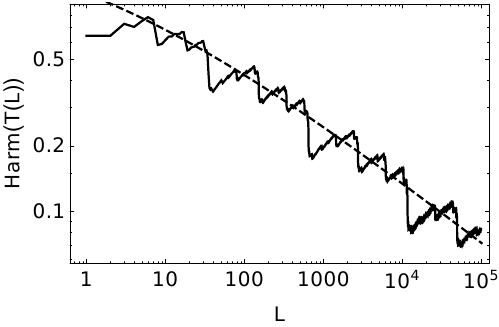}
	\label{fig.transmission}
	\caption {\small {Mean $E=0$ transmission coefficient as a function of
length L of the Fibonacci chain. The dashed and continuous lines show the analytical prediction of Eq.\ref{eq.harmonic} and  numerical results for $n=27$ ( N= 196418 atoms) respectively. Figure reprinted from \cite{mace2017critical}.}}
\end{figure}

One can also compute the harmonic mean of the transmission over a chain of $L+1$ sites. The harmonic mean which is more adapted than the arithmetic mean for systems with very large fluctuations is defined by 
\begin{eqnarray}
\langle \mathcal{T}\rangle_L = \left( L^{-1} \sum_i^L \mathcal{T}_i^{-1}\right)^{-1}
\end{eqnarray}
The scaling of $\langle \mathcal{T}\rangle_L$ with system size $L$ can be expressed analytically using the height distribution Eq.\ref{eq:heightP}, which yields
\begin{eqnarray}
\langle \mathcal{T}\rangle_L \sim 2\left(1+\frac{L}{L_0}^\phi\right)^{-1}
\label{eq.harmonic}
\end{eqnarray}
where the exponent $\phi$ can be expressed in terms of $\kappa$, the golden mean $\tau$ and the maximal eigenvalue of a certain product of generalized inflation matrices (see \cite{mace2017critical} for details). This expression predicts a power law decay of the mean transmission as a function of distance $L$, in accord with the numerical data, as seen in Fig.\ref{fig.transmission}. Note that there is no contradiction between this power law and the fact that the chain can be transparent for very long distances. The apparent paradox is explained by noting that the power law represents an $average$ behavior, while for a given system there are large fluctuations around the mean (as already observed by Sutherland and Kohmoto \cite{PhysRevB.36.5877resistivitykohmoto}), whereas in a $given$ chain, the transmission coefficient can be (and is) equal to 1 for certain sites.

{\subsection{Landauer approach}} 
The Landauer formula relates the resistivity $\rho(n)$ to the transfer matrix $T_n$ of a system of $n$ sites as follows
\begin{eqnarray}
\rho_n =\frac{1}{4}\left(T_n^T T_n - 2\right)
\end{eqnarray}
where $T^T$ denotes the transpose of the transfer matrix. Using the trace map techniques described in Sec.\ref{sec.overview}, Sutherland and Kohmoto \cite{PhysRevB.36.5877resistivitykohmoto} studied the behaviors near the band edge and the band center.  They showed that the resistance grows no faster than a power law of the system size. They pointed out as well that there is a wide distribution of powers governing its growth with system size, and this leads to very large fluctuations of the resistance. They speculated, finally, that this behavior is qualitatively also to be expected for other energies in the band. Their conjecture as to power law behavior of the resistivity was proved by Iochum and Testard \cite{iochum}.

For comparison with measurements, it is pertinent to consider the average resistance where the average is taken over states lying within some energy interval. The energy interval chosen should depend on factors such as the temperature, or energy-scale corresponding to inelastic scattering, disorder conditions, etc. For a system of length $L$ and some appropriately chosen $\Delta E$, 
\begin{eqnarray}
\overline{\rho}(L)=\frac{1}{n(E)\Delta E}\sum_E\rho(L,E_i)
\end{eqnarray}
defines an average resistance which was studied for the FC by Das Sarma and Xie using the Landauer formalism \cite{PhysRevB.37.1097dassarma}. The result is a power law behavior for the resistivity, $\overline{\rho} \sim L^{-a}$. The authors noted that the power law holds for other fillings, provided that the Fermi level is not close to a large gap.

\subsection{Kubo-Greenwood approach}

\begin{figure}[htp]
	\centering
	\includegraphics[width=.45\textwidth]{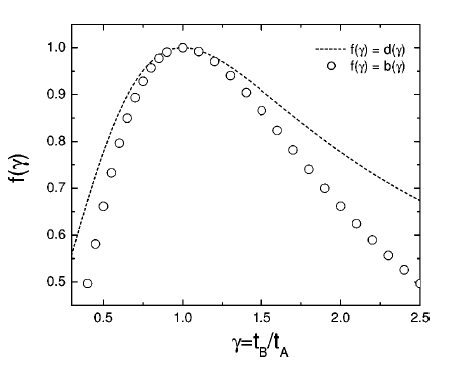}
	\caption{\small{Plot of the conductivity exponent $b$ (circles, for the definition see text) and the density of states exponent $d$ (following the analytic formula \cite{kohmoto1987critical}, dashed line) as a function of the hopping ratio $\gamma=t_A/t_B$ (figure reproduced from \cite{PhysRevB.70.144207sanchezwang} with permission}}
	\label{fig.sanchezwang}
\end{figure}

Sanchez et al \cite{PhysRevB.64.174205sanchez,PhysRevB.70.144207sanchezwang} developed an RG approach for the conductivity starting from the Kubo-Greenwood formula:
\begin{eqnarray}
\sigma(\mu,\omega,T)= \frac{2e^2\hbar}{\pi m^2V}\int dE \frac{f(E)-f(E+\hbar\omega)}{\hbar\omega} \nonumber \\ \mathrm{Tr}\left(\mathrm{Im} G^+(E+\hbar\omega)\mathrm{Im}G^+(E)\right)
\end{eqnarray}
where V is the volume of the system, $G^+(E)$ is the retarded
one-particle Green’s function, $f$ is the Fermi-Dirac distribution, with Fermi energy
$\mu$ for temperature $T$.
Applying the RG method to the Kubo-Greenwood formula they could study the AC and DC conductivities for very large systems  \cite{PhysRevB.70.144207sanchezwang}. In the hopping problem, for half-filling, Sanchez and Wang found that the scaling exponents of the conductivity and of the density of states have a similar dependence.  Fig.\ref{fig.sanchezwang} shows the parameter $b$ (defined by $\sigma \sim b^{-n/6}$) and the parameter $d$ (defined by $DOS \sim d^{-n/6}$) plotted against $t_A/t_B$. The authors concluded that these results show the existence of an Einstein relation, $\sigma \sim (dN/dE) D$, wherein the conductivity and density of states are related by the diffusivity $D$. 

\subsection{Noninteracting many body metallic and insulating states }
\begin{figure}[htp]
	\centering
	\includegraphics[width=.45\textwidth]{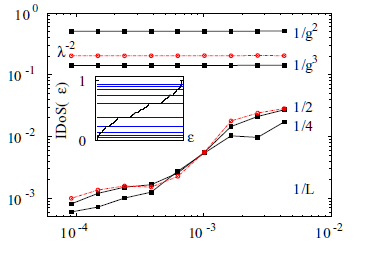}
	\caption{\small{Scaling of the many body localization length $\Lambda$ with system size $L$, for different band fillings. One sees a  metallic trend ($\Lambda$ decreasing with $L$) for fillings of 1/2 and 1/4, and an insulating behavior ($\Lambda$ constant with $L$) for fillings $g^{-2}$ and $g^{-3}$ where $g$ is the golden mean. Red (grey) and black symbols correspond to curves for hopping ratios equal to 0.5 and $\omega$ respectively. (Figure reproduced with permission from \cite{PhysRevB.94.214204varma}).}}
	\label{fig.varma}
\end{figure}

Even in the absence of electron-electron interactions, many body effects can play an important role in quasicrystals. Varma et al considered such many body effects -- due solely to the Pauli principle -- on the conductance of Fibonacci chains \cite{PhysRevB.94.214204varma}. They computed the effective localization length $\Lambda$ for various band fillings, using a formalism developed by Kohn in his theory of the insulating state \cite{PhysRev.133.A171kohn}. This localization length, which is related to the real part of the conductivity tensor, is more sensitive to spectral gaps, and to transport properties than the more familiar single particle localization length defined in terms of the decay of the envelope of a given wavefunction. In a conductor, this many body localization length scales to infinity, while in an insulator, this quantity saturates with the system size. Varma et al showed that in the hopping model, there are special values of the band filling corresponding to an insulating state. These fillings correspond to values of IDOS of $\omega^2$ and $\omega^3$ correspond to the positions of the large gaps. This is shown in Fig.\ref{fig.varma} where $\Lambda^{-2}$ is plotted versus inverse system size $1/L$. The fillings $\frac{1}{2}$ and $\frac{1}{4}$ for which the trend is decreasing with increasing system size most likely correspond to a metallic state.

\section{Disorder and boundary effects}\label{sec.perturbations}
The question of perturbations and their effects  on critical states has been much discussed. The interesting conceptual problem raised has, of course, experimental implications. Disorder can be expected to play an important role for the electronic transport in real quasicrystals for the following reason. We have seen in the previous section that, for a perfect FC, transport at $T=0$ can be described by different power laws, and can either scale towards an insulating or to a metallic state in the thermodynamic limit, depending on the Fermi energy. Adding disorder of any form -- be it structural defects or chemical substitutions or phonons -- introduces a cut-off timescale, $\tau$, whose value depends on disorder. One can therefore have a conductance which is finite in a quasicrystal that is weakly disordered. For increasing disorder, one expects that the quantum interference phenomena which lead, in a perfect quasicrystal to the multifractality of the density of states and the multifractality of states, would be progressively suppressed. Combined, these effects might contribute to improving transport as disorder increases, however, these are still open questions.

The problem of a single impurity in a FC was studied by several authors. The trace map method was used to compute localization lengths of an impurity state represented by a delta-function potential in \cite{PhysRevB.59.11315naumis99}.  The effect of structural defects, namely a single phason defect (in which a pair of bonds is locally exchanged, as for example $AB \rightarrow BA$ has been studied \cite{PhysRevB.54.15079naumisaragon}. This study concluded that the presence of a single impurity affects all of the states and leads to an increase of the fractal dimension of the spectrum. A weak form of structural disorder was considered in \cite{PhysRevB.61.1043pimentel}, by allowing randomness in the substitution rules for building chains. The conclusion reached in this and a later study  \cite{PhysRevB.70.205124huanghuang} is that this type of disorder is irrelevant, in that the Lyapunov exponents of states were not changed. To modify the critical states of the pure system, the disorder must break some symmetries of the Fibonacci Hamiltonian as in the models we discuss in the following section.

\subsection{Finite systems and approach to Anderson localization} \label{subsec.disorder}
We now focus on the effects of adding a finite bulk disorder to the two models under discussion, Eqs.\ref{eq.hophamilt} and \ref{eq.diaghamilt}.
Quite general rigorous arguments show that, in one dimensional models an infinitesimal disorder leads to localized states \cite{delyonPhysRevLett.55.618}. This has been confirmed by a variety of numerical studies \cite{PhysRevB.59.11315naumis99,PhysRevB.35.6034riklund}.
Das Sarma and Xie \cite{PhysRevB.37.1097dassarma} studied the effect of randomness in a Fibonacci QC using a scattering model for a system in which the scatterers of constant height were placed in a Fibonacci sequence of two distances $a$ and $b$. The system was coupled to leads and conductance computed using the Landauer formula $G= \frac{2e^2}{\hbar}T/(1-T)$, where $T$ is the transmission coefficient of $F_n$ site system. They reported that, while disorder (in the positions of the scatterers) eventually localizes all states, small disorder did not change the physics qualitatively. Introducing a shuffling of the sequence of Kronig-Penney type scatterers leads as well to localization of states and consequently an exponential decay of the conductance. 

\begin{figure}[htp]
	\centering
	\includegraphics[width=.5\textwidth]{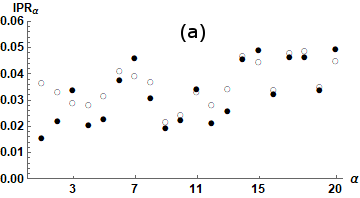}
		\includegraphics[width=.45\textwidth]{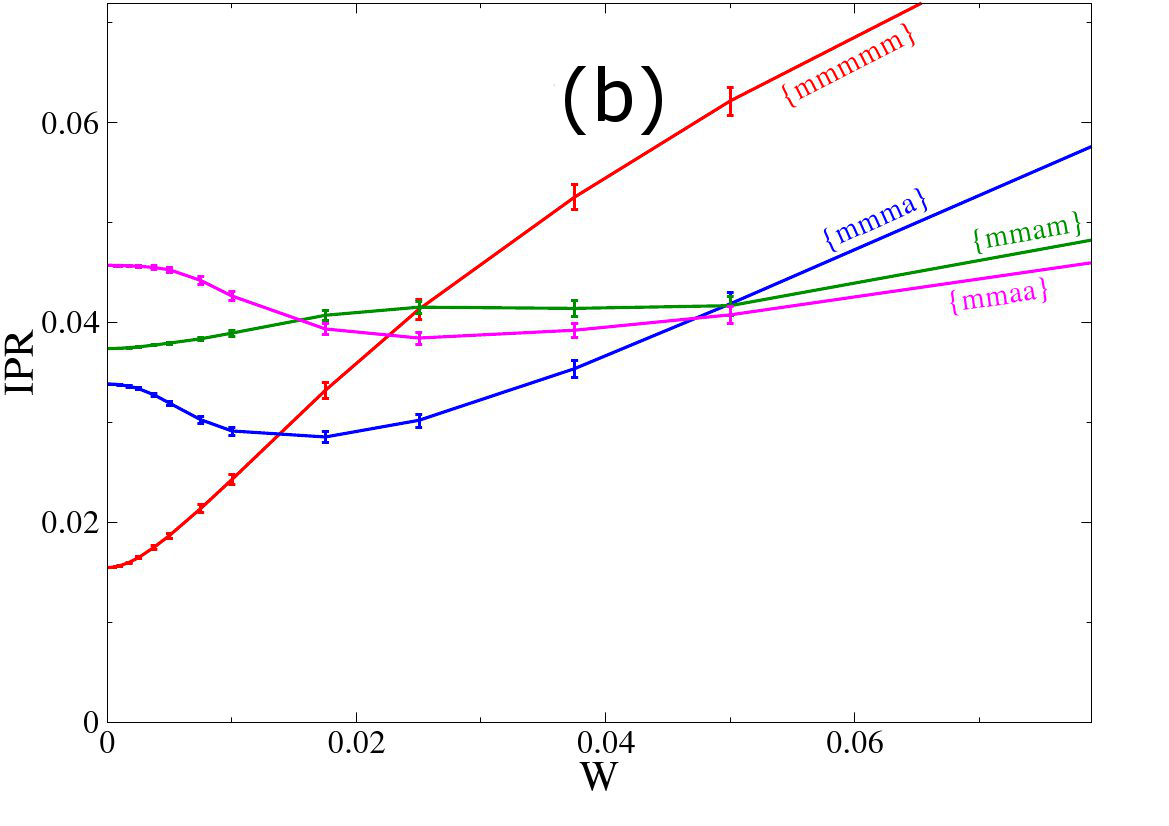}
	\caption{\small{(a) IPRs plotted for the lowest 20 states of a $n=10$ Fibonacci chain for $\rho=0.5$. Filled circles give IPR values of the pure chain while open circles give the sample averaged IPR values for weak disorder ($W=0.05t_B$). States with $\alpha=3,7,..$ indicated are those showing an anomalous behavior. (b) Plot of the averaged IPRs versus disorder strength $W$ for four low-lying states of $n=10$ approximant. The RG path is indicated next to each curve. Figure reproduced from \cite{PhysRevB.99.054203jeena}}}
	\label{fig.iprdiff}
\end{figure}

There is no doubt that sufficiently large disorder strength leads to strong localization. Interestingly, however, the approach to localization can be quite complicated and state-dependent. It has been shown in \cite{PhysRevB.99.054203jeena} that there are interesting crossover phenomena going from the pure system to the localized system as the strength of disorder is increased. The authors studied the change of critical states of the pure hopping model when hopping amplitudes are randomly perturbed from their initial values $t_B$ and $t_A =\rho t_B$. While most states tend to become more and more localized as the disorder strength is increased, some states go the other way initially, becoming delocalized before turning over and localizing like the other states. This can be seen in Fig.\ref{fig.iprdiff}a which shows the change of the IPR for the 20 lowest lying states $\alpha=1,..,20$ of the $n=10$ approximant taking a hopping ratio of 0.5. The IPR values for the pure chain are shown by filled circles, while the disorder averaged values for weak disorder $W=0.05t$ are shown by open circles. It can be seen that, for most of the states, disorder results in increased IPR, as expected. However some states behave anomalously  -- these states are indicated $\alpha=3,7,11,15,...$ in the figure. The anomalous behavior is observed for arbitrarily large system size, albeit for weaker disorder for longer approximants. 

When the averaged IPR of each state is plotted versus disorder strength $W$, as is shown for four low-lying states in Fig.\ref{fig.iprdiff}b), the anomalous states in Fig.\ref{fig.iprdiff}a) show a marked initial decrease of IPR, followed by an upturn. 
The perturbative RG method provides an explanation of the observed behaviors in terms of the renormalization path of the states. Fig.\ref{fig.iprdiff}b) shows the RG path of each of the states, and one sees that the band edge state $\alpha=1$ (shown in red, RG path: $mmmmm$) has a monotonic increase of the IPR. In contrast, the state $\alpha=3$ (shown in blue, RG path: $mmma$) has a minimum of the IPR -- the state first delocalizes under weak disorder before the upturn sets in. One can show that the states which have ``atomic" character in the final step of RG have such re-entrant localization behavior and that they occur all through the spectrum. In simple terms, using the picture for small values of $\rho$, for atomic states IPR tends to decrease since disorder acts to ``smear" the wavefunction onto neighboring sites. In contrast, for a wavefunction composed of molecular states, the change of IPR has the opposite sign (for details see \cite{jagannathanEPJB}).  

\begin{figure}[htp]
	\centering
	\includegraphics[width=.45\textwidth]{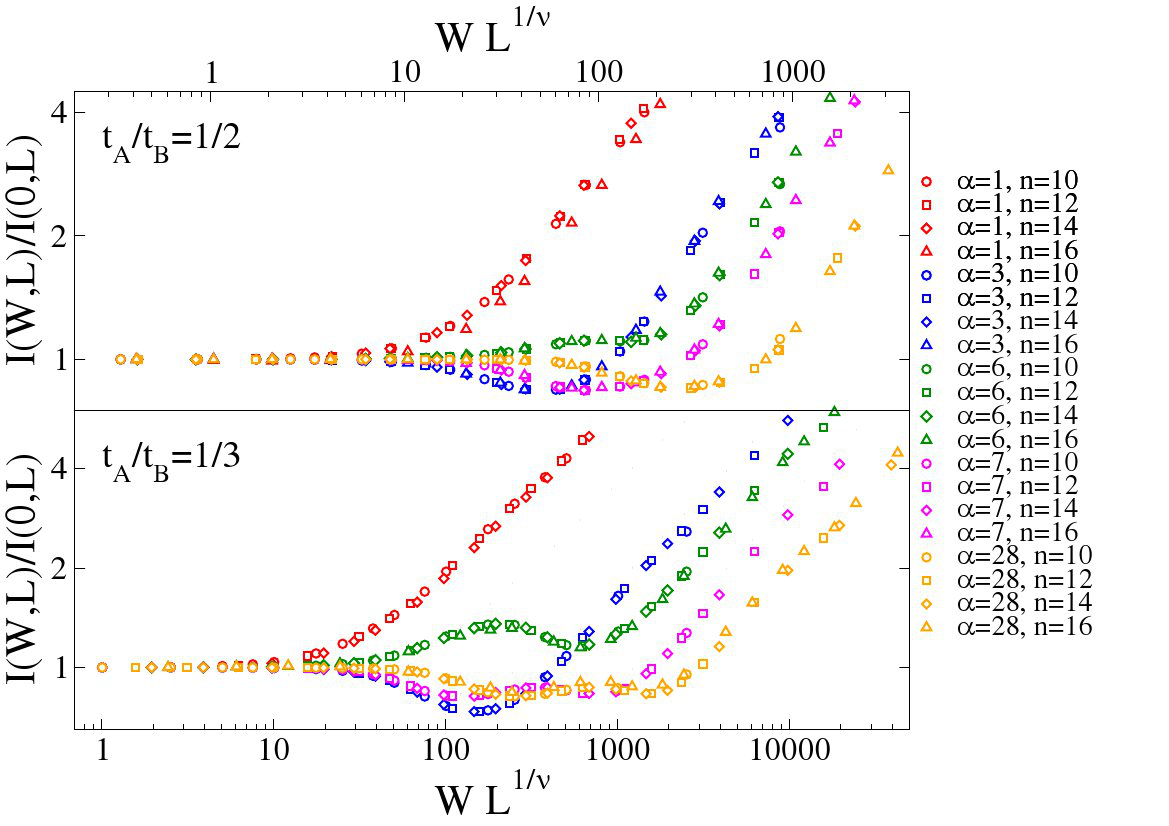}
	\caption{\small{ Data for the normalized IPR of several low lying states plotted versus the scaling variable $WL^{1/\nu}$ showing the data collapse for different disorder and system sizes. The top and bottom figures show results for two values of the hopping ratio $t_A/t_B$ --  note the changes in the scaling functions, which are nonuniversal. (from \cite{tarziaEPJ})}}
	\label{fig.scaling} 
\end{figure}

The changes in the IPRs can be described in terms of scaling functions -- i.e. for a $given$ state one can collapse the data for different system sizes $L$ and different disorder strengths $W$ onto a single curve. The finite size analysis in \cite{PhysRevB.99.054203jeena} showed that despite their different approaches to localization, ALL states are described by a single critical exponent, $\nu$. The value of $\phi$ depends on the ratio $t_A/t_B$ and was found  numerically from finite size scaling plots as in Fig.\ref{fig.scaling} --  $\nu=0.53$ for $\rho=0.33$. The  re-entrant behavior of the IPR can be explained in terms of the perturbative RG theory (see \cite{PhysRevB.99.054203jeena, tarziaEPJ} for details). However the phenomenon seems to be more generic, and the re-entrance behavior is observed for the diagonal model, as well as for other types of generalizations of the Hamiltonian.

\subsection{The proximity effect} \label{subsec.proximity}
It is well-known that it is possible to induce superconducting correlations in a non-interacting conducting system (N) by coupling it to a superconductor (S) -- the so-called proximity effect. The proximity effect provides a way to observe experimentally the unique properties of critical states in the FC. A first step consists of examining the proximity induced local pairing order parameter (OP) as a function of distance from the N-S interface. 

\begin{figure}[htp]
	\centering
	\includegraphics[width=.4\textwidth]{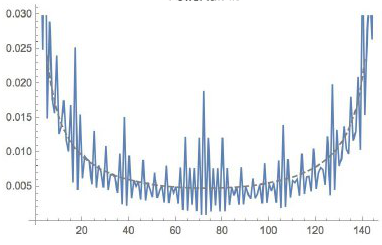}
	\caption{The superconducting order parameter $\Delta(i)$ in a Fibonacci chain placed in contact with a BCS superconductor at both ends, plotted versus position showing power law decay ($t_A/t_B=0.8$, decay exponent=0.6)	}
	\label{fig.proximity} 
\end{figure}

Connecting the FC to a bulk superconductor and using a mean field theory, Rai et al \cite{rai2019} computed the  distribution of $\Delta_i$, the induced local superconducting order parameter on site $i$. There are very large spatial fluctuations of $\Delta_i$, due precisely to  critical states. Fig.\ref{fig.proximity} (top) shows the profile of the order parameter (OP) as a function of the site number. One sees here the characteristic multifractal properties reflected in the variations of the order parameter. Fitting the average curve obtained by changing the phason angle parameter $\phi$, one sees that the OP decays as a power law in the distance from the N-S interface \cite{rai2020}. The power, which varies with $t_A/t_B$, is expected to depend both on the exponent of the density of states, and the averaged fractal dimensions of the wavefunctions near the Fermi energy, here taken to be at $E=0$. 

\begin{figure}[htp]
	\centering
	\includegraphics[width=.45\textwidth]{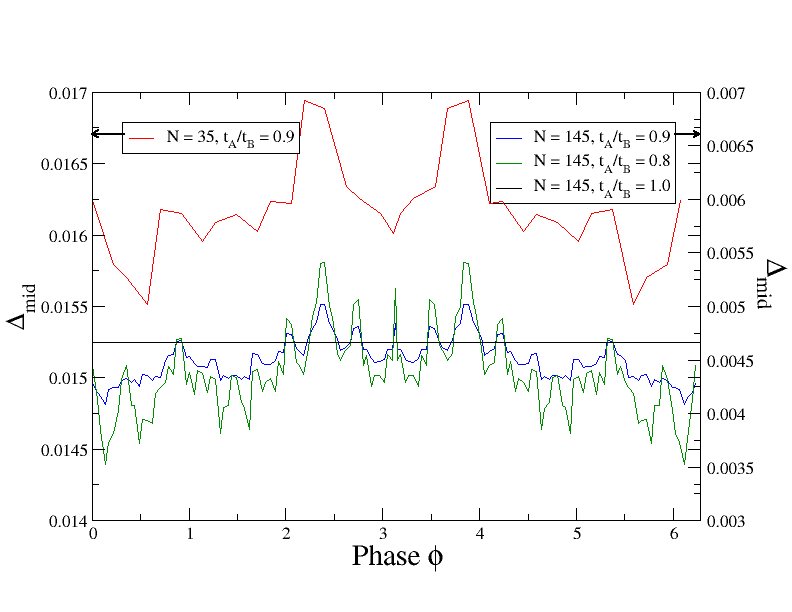}
	\includegraphics[width=.45\textwidth]{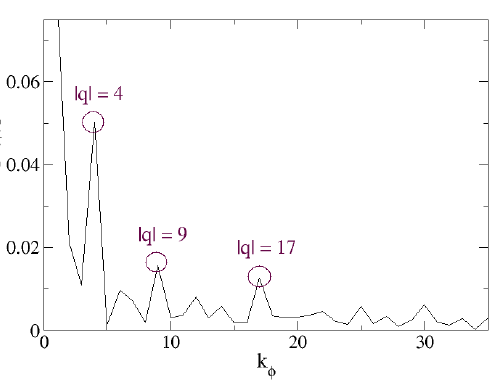}
	\caption{(top) The superconducting order parameter $\Delta(mid)$ at the mid-point of a Fibonacci chain versus the phason angle $\phi$ showing oscillations. (bottom) Fourier spectrum of the above plot, showing the main periods showing the peaks at values equal to topological labels $\mathrm{q}$ . \small{(reproduced with permission from \cite{PhysRevB.100.165121rai})
}}
	\label{fig.proximity2} 
\end{figure}

Along with the other states, edge modes contribute to the induced order parameter on each site. In fact, if one cycles through chains as a function of the phason angle parameter $\phi$ in Eq.\ref{eq.chij}, the induced order parameter $\Delta_i$ at a given site oscillates and the periods are just the topological numbers of gaps close to the Fermi energy \cite{PhysRevB.100.165121rai}. This can be seen in the  Fig.\ref{fig.proximity2}  which shows the variations of the OP at the midpoint of the chain as a function of the phason angle $\phi$. Two different chains are shown to emphasize that the basic periods do not change when going from smaller to larger systems, only additional periods appear. The lower figure shows the power spectrum of the oscillations, and the periods which are present in the curve of $\Delta_{mid}$. One sees the periods $4, 17, 9, 21, 6$, -- these correspond precisely to the $\mathrm{q}$ values of the largest gaps near the Fermi energy, in this case situated in the band center, $E_F=0$.

\section{Generalized Fibonacci models}\label{sec.mixed}
\subsection{{Phonon models}}  
\begin{figure}[htp]
	\includegraphics[width=.45\textwidth]{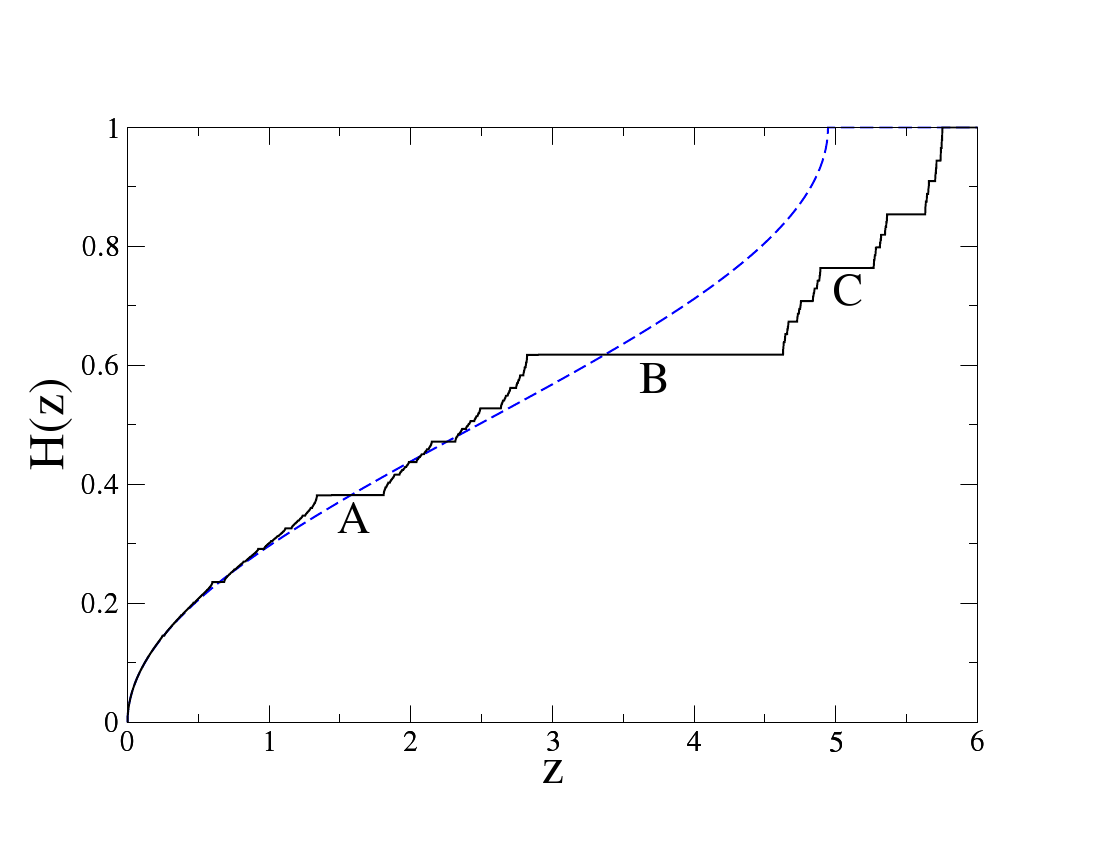}
	\caption{\small{Plot of integrated phonon density of states (IDOS) $H(z)$, versus the eigenvalues $z$ of the Laplacian operator on the Fibonacci chain for a coupling ratio 0.5 (from \cite{luckpetritis} reproduced with permission). The three main plateaux are labeled A, B and C. The dashed line represents the IDOS for the periodic system based on the average lattice,showing the similarity of the two curves for small $z$.}}
	\label{fig.phononDOS} 
	
\end{figure}

Phonon modes in a quasiperiodic chain can be studied by considering the set of equations
\begin{eqnarray}
H\psi_n = K_{n-1}\psi_{n-1}+K_{n}\psi_{n+1} \nonumber \\
-(K_{n-1}+K_{n})\psi_{n} =  E \psi_n
\end{eqnarray}
where $\psi_n$ denotes the displacement of atom $n$ of mass $m$  with respect to its equilibrium position, and the couplings $K_n$ can take two values,  $K_A$ or $K_B$. One can consider, alternatively, another version of the model in which the masses $m$ can vary, and the couplings are constant. The operator $H$ generalizes the discretized Laplacian operator, and its eigenvalues yield the frequencies of phonon modes.
The phonon problem is tackled using the methods we have seen already for the closely related electron problem \cite{PhysRevLett.50.1870kadanoff,luckpetritis,PhysRevLett.50.1873ostlund,PhysRevB.34.2207norirodri,PhysRevB.33.4809luodagaki,PhysRevB.39.2670ashraff}. The trace map equation is the same as in the electron problem, namely Eq.\ref{eq.tracemap}, and the same kind of analysis applies. The spectrum of energies $E$ has a Cantor-set structure as can be seen in Fig.\ref{fig.phononDOS} (taken from \cite{luckpetritis}) which shows the IDOS versus energy (these quantities are denoted in the figure by $H$ and $z$). In contrast to the electronic case, for phonons, the scaling is non-uniform as a function of energy, and gaps become very small as $E$ tends to zero. At the lowest frequencies, the integrated density of states, which is plotted in Fig.\ref{fig.phononDOS} looks almost indistinguishable from that of a periodic chain where the coupling is given by the appropriately defined average value of the two Fibonacci couplings. Like the periodic chain, the IDOS has a van Hove singularity,  $IDOS \sim \sqrt{E}$, for small $E$. This behavior seems to indicate at first glance that the long wavelength Goldstone modes are robust with respect to the quasiperiodic modulation. However Luck and Petritis  presented a rigorous argument to show that the spectrum does not have any absolutely continuous component, even for frequency tending to zero.

The gap labeling theorem is seen to hold, as expected, and the letters A, B and C indicate three important gaps having gap labels given by the three smallest Fibonacci numbers \cite{luckpetritis}. The IDOS at these plateaux are given by $H_k = 1-\omega^k$ for $k=1,2$ and 3 where $\omega$ is the inverse of the golden mean.  Luck and Petritis showed that the upper edge of the spectrum is described by a new  6-cycle $a \rightarrow -b \rightarrow -a \rightarrow b \rightarrow -a \rightarrow -b$. This observation was then used to show that, close to the upper edge, the quantity $1-N(E)$ where $N$ is the IDOS has a power law modulated by log-periodic oscillations. 
Ashraff and Stinchcombe \cite{PhysRevB.39.2670ashraff}  computed the dynamic structure factor for the Fibonacci chain. Quantum diffusion properties in this model have been studied \cite{kohmotobanavar,darmandel} and found to share electronic properties of multifractal structure and log periodic oscillations. For a more detailed discussion on phonon modes in Fibonacci quasicrystals we refer the reader to the corresponding chapter in \cite{deboissieubook}.

\subsection{Mixed Fibonacci models}
The term mixed models is used to denote a general member of the family of models Eq.\ref{eq.hamilt}, where diagonal and off-diagonal quasiperiodic modulations are both present. These are relevant for experiments, for example, as real systems can be expected to have both forms of quasiperiodicity. 

Many of the techniques, including the powerful transfer matrix analysis can be extended to mixed models. 
Macia and Dominguez-Adame \cite{macia1996,maciaerratum} considered a mixed model in which the A and B sites have onsite energies of $\alpha$ or $\beta$, following a Fibonacci sequence. The hopping amplitudes are assumed to have two possible values, $t_{AB}=t_{BA}$ and $t_{AA}=\gamma t_{AB}$. The initial step consists of defining the basic transfer matrices. Choosing, without loss of generality, units such that $\beta=-\alpha$ and $t_{AB}=1$ one obtains four different transfer matrices in this model as follows :
\begin{eqnarray}
X&=&\begin{bmatrix} (E+\alpha) & -1  \\
	1 & 0 
	\end{bmatrix}, 
	Y=\begin{bmatrix} (E-\alpha)/\gamma & -1/\gamma  \\
	1 & 0 
	\end{bmatrix}, \nonumber \\
	W&=&\begin{bmatrix} (E-\alpha) & -\gamma  \\
	1 & 0 
	\end{bmatrix},
	Z=\begin{bmatrix} (E-\alpha) & -1  \\
	1 & 0 
	\end{bmatrix}
\end{eqnarray}
Macia and Dominguez-Adame showed that after renormalization the global transfer matrix in this model has a structure identical to the one for the Fibonacci sequence for the diagonal model,   Eq.\ref{eq.transfermatrix}. This can be seen by defining blocks of sites via $T_A=ZYX$ and $T_B=WX$. They showed that in finite chains the energies of certain transparent states i.e. with transmission coefficients of unity are of the form
\begin{eqnarray}
E(k)= \pm \sqrt{\alpha^2+4\cos^2(k\pi/N)}
\end{eqnarray}
with $\vert\alpha\vert \leq 2$ and $k$ an integer such that $N\tau =k\pi$ with $k=0,1,...$. The states with such energies $E(k)$ were confirmed as being extended states via a multifractal analysis. 

\begin{figure}[htp]
	\centering
	\includegraphics[width=.5\textwidth]{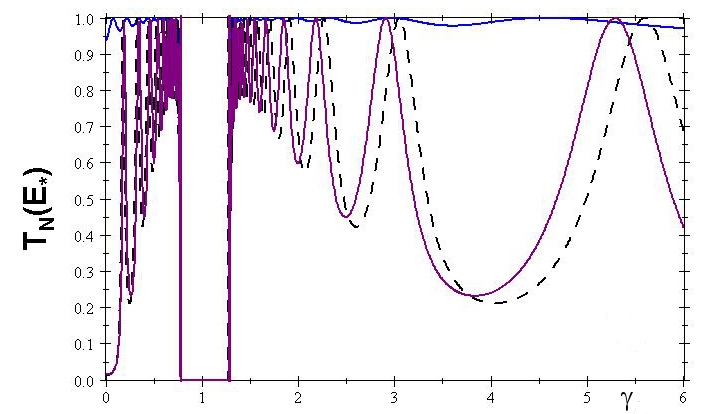}
	\caption{\small{Transmission coefficients at resonance energies $E_*$ plotted versus $\gamma$ (see text for definitions) for several mixed models. The Fibonacci case is shown by the dashed line. The purple and blue curves represent the silver mean and period doubling models respectively (figure adapted with permission from \cite{macia2017})}}
	\label{transparentstates.fig} 
	
\end{figure}

For this set of models one can determine the resonance energy $E_*$ -- for which the two elementary transfer matrices of the chain commute -- as a function of the parameters. Fig.\ref{transparentstates.fig} adapted from \cite{macia2017} shows the transmission coefficient $T_N(E_*)$ plotted as a function of resonance energy $E_*$ versus the hopping amplitude $\gamma$  with fixed diagonal term strength $1/2$ (both in units of $t_{AB}$) for three different models. The dashed line represents results for the Fibonacci model for $N=41$ atoms, while the blue and purple lines correspond to silver mean quasicrystal and the period doubling cases respectively. One sees that transparent states occur throughout the spectrum. 

Sire and Mosseri have investigated a model for approximant chains in which the hopping takes two values, $t_A$ or $t_B$, according to the Fibonacci sequence \cite{siremoss90}.  The onsite potentials whose values depend on the nature of bonds to the left and right, were taken to be $V_{AA}=-\lambda/2$ and $V_{AB}=V_{AB}=\lambda/2$. The model thus has two parameters $\lambda$ and $\rho$. By considering $n$th generation approximant chains, Sire and Mosseri showed that there are gap closings and quasi-extended states for two different families of solutions. In particular, they showed that as $n\rightarrow \infty$ there are Bloch-type extended wavefunctions that can be described in terms of a wave-vector $k \in (0,\pi)$. The energies of such states are distributed throughout the band, with the band edges corresponding to $k=0$ and $k=\pi$. Similar conclusions as to the existence of such extended Bloch-type states were reached by Kumar and Ananthakrishna \cite{PhysRevLett.59.1476kumar,kumar90}.
Extended states with periodic envelopes may exist in mixed systems can exist even for certain disordered cases, as reported in \cite{PhysRevB.58.739gong}.

\subsection{Interference and flux dependent phenomena}  Transmission properties of chains of Aharonov-Bohm rings of two different sizes, and connected in a Fibonacci sequence, have been studied \cite{PhysRevB.68.195417roemer}. The transport properties now become flux dependent. It was observed that transmission decreases as a power law in the number of rings and that there are resonant states for specific flux values. An RG analysis using the Landauer formalism and the trace map method shows that the transmission coefficient possesses a self-similar structure  \cite{PhysRevB.75.115130aharanovbohm}.

\section{Other quasiperiodic chains}\label{sec.other}

Related models of particular interest include a class of aperiodic 1D chains which can be obtained by generalizing the substitution rules we have introduced in Sec.\ref{subsec.inflation}. Higher dimensional lattices are briefly described. 

\subsection{ Aperiodic substitutional chains} Although we have focused on a single quasiperiodic system, described by the golden mean $\tau$, many of the methods used are generalizable to other irrationals. The nature of the irrational number (algebraic or not) is of primary importance for the geometric properties and in consequence for the electronic properties as well. As we have said, quasicrystals are a special class of structures based on  Pisot numbers. The so-called metallic means which are solutions of the equation $x^2-nx-1=0$ (n=1,2,..) belong in this category. Of this series, the two best studied members are the gold (n=1) and silver (n=2) mean quasicrystals, the latter also called the octonacci chain. These chains which have inflation properties and electronic structures analogous to those of the Fibonacci chain are reviewed in \cite{PhysRevB.62.15569yuan,PhysRevB.85.224205thiemschr, thiemschrJPC,thiemEPJB2011}.  A study of the multifractal exponents for the central $E=0$ state for metallic mean chains is done in \cite{mace2017critical}. 
Energy spectra of generalized Fibonacci-type quasilattices having self-similar as well as quasiperiodic structure were studied in \cite{PhysRevB.55.2882sritra}. A gap labeling theorem is shown to exist in these cases.

\begin{figure}[htp]
	\centering
	\includegraphics[width=.45\textwidth]{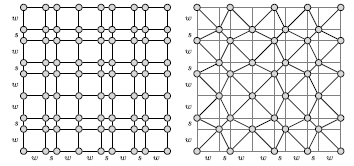}
	\caption{\small{Schemas of 2D product lattices hosting two different types of hopping models -- (left) direct product Hamiltonian with $t_A$ (long bonds)and $t_B$ (short bonds) (right) Labyrinth model (hopping along one of the diagonals of each plaquette)  (from \cite{thiemschrJPC} reproduced with permission)}}
	\label{fig.labyrinth} 
\end{figure}
Some well-known aperiodic, but not quasiperiodic, systems which can be obtained by substitution rules are the Thue-Morse, the period-doubling and Rudin-Shapiro sequences. We refer readers to \cite{Macia2005} for a discussion of their  electronic properties. 

\subsection{Products of chains} The 1D chain can be used as the basis for extensions to arbitrary dimensions $d$. Taking $d=2$, for example, the direct product $C_n \times C_n$ of two Fibonacci approximants aligned along $x$ and $y$ axes forms a 2D lattice of squares and rectangles \cite{lifshitzJAI}. As can be seen in Fig.\ref{fig.labyrinth}a) its connectivity is that of the square lattice. The energy spectrum and wavefunctions for tight-binding models on these direct product lattices have been studied. In the pure hopping vertex model, electrons can hop along the two directions with amplitudes $t_A$ or $t_B$. The Hamiltonian is separable into two independent Fibonacci chain problems. The energies are the sum of two 1D energies $E_{ij}=E_i+E_j$, and the corresponding wavefunctions given by the product $\psi_{ij}(x,y)=\psi_i(x) \psi_j(y)$ where $E_i$ and $\psi_i$ are solutions to the 1D problem. The properties of the spectrum  depend on the value of $\rho$. The spectrum of the $d=2$ product lattice is purely singular continuous for $\rho<\rho_1$ where $\rho_1\approx 0.6$ in the product lattice. For $\rho>\rho_1$, the spectrum has a continuous part \cite{sireEPL,lifshitz2008,thiemschrJPC}.  

\begin{figure}[htp]
	\centering
	\includegraphics[width=.5\textwidth]{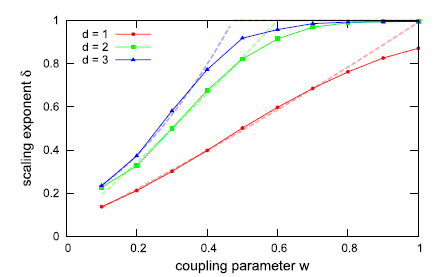}
	\caption{\small{Plot of the autocorrelation function exponent $\delta$ as a function of the strength of quasiperiodic modulation (the variable $w=t_A/t_B$ corresponds to $\rho$ of our text) for product lattices of dimensions 1,2 and 3 (from \cite{thiemschrJPC} reproduced with permission)}}
	\label{fig.deltaexpo} 
\end{figure}

The labyrinth model \cite{siremosssadoc}, a 2D variant based on the direct product of chains (see Fig.\ref{fig.labyrinth}b), also has properties derivable from the 1D solutions. 
The generalized dimensions describing multifractal properties of wavefunctions in $d$ dimensional product lattices was investigated in \cite{PhysRevB.62.15569yuan, 
thiemEPJB2011}. The exponents for $d$-dimensions are simply proportional to the 1D exponents, $D^{\psi,d}_q = d D^{\psi,1}_q $. Dynamical exponents have been computed for these lattices \cite{zhongmoss,PhysRevB.85.224205thiemschr}. The return probability exponent shows a $d$ dependence, $\gamma^{d}=d\gamma^{(1)}$. The diffusion exponent $\beta$ is expected according to theory to be constant as the dimensionality $d$ increases and this is indeed found numerically, as can be seen in Fig.\ref{fig.betavals}. The autocorrelation function exponent $\delta$ depends on the dimensionality, and for higher $d$ increases faster as a function of the modulation strength parameter $w=t_A/t_B$ as can be seen in Fig.\ref{fig.deltaexpo}. A complete account of the electronic properties of such $d$-dimensional generalizations is given in \cite{thiemschrJPC}.

{\it {$d$-dimensional tilings}}  The spatial connectivity of the product lattice systems is simple in that they have an underlying average periodic structure:  $d$-dimensional hypercubic lattices. The electronic properties of such product lattices are seen to be ``inherited" from the parent chains. The situation is different for other quasiperiodic tilings. For 2D and 3D tilings, such as the Penrose tiling, spectra and wavefunctions remain difficult to compute analytically, with the exception of the ground state \cite{mace2017critical}. As for the Fibonacci quasicrystal, nontrivial topological properties are to be expected in these higher dimensional cases. The possibility of higher order topological insulators based on the Penrose and octagonal tilings has been discussed, for example, in  \cite{PhysRevLett.116.257002topol,PhysRevLett.124.036803topol}.

\section{Interactions and quasiperiodicity }\label{subsec.interactions} The topic of interacting quasiperiodic systems requires a separate review. This section will be restricted to giving brief outlines of some of the main contributions, along with a non-exhaustive list of references. 

The effects of quasiperiodic perturbations in interacting fermionic chains was investigated by Vidal et al for continuum models using renormalization group in \cite{PhysRevLett.83.3908vidalprl,PhysRevB.65.014201vidalprb}. Considering in particular the case of metallic mean chains, they found that there was a metal-insulator transition \cite{PhysRevB.65.014201vidalprb} for repulsive interactions. Hiramoto \cite{hiramotohartree} did a Hartree-Fock analysis to study the effect of weak interaction U. The study showed that the singular continuous single particle spectrum persists in the presence of interactions, in contrast to the critical Harper model where the singular continuous behavior is destroyed by U. 

In a study of  the Hubbard model on a Fibonacci chain by weak-coupling renormalization group and density matrix renormalization group methods \cite{PhysRevLett.86.1331hida}, Hida showed that, for the diagonal Fibonacci model, weak Coulomb repulsion is irrelevant in the sense of RG and the system will behave as a free Fibonacci chain. For strong Coulomb repulsion the system becomes a Mott insulator and, in the spin sector, can be modeled in terms of a uniform Heisenberg antiferromagnetic chain. For the off-diagonal case, he obtained a Mott insulator with a low energy sector that could be described in terms of a Fibonacci antiferromagnetic Heisenberg chain. Gupta et al \cite{hubbardfibo} studied the DC electrical conductivity for half filling, using Hartree-Fock mean field theory, to see the interplay of interactions and quasiperiodicity. They concluded that, while each of these factors taken individually tend to $decrease$ the conductivity, there may be enhancement of the conductivity due to competition between them.

\begin{figure}[htp]
	\centering
	\includegraphics[width=.5\textwidth]{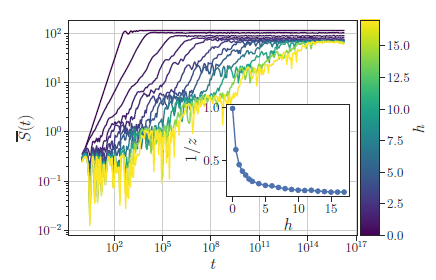}
	\caption{\small{Entanglement entropy as a function of time for different values of the strength of quasiperiodic modulation $h$ (black curve: periodic chain, yellow (light grey) chain: strongly quasiperiodic chain). The inset shows the dependence of the power $z$ on $h$ (reproduced with permission from \cite{scipost})}}
	\label{fig.entanglement} 
	
\end{figure}

The evolution of multifractality in an interacting fermion chain was studied in \cite{scipost}. Contrary to naive expectations, these authors found that adding repulsive interactions did not lead to enhanced delocalization.  Fig.\ref{fig.entanglement} shows the half chain von Neumann entropy plotted against time for different strengths of the quasiperiodic potential (controlled by a parameter $h$, with $h=0$ being the periodic case, and $h=1$ the strongly quasiperiodic chain). For a periodic chain, (black curve) the entanglement
entropy grows as a power-law in the time, $ S(t)\sim t^{1/z}$. The exponent $z$ increases as the strength of the quasiperiodicity $h$ is increased. This further confirms that the free Fibonacci chain is intermediate between the free delocalized state and
an Anderson localized state, from the point of
view of its transport properties. One also sees log-periodic oscillations superposed on the power-law behavior, especially visible for the yellow curve (strong quasiperiodicity limit). 

\subsection{Heisenberg and XY chains}\label{subsec.spinchains}
The properties of spin chains with Fibonacci couplings have been investigated by a number of methods. Hida analyzed the spin 1/2 Heisenberg model using density matrix renormalization group finding that quasiperiodic modulations are relevant in this case and that the ground state is in a different universality class from that of the XY chain (or free particle) problem \cite{hermisson,hidaJPSJ}. The entanglement entropy $S$ of aperiodic critical chains  was studied by Igloi et al \cite{Igl_i_2007igloi}. For these chains, the half chain entanglement increases as $S \sim \frac{c}{3} \ln_2 L + cst$, where $L$ is the chain length and $c$ is the central charge (=0.5 for the periodic case). They found that for Fibonacci XY chains the quasiperiodic modulation is marginal, in that the central charge $c$ in this case is non-universal and depends on $\rho=J_A/J_B$ (the ratio of spin-spin couplings).  For Fibonacci Heisenberg chains, the quasiperiodicity is strongly relevant and the prefactor is given by $c(0)\approx 0.8$.

{\subsection{Many body localization}} The question of many body localization due to quasiperiodic potentials was raised in \cite{PhysRevB.87.134202iyer,PhysRevLett.119.075702khemani}. These studies asked whether the MBL transitions are different in the presence of quasiperiodic potentials compared to random potentials, and if so in what ways. The MBL transition in the quasiperiodic AAH model, which can be experimentally realized in interacting boson and fermions cold atoms systems, has been studied, and shown to lead to a new type of ``non-random" universality class \cite{PhysRevLett.119.075702khemani}. It is interesting to ask whether there are any significant differences between many body localization in AAH and in Fibonacci chains. Details of the transition have been discussed in \cite{scipost,PhysRevB.100.085105varmaznid}. Density-density correlations at infinite temperature have been investigated using the dynamical quantum typicality approach. This allows to study the evolution of dynamical properties of the Fibonacci model as interaction strength is ramped up, and the crossover to a MBL state.

{\subsection{Anomalous diffusion properties}} Settino et al \cite{PhysRevB.101.144303logullo} have studied dynamics in interacting aperiodic many body systems including the FC . They showed that, for the onsite Fibonacci model, the singular continuous spectrum for the non-interacting problem remains and induces an anomalous dynamics.
 Lo Gullo et al  \cite{PhysRevE.96.012111logullo} studied aperiodic discrete time quantum walk problems, relevant in quantum computing, and which could be realized using optical fibers \cite{borrelli}. They computed the energy spectra and the spreading of an initially localized wave packet for different cases, finding that in the case of Fibonacci and Thue-Morse chains the system is superdiffusive, whereas for Rudin-Shapiro, another substitutional chain, it is strongly subdiffusive. They propose that the different dynamics are linked to the nature of the spectra in the two cases -- singular continuous for the former, discrete for the latter.

\section{Experimental systems} \label{sec.experimental}
\begin{figure}
		\includegraphics[width=0.35\textwidth]{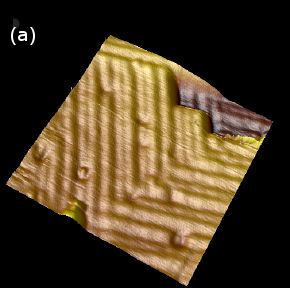}
		\includegraphics[width=0.35\textwidth]{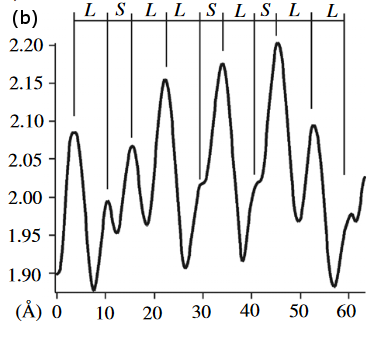}
\caption{a) STM image of a $100 \times 100$A square zone of copper atoms deposited on the 5-fold surface of i-AlPdMn b) height profile between points marked X showing an alternating sequence of distances L(7.3 A) and S(4.5A) (figure reproduced from \cite{mcgrath}}
\label{fig.copper}
\end{figure}

Fibonacci sequences occur $naturally$ in 3D icosahedral quasicrystals and also in dodecagonal quasicrystals. These  are structures which are based on the golden mean. Fig.\ref{fig.copper}a which shows an STM image of copper adatoms deposited on an icosahedral AlPdMn quasicrystal \cite{mcgrath} provides a striking illustration of this connection. As the height profile in Fig.\ref{fig.copper}b shows, the distances between columns is a Fibonacci sequence. These rows of aperiodically spaced layers are coupled to the bulk, and the resulting Hamiltonians are likely to be fairly complicated. For experimental investigations of the 1D model, it is therefore useful to fabricate artificial systems in order to study the Fibonacci chain, and we will now describe a few such systems.

The off-diagonal tight-binding Fibonacci model has been experimentally realized in a polaritonic gas in a quasi-1D cavity \cite{tanese2014fractal,baboux2017measuring}. Some of the theoretical predictions for the energies and the eigenmodes of this system were observed. The discrete scale invariance of the spectrum and the gap labeling theorem were thus experimentally verified. The existence of gap states was checked, and their spatial structure mapped out. Most interestingly, their topological winding numbers could be experimentally measured by varying the phason angle $\phi$. 

Optical waveguides fabricated by using a femtosecond laser beam to inscribe quasiperiodic spatial modulation in a bulk glass have been studied in \cite{PhysRevLett.109.106402kraus}. Injecting light into these allowed to study propagating and localized modes, and to demonstrate the topologically protected edge modes predicted by theory. Topological pumping of photons was demonstrated in \cite{PhysRevB.91.064201zilberberg}, where the topological equivalence between the Harper and the Fibonacci models was experimentally verified. Related subjects of recent investigations, outside the scope of this review, are higher order topological insulators using quasicrystals \cite{PhysRevLett.124.036803topol} and topological quantum computation using Fibonacci anyon chains \cite{PhysRevLett.98.160409feiguin, PhysRevB.101.075104chandran}.

Merlin et al\cite{PhysRevLett.55.1768merlin,PhysRevB.36.4555merlin} fabricated semiconductor superlattices following a Fibonacci sequence and studied their  properties using Raman spectroscopy. The samples were composed of two types of films : 27 nm-thick layers of GaAs and 43 nm-thick layers of GaAlAs (such that the ratio of thicknesses is close to the golden mean). These layers were piled on top of each other in a quasiperiodic sequence along the $z$ direction. The Raman frequency shifts were compared for periodic and for quasiperiodic structures. In addition to the acoustic phonons in this system, plasmon-polariton modes are argued to play an important role in \cite{cottam}, where their Raman cross-sections for the samples are discussed. Hawrylak et al \cite{PhysRevB.36.6501quinn} have computed the plasmon spectrum for such superlattices, described its scaling properties and explicitly computed the $f(\alpha)$ spectrum. For a wide-ranging discussion of photonic and phononic heterostructures see the review by \cite{SteurerSutter}.

A variety of photonic structures can be obtained by, for example, coupling single mode waveguides to form lattices, through dynamical coupling or using nonlinearities. It can be useful to consider these in terms of effective models in a higher dimensional ``synthetic space" in which the external parameters play a role of extra dimensions. These systems offer in particular a means to investigate the topological properties of the Su-Schreiffer-Heeger (SSH) and the AAH models \cite{yuanreview}, and leave open the possibility of extensions including the Fibonacci chain.  

Metamaterials in the nanoscale also provide a wide array of possibilities for aperiodic structures. Optical transmission spectra of photonic band-gap Fibonacci quasiperiodic nanostructures composed of both positive (SiO2) and negative refractive index materials are discussed by de Medeiros et al \cite{medeiros}. In another direction, there are possibilities to make quasiperiodic sequences in 1D biomaterial (DNA-based) systems and study the consequences for transport \cite{PhysRevE.71.021910vasconselos}.

Magnetic multilayers composed of Fe/Cr layers and studied by MOKE (magneto-optic Kerr effect) and FMR (ferromagnetic resonance) should display self-similar magnetization versus curves and  interesting thermodynamic signatures \cite{cottamFeCr}. Another promising system is composed of epitaxially grown layers of Fe and Au using ultra-high-vacuum vapor deposition \cite{suwa} which were theoretically predicted to have anomalous magnetoresistance \cite{PhysRevB.85.224416machado}. 

Fibonacci nanowire arrays have been studied recently by Lisiecki et al who discuss their magnonic properties and interest for possible applications \cite{PhysRevApplied.11.054061magnon}.

Quantum dots can be used to make artificial crystals and quasicrystals \cite{PhysRevLett.65.361kouwenhoven}. They have already been used to study magnetotransport in a periodic crystal. It may therefore be possible to study transport in artificial Fibonacci chains made with quantum dots.

On a macroscopic length scale, microwave propagation in dielectric resonators have been used with success to simulate the tight-binding model for graphene \cite{PhysRevB.88.115437mortessagne}. Preliminary work \footnote{F. Pi\'echon and F. Mortessagne, private communication.} shows that this may provide an extremely versatile system in which to study electronic properties of Fibonacci chains, including effects of various forms of disorder or interactions.

Cold atoms in optical potentials constitute a particularly fertile ground to realize quasiperiodic models and study their properties under controlled conditions. A number of recent theoretical studies have thus looked at generalizations of the tight-binding models that are relevant to cold atom experiments. 
The discrete-valued Fibonacci potential is more difficult to realize experimentally, compared to the Harper model, which  can be realized by applying an incommensurate laser potential \cite{PhysRevA.75.061603inguscio,PhysRevLett.98.130404fallani}. Recently, however, Singh et al \cite{PhysRevA.92.063426paramesh} proposed a means of realizing generalized Fibonacci models on chains based on the cut-and-project method, in a 2D optical lattice. If realized, this would provide opportunities to study experimentally multifractal states and probe the multiscale dynamics in the Fibonacci quasicrystal. An interesting direction discussed in \cite{PhysRevB.94.035131nussinov} concerns emergent quasiperiodic structures in interacting systems. The authors show that fractional quantum Hall systems for irrational filling fractions can result in quasiperiodic electronic configurations which include, in 1D, the case of an emergent Fibonacci ordering. The authors argue that these could be realized with ultracold Rydberg atoms on optical lattices. Such structures, although not expected to be stable under disorder or quantum/thermal fluctuations, could nevertheless persist on intermediate length scales in the form of ``quasicrystalline puddles".

\section{Summary and Outlook} \label{sec.conclusions}
We have introduced some of the main concepts and techniques relevant to the 1D Fibonacci tight-binding Hamiltonians. This class of model is important from a fundamental viewpoint in its own right, but also as a starting point for understanding higher dimensional quasicrystals. The fascinating topological characteristics of the 1D models arising from their ``hidden dimension" can be generalized to higher dimensional quasicrystals. Thus, just as the Fibonacci chain has topological properties inherited from a parent 2D system, similarly, certain 2D quasicrystals have been shown to have topological invariants corresponding to a 4D quantum Hall system \cite{krausPRL} with associated edge modes which are symmetry protected.

Interpolating between the AAH and Fibonacci models  offers a possibility of realizing topological pumps, as in the optical waveguide systems of \cite{PhysRevB.91.064201zilberberg}. More recently it was shown that polaritonic quasicrystals, also allow this type of tunability \cite{blochZilber}. For these systems one can control localization-delocalization transitions for specific bands, suggesting the possibility of their use in selective band-pass filters.

One of the distinctive characteristics of the quasicrystal is the existence of multifractalities as a function of the energy, of the space coordinates and of temporal correlations. We have discussed these properties, along with explicit calculations for specific examples. We have mentioned a few consequences of these multifractal states for physical properties : transport, disorder induced localization/delocalization and the proximity effect. Critical states persist in higher dimensional models. There exists notably an exact solution for ground states which is a 2D analogue of the $E=0$ solution on the FC that was discussed here \cite{kaluginkatz, mace2017critical}. 

The Fibonacci family of models is interesting from the point of view of applications as well : in electronic devices or for their mechanical properties. Although we focused here on electronic properties, there are many interesting and closely related problems for electromagnetic wave modes in aperiodic media. The unique optical reflectivity properties of aperiodic multilayers suggest applications as perfect mirrors with omnidirectional reflectivity for all polarizations of incident light over a wide range of wavelengths \cite{Axel_2010}, and more generally in nanodevices \cite{SteurerSutter,maciawaves}. The experimental possibilities to create phononic systems suggest their use in thermal and acoustic shields, or in acoustic lenses, for example. In superconductors, the critical currents versus field of quasiperiodically spaced pinned vortices were computed by \cite{PhysRevB.74.024522misko} with a view to applications, and it was observed that for a 1D Fibonacci vortex array, the critical currents have a self-similar structure. More speculatively, one can conceive of applications on larger scale suggested by some of the properties discussed here. The large number of spectral gaps generically present in this family of models suggests the possibility of dissipating ocean waves and even reflecting tsunamis on the shoreline using quasiperiodic arrays of scatterers. Seismic barriers against propagating Rayleigh waves (surface seismic modes) using, for example quasiperiodic trenches, are another interesting possibility.

The models discussed here hopefully already provide a useful framework for understanding 1D quasiperiodic systems. Many questions still remain, however. The multifractal dynamical properties of quasiperiodic chains remain to be elucidated in more detail both theoretically and by experiment. The effects of interactions and their interplay with quasiperiodicity is an important problem requiring more investigations. The extension to higher dimensional quasicrystals is a major challenge. As far as understanding and controlling physical properties of quasicrystals is concerned, we have only scratched the surface.

\begin{acknowledgements}
I gratefully acknowledge many valuable discussions over the years with Michel Duneau, Pavel Kalugin, Jean-Marc Luck, Nicolas Mac\'e and Fr\'ed\'eric Pi\`echon.  
\end{acknowledgements}

\bibliography{Bibliography}

\end{document}